\documentclass[aps,prd,twocolumn,superscriptaddress,nofootinbib]{revtex4-2}

\usepackage{graphicx}% Include figure files
\usepackage{dcolumn}% Align table columns on decimal point
\usepackage{bm}% bold math
\usepackage{hyperref}% add hypertext capabilities
\usepackage[mathlines]{lineno}% Enable numbering of text and display math
\usepackage{amsmath}
\usepackage{color}
\usepackage{xspace}

\newcommand{\todooptional}[1]{} %
\DeclareMathOperator{\sinc}{sinc}

\usepackage{lineno}

\begin{document}
% \linenumbers
\title{\textbf{The Atacama Cosmology Telescope: DR6 Constraints on Isotropic Time-Dependent Cosmic Birefringence from Axion-like Particles} 
}% 

\author{Zachary B. Huber}
 \affiliation{Department of Physics, Cornell University, Ithaca, NY 14853, USA}
 %\orcidlink{0000-0003-4573-4094}

\author{Huajian Yang}
 \affiliation{Department of Physics, Cornell University, Ithaca, NY 14853, USA}
 %\orcidlink{0009-0004-7639-1025}

\author{Steve K. Choi}
 \email{Contact author: schoi@ucr.edu}
  \affiliation{Center for Experimental Cosmology and Instrumentation, Department of Physics and Astronomy, University of California, Riverside, CA 92521, USA}
  %\orcidlink{0000-0002-9113-7058}

\author{Erminia Calabrese}
  \affiliation{Department of Physics and Astronomy, School of Physical, Chemical and Environmental Sciences, Cardiff University, The Parade, Cardiff, Wales, CF24 3AA}
  %\orcidlink{0000-0003-0837-0068}

\author{Vera Gluscevic}
  \affiliation{Department of Physics and Astronomy, University of Southern California, Los Angeles, CA 90089, USA}
  %\orcidlink{0000-0002-3589-8637}

\author{Sigurd Naess}
  \affiliation{Institute for Theoretical Astrophysics, University of Oslo, Norway}
  %\orcidlink{0000-0002-4478-7111}

\author{Michael D. Niemack}
  \affiliation{Department of Physics, Cornell University, Ithaca, NY 14853, USA}
  \affiliation{Department of Astronomy, Cornell University, Ithaca, NY 14853, USA}
  %\orcidlink{0000-0001-7125-3580}

\author{Lyman A. Page}
  \affiliation{Joseph Henry Laboratories of Physics, Jadwin Hall, Princeton University, Princeton, NJ 08544, USA}
  %\orcidlink{0000-0002-9828-3525}

% \author{Bruce Partridge}
%   \affiliation{Department of Physics and Astronomy, Haverford College, Haverford, PA 19041, USA}
%   %\orcidlink{0000-0001-6541-9265}

\author{Alexander van Engelen}
  \affiliation{School of Earth and Space Exploration, Arizona State University, Tempe, AZ 85287, USA}

\author{Edward J. Wollack}
  \affiliation{NASA/Goddard Space Flight Center, Greenbelt, MD 20771, USA}
  %\orcidlink{0000-0002-7567-4451}

% \author{ACT Collaboration}
%   \noaffiliation

\date{\today}% It is always \today, today,
             %  but any date may be explicitly specified

\begin{abstract}
Axion-like particles (ALPs) with sufficiently low masses are dark matter candidates because they could account for much or all of the observed dark matter energy density and help explain observations of galactic substructure. Such pseudoscalar fields could couple to photons, causing a time-dependent rotation of the polarization of the light with an oscillation frequency equal to the ALP mass. We use measurements of the polarization of the cosmic microwave background (CMB) by the Atacama Cosmology Telescope (ACT) between 2017 and 2022 to search for this effect using an EB power spectrum estimator. By probing oscillation periods from days to years, the data constrain the coupling between ALPs and photons for ALP masses between $2.5\times 10^{-23}$ eV and $4.8\times 10^{-20}$ eV. We find no evidence for such a signal and set a 95\% upper bound on the axion-photon coupling constant, $g_{\phi\gamma} < 1.1\times10^{-11}\left({m_\phi}/{10^{-21}\,\mathrm{eV}}\right)\,\mathrm{GeV}^{-1} .$ This limit assumes that ALPs make up all of the dark matter and that the ALP field amplitude is stochastically distributed, which leads to more conservative bounds than the deterministic-field assumption used in some previous CMB polarization analyses. Among searches for ALP-photon coupling using CMB polarization, we set the strongest bounds to date under the stochastic-field assumption. This work develops a methodology to analyze over 10,000 time-sliced ``depth-1'' polarization maps from ACT, a data product that will be standard for time-domain astrophysics with the Simons Observatory, enabling searches with a higher sensitivity and observing span.

\end{abstract}

\maketitle

\section{Introduction}
\label{sec:introduction}

Measurements of the cosmic microwave background (CMB) \cite{Planck2018_params_2020} and the large scale structure of the universe \cite{Alam_BOSSDR12clustering_2017} provide ample evidence for an additional matter-like energy density in the universe beyond ordinary baryons: dark matter. Understanding the nature of this component is a major priority for modern cosmology and particle physics. 
% \draftnote{Could cite \cite{Bertone_DMreview_2005} if a generic DM review is needed.}

Among the many models attempting to explain dark matter in terms of a new particle beyond the Standard Model, much effort has gone into articulating and testing theories that include an ultralight axion-like particle (ALP). The term axion was initially associated with the QCD axion \cite{Weinberg_QCDaxion_1978, Wilczek_QCDaxion_1978}, a proposal to resolve the strong CP problem in quantum chromodynamics. While the QCD axion would require specific properties and couplings to the Standard Model, more generic ALP models with similar behavior but a wide range of possible masses and coupling strengths appear in theories beyond the Standard Model \cite{Svrcek_Witten_axion_stringtheory_2006, string_axiverse_2010}. With sufficiently large occupation numbers, even ALPs with masses many orders of magnitude below an eV could account for all the dark matter in the universe (see Ref.~\cite{marsh_axion_2016} and its references for a review of models and cosmological implications).

The standard QCD axion would have a mass larger than approximately $10^{-12}$ eV. A wide variety of experiments test axions and ALPs through their coupling to photons or other Standard Model particles. Selected examples include converting solar ALPs to photons with strong magnetic fields \cite{anastassopoulos_castlimit_2017}, light-shining-through-walls experiments \cite{alps_lightthroughwall_2010, osqar_lightthoughwall_2015}, microwave cavity searches \cite{admx_alp_cavity_2024}, and vacuum birefringence searches \cite{ejlli_pvlas_2020}.

If the ALP mass is much smaller than the scale of the QCD axion mass, the de Broglie wavelength becomes astrophysical in size, leading to a variety of effects. For ALP masses less than roughly $10^{-27}$ eV, the particle wavelength spans cosmological scales; such ALPs would function as dark energy instead of dark matter, but they are strongly constrained by studies of their effect on the CMB power spectrum \cite{hlozek_cmb_2018, hlozek_cmb_2015} (see also Section 6.3 of Ref.~\cite{calabrese_actdr6_models_2025}). The parameter space of masses several orders of magnitude more massive has received intense interest because ALPs with masses of $10^{-23}$--$10^{-21}$ eV could function as fuzzy dark matter (FDM) \cite{Hu_FDM_2000,Hui_fdm_2017}. FDM models modify structure formation on small scales through the wave-like effects of ALPs with de Broglie wavelengths of order a kiloparsec, making observations of galactic substructure a sensitive probe of such models. More recently, several works have argued for a stronger lower bound on the ALP mass for ALPs comprising all of the dark matter, with the precise bounds depending on the production scenario and the observable considered \cite{Chathirathas2025, Nadler2026, GarciaGallego2026, DalalKravtsov2022}.

One way of constraining ultralight ALP models involves measuring small-scale structure formation over time. Example observables include the effect of ALPs on Lyman-$\alpha$ emission \cite{irsic_alp_lymanalpha_2017}, the populations of Milky Way satellite galaxies \cite{des_alp_milkywaysatellites_2021}, and galaxy rotation curves \cite{bar_gal_rotation_curves_2022}.

Axion-photon interactions can convert axions into photons or vice versa. This effect is studied in direct detections experiments like the CERN Axion Solar Telescope (CAST) that search for conversions between solar axions and photons \cite{anastassopoulos_castlimit_2017,cast_axionlimit_2024} as well as various astrophysical measurements including searches for excess gamma rays from ALPs produced in supernovae \cite{payez_sn1987a_gamma_2015} or alterations to the X-ray spectra of active galactic nuclei (AGN) \cite{marsh_xray_m87_2017} or quasars \cite{sisk-reynes_xray_quasar_2022}, to name only a few.

Another common method in this mass range is to search for ALP-induced birefringence effects across cosmic scales. Any generic Chern-Simons coupling between a light pseudoscalar boson and electromagnetic fields generates rotation in linearly polarized light as it interacts with the ALP field \cite{Carroll1990}. The absence of the rotation of light from polarized sources in regions rich in dark matter places constraints on how strongly ALPs couple to photons. Example studies in this category include pulsar timing and polarization \cite{caputo_alp_pulsars_2019,castillo_alp_pulsars_2022,xue_ppta_alp_2024}, the polarization of AGN jets \cite{ivanov_alp_agn_pol_2019}, and the polarization variability of polarized millimeter-wave sources like Tau A \cite{polarbear_taua_2024}. Many of the methods cited here involve extensive modeling of the source environment and Milky Way. CMB polarization provides a complementary probe with different astrophysical systematics.

In particular, the CMB is a clean source for studies of the interaction between ALPs and photons, as pointed out by Fedderke, Graham, and Rajendran \cite{fedderke_axion_2019} (hereafter referred to as F19). The authors discuss two main effects of ALP-photon interactions on measurements of CMB polarization: the ``washout" effect and the ``AC oscillation" effect. In the former, variations in the ALP field strength over the period of the decoupling of the CMB can cause a range of small rotation angles that sum incoherently and reduce the overall amplitude of the CMB polarization. This small amplitude reduction would appear in both the TE and EE spectra, and F19 used Planck data to tightly constrain this effect.

The AC oscillation effect refers to a rotation of the Q/U or E/B polarization angle as a function of time due to the oscillation of the ALP field in the Milky Way. Because the time-dependent signal arises from the local ALP field at the observer, which is common to all lines of sight, this polarization angle oscillation should be both time-dependent and isotropic, distinguishing this measurement from searches for constant cosmic birefringence, whether isotropic or anisotropic, as described in Refs.~\cite{louis_DR6spectra_params_2025, namikawa_act_aniso_biref_2020,kramer2026}. It should also be coherent across the timescales of CMB experiments and have the same phase across the sky. The frequency of the oscillation is directly related to the ALP mass, and the period of oscillation falls between a few hours to a few years for ALP masses between $5\times10^{-18}$ and $1\times10^{-23}$ eV. These time scales align well with regular measurements of CMB polarization by active ground-based experiments, and constraints on ALP-photon coupling in this mass range have been placed by BICEP/Keck \cite{bicepkeck_xii_2022,bicepkeck_xiv_2022} (hereafter referred to as BXII and BXIV, respectively), SPT-3G \cite{spt_alp_2022} (hereafter referred to as S22), and POLARBEAR \cite{polarbear_axionCMB_2023} (hereafter referred to as P23) using this method. Of these, BXII, BXIV, and S22 report deterministic limits, assuming a fixed local ALP field amplitude, while P23 presents stochastic limits, marginalizing over the unknown amplitudes of individual wave modes in the local ALP field, which generally results in a more conservative constraint on the coupling.

This paper presents our search for this effect in polarization maps from data release 6 (DR6) of the Atacama Cosmology Telescope (ACT). In Section~\ref{sec:act}, we introduce our data in greater detail. Section~\ref{sec:method} outlines the analysis method, Section~\ref{sec:nulls} describes tests for systematic effects in our analysis and data, and Section~\ref{sec:results} provides both our stochastic and deterministic constraints on rotation angle oscillations and ALP-photon coupling. We conclude with a discussion of the implications of our results in Section~\ref{sec:conclusion}. Appendix~\ref{sec:derivations} gives the derivations of our estimator and analytic variances, while Appendix~\ref{sec:fisher_info} presents the simulations and calculations used to derive the stochastic limits.

\section{The Atacama Cosmology Telescope}
\label{sec:act}
The Atacama Cosmology Telescope was a six-meter, off-axis Gregorian telescope that observed from 2007 to 2022 on Cerro Toco in the Atacama Desert in northern Chile. ACT observed with three different cameras over its lifetime, culminating in the Advanced ACTPol detector arrays which were installed in 2016--2017. The intensity and polarization measurements made with this receiver from 2017--2022 cover over 19,000 square degrees on the sky at roughly arcminute resolution and were recently released as ACT Data Release 6 (DR6). The data collection, selection, and mapmaking procedure for DR6 is described in detail in Ref.~\cite{naess_actdr6_maps_2025} (hereafter referred to as N25).

The DR6 data were collected with three modular optics tubes known as polarization-sensitive arrays (PAs): PA4, PA5, and PA6\footnote{PA1, PA2, and PA3 refer to the optics tubes for the earlier ACTPol receiver. The results for ACTPol were released in DR4 and DR5.}~\cite{Henderson2016, Ho2017, Choi2018}. Each PA contained dichroic transition-edge sensors with passbands spread across three frequency bands centered at 90, 150, and 220 GHz. We will refer to these as the f090, f150, and f220 bands. PA4 covered f150 and f220, and PA5 and PA6 covered f090 and f150. While PA4 and PA5 observed for the full duration of DR6, PA6 was replaced at the beginning of 2020 with PA7, which observed in bands centered at 30 and 40 GHz~\cite{Li2021}.

In order to search for time-varying birefringence due to the impact of ALPs on CMB polarization, we need to group the data into measurements over short timescales. While such a measurement could be made by examining the time-ordered data directly (as originally envisioned by F19), we work instead with the short duration maps known as ``depth-1 maps" (read ``depth one") that were already produced for other astrophysical time-domain analyses with DR6 \cite{orlowski2024,biermann2025}. As described in Section 3.7 of N25, these maps constitute the data from a single continuous scan at constant elevation while the sky drifts over the telescope. Depending on the details of the observation and data cuts, the depth-1 maps range in their duration and sky area from as little as 15 minutes covering around 100 sq.~deg.~to several hours of data covering around 2900 sq.~deg. The maps cover varying portions of the larger ACT DR6 footprint given in Fig.~1 of N25, effectively sampling a wide range of locations and shapes on the sky. An example depth-1 total intensity map is shown in Fig.~14 of N25. 

Since we expect the ALP polarization rotation signal to be isotropic and coherent for the mass range we consider, the different shapes and sizes of the depth-1 maps across a wide sky area are not a problem. We expect this signal to appear primarily in CMB-dominated bands, so we restrict our analysis to f090 and f150 depth-1 maps. See N25 for further details about the depth-1 mapmaking and data selection.

\section{Method}
\label{sec:method}

Our analysis processes more than 20,000 ACT depth-1 maps with $\sim 0.5'$ pixel size. In previous ACT analyses, only the temperature depth-1 maps have been used extensively, through map-space analyses. Extending this to power spectrum analyses (in polarization) requires a method that is fast enough to handle the full data volume while filtering the usual contaminated Fourier modes in each map. With this constraint, we aim to measure the time variation of a polarization rotation angle through an EB power spectrum estimator. Unlike the maps used in previous ACT cosmological analyses, however, each depth-1 map contains only a single scan and therefore cannot be divided into further temporal data splits, so a naive auto-spectrum estimator for each observation is vulnerable to noise bias. We therefore construct a new estimator based only on cross-correlations: the difference of two EB cross-correlations with a reference map, which is given by the full coadded map over the full duration of DR6. As shown below, this estimator is sensitive to the {\it angle difference} between each depth-1 map and the reference map, allowing us to track that relative angle over time. Of equal importance, it also allows us to derive an analytic expression for the uncertainty of the estimator itself.

\subsection{Polarization Rotation}
\label{sec:pol_rot}

Equation 54 of F19 gives the full form of the AC oscillation effect for the CMB in the presence of an ALP field that couples to photons via a Chern-Simons term in the Lagrangian. The time-varying polarization rotation angle is the argument of the complex exponential and has the form,
\begin{equation}
    \theta = \frac{g_{\phi\gamma}}{2}\phi_0\cos(m_\phi t + \alpha),
    \label{eq:rotation_angle}
\end{equation}
where $g_{\phi\gamma}$ is the ALP-photon coupling constant, $\phi_0$ is the local ALP field amplitude, $m_\phi$ is the mass of the ALP field in natural units with $\hbar=c=1$, and $\alpha$ is the phase of field oscillations. We parameterize this rotation in terms of an amplitude $A$, frequency $f$, and phase $\alpha$:
\begin{equation}
    \theta = A\sin(2\pi f t + \alpha).
    \label{eq:sine_parameterization}
\end{equation}
For each depth-1 map, we measure a rotation angle, generating a timestream covering over five years of data. Sine waves of the form of Eq.~\ref{eq:sine_parameterization} are fit to this timestream to extract the best fit amplitude and phase for many different frequencies, with each frequency corresponding to a single ALP mass. The measured oscillation amplitude determines the level of ALP-photon coupling.

\subsection{Angle Estimator}
Because the depth-1 maps vary considerably in size, shape, and location on the sky, we calculate the rotation angle for each depth-1 map\footnote{For convenience, we will speak of each depth-1 observation that generates a T, Q, and U map as a ``depth-1 map" except when we need to refer to one of the individual components directly. The same convention is used for the reference maps.} relative to a reference map using an EB power spectrum estimator of the form
\begin{equation}
    \Delta\widehat{C}_{\ell}^{EB} = \widehat{C}_{\ell}^{' E_{1} \times B_{2}} - \widehat{C}_{\ell}^{' E_{2} \times B_{1}},
\end{equation}
where primes denote rotated quantities calculated from the data, as opposed to `primordial' or fiducial spectra without any rotation. For our analysis, index 1 is a depth-1 map, and index 2 is the reference map. We use per-frequency coadded Q and U maps of all night-time DR6 data as our reference maps, which have negligible data overlap and correlation with each depth-1 map. These reference maps are masked to cover the same footprint as each depth-1 map for calculating the cross-spectra.

This estimator is proportional to the difference between the rotation angles in the two maps given by
\begin{equation} \label{eq:estimator}
    \Delta\widehat{C}_{\ell}^{EB} = C_{\ell}^{EE} \sin{[2(\theta_1 - \theta_2)]},
\end{equation}
where $\theta_1$ and $\theta_2$ are the rotation angles in maps 1 and 2, respectively, and $C_{\ell}^{EE}$ is the fiducial EE spectrum from Planck 2018 cosmology. This relation holds under standard cosmological assumptions with $C_{\ell}^{EE} \gg C_{\ell}^{BB}$ and intrinsic $C_{\ell}^{EB} = 0$. The full derivation is found in Appendix~\ref{sec:estimator_deriv}. The AC oscillation that we seek to measure is insensitive to a constant offset in the polarization rotation angle, so measuring the relative angle between each depth-1 map and the reference map is sufficient. This estimator also allows the covariance to be worked out analytically, without the computational expense of large simulation suites. The covariance is derived in Appendix~\ref{sec:cov_deriv} in terms of all measured power spectra. 

Simulations with a representative noise level from our depth-1 maps confirm that this method can recover an injected angle within the uncertainty of the measurement.

%Since each depth-1 map contains only a single scan, there is no way to split our data to provide an unbiased estimate of $C_{\ell}^{EB}$ if we were calculating a standard EB rotation angle from each depth-1 map alone. Adding a reference map allows us to take cross-spectra that eliminate this noise bias at the expense of calculating the angle relative to the reference map instead of an absolute rotation angle for each depth-1 map. The AC oscillation effect described in F19 is not sensitive to a constant offset in the polarization rotation angle, however, so measuring the timestream of a relative angle is sufficient for our needs. %(\draftnote{does it help us at all with mitigation of systematic instrumental effects? Should we be concerned at all about whether there are different angles across different parts of the coadd map? - remove comment MDN}). 
%Simulations with a representative noise level from our depth-1 maps confirm that this method can recover an injected angle within the uncertainty of the measurement.

\subsection{Power Spectra Calculation}
\label{sec:spectra_calculation}
% \idea{It could be a good figure to show the various stages of this process. That would also be a chance to show what some real depth-1 maps and masks look like.}

We estimate the various power spectra used in the estimator and its covariance using a flat sky, pseudo-$C_{\ell}$ method. At the angular scales considered here ($\ell>1000$), the flat-sky approximation is valid and speeds up the analysis of the more than $20{,}000$ depth-1 maps~\cite{Bernardeau2011}. The use of the Plate Carrée (CAR) pixelization results in a declination dependence in the mapping between Fourier wavenumber and spherical multipole. We recover unbiased power spectra in simulations for the $\ell$ range, binning, and depth-1 footprints used here. The flat-sky treatment also allows us to filter ground contamination directly in Fourier space, as was done in the ACT DR4 and DR6 analyses~\cite{Choi2020,louis_DR6spectra_params_2025}. Each depth-1 Q and U map is loaded into memory and tapered with a mask that goes smoothly to zero in a 0.5 degree radius around any missing pixels and the edge of the map to prevent ringing in the Fourier transform. Prior to tapering, the Planck PR3 HFI galaxy mask that masks 70\% of the galactic plane is applied as well. The coadded reference maps have the brightest point sources removed during the mapmaking process, and we do not apply a further point source mask in our analysis as bright polarized point sources are rare. %\draftnote{Should we quantify this?}$Any depth-1 map containing a point source was likely eliminated by the data cuts described below. 

Some depth-1 maps fall entirely in the region that is masked by the galaxy mask, and spectra are not calculated for these maps. The maps that are not entirely eliminated by the mask and tapering are converted to Fourier space E and B modes and filtered. The filter is the same as that applied in DR4 and DR6 ACT power spectra analyses to eliminate ground pickup by removing Fourier modes (k-space) with $|\ell_x|<90$ and $|\ell_y|<50$. For each depth-1 map, the corresponding region is selected in the reference maps, which are also masked and filtered in the same way. 

After filtering and Fourier transforming back to real-space Q and U maps, we apply a per-pixel inverse variance weight, proportional to the hit map, with a 0.5 deg edge taper to compute the power spectra. %The inverse variance map is normalized prior to being applied. Occasionally, the inverse variance maps contained large, unphysical outliers, so the normalization was done to the 95th percentile of the inverse variance map. Any pixel above the 95th percentile was set to one.
%Finally, these maps were converted once more to Fourier space. The power spectra were 
These cross correlations are calculated for all relevant pairs of maps by binning radially in Fourier space with width $\Delta\ell = 400$ over the ell range $\ell \in [1000,3000]$. This minimum $\ell$ cut helps ensure the robustness of our measurements by matching the conservative baseline multipole range used in the main ACT DR6 analyses, for which all null tests passed (see Table 1 in Ref.~\cite{louis_DR6spectra_params_2025}). The maximum $\ell$ was chosen by balancing the need for at least several bins with the observed residual white noise level at high multipoles. The relatively large bin size was determined by a series of tests that minimized any correlations between bins in the measure EB spectra to a negligible level.

Several analytic corrections were applied to the binned pseudo spectra to account for instrumental effects and biases introduced by our process. We divide each spectrum by the same filtering transfer function, $t_b^2$, described in Section 6.6 of Ref.~\cite{Choi2020} to account for our k-space filtering. %This transfer function is calculated for each ell and then binned with the same binning scheme as our spectra. 
We also divide each spectrum by a window factor, $w_2$, to correct for the effects of masking and inverse variance weighting (cf.~equation A32 in Ref.~\cite{hivon_master_ps_2002}). This factor is the mean of the product of the inverse variance weighted mask for the two maps that go into the power spectrum calculation. The geometric contribution from the shape of the mask is common for every spectrum for the estimator and its uncertainty, but the inverse variance weighting contribution varies depending on whether it is a depth-1 auto spectrum (two copies of the depth-1 inverse variance map), a cross spectrum between the depth-1 map and the reference map (one copy of the depth-1 inverse variance map and one of the reference inverse variance map), or a reference map auto spectrum (two copies of the reference inverse variance map). This correction ranges between 
$1.2\times 10^{-4}$ and $0.66$
% 0.1 and 1.0 
%\draftnote{we need the exact number here}
for each spectrum. Finally, we correct the spectra with the appropriate beam window functions used for the DR6 power spectrum analysis \citep{louis_DR6spectra_params_2025}. The cumulation of corrections are described by 
\begin{equation}
   \widehat{C}_{b} = \frac{\widetilde{C}_{b}}{w_2 B_{b}^2 t_b^2},
\end{equation}
where $\widetilde{C}_{b}$ is the binned pseudo-spectrum, $w_2$ is the window factor, $B_b$ is the normalized beam window function, and $t_b$ is the analytic filtering transfer function. The full pipeline is tested on signal (CMB) only simulations to confirm it could recover unbiased power spectra for a range of depth-1 footprints and masks.%{\color{red} (pretty sure Zach had done this)}

Using these unbiased spectra, the EB estimator (Eq.~\ref{eq:estimator}) and diagonal elements of the covariance matrices are constructed. The covariance term has a mode counting factor, $\nu_b$, which is corrected for the reduction in total modes due to k-space filtering, masking, and inverse variance weighting. This factor begins with the total number of Fourier-space pixels in each bin and is multiplied by the k-space filter transfer function, $t_b$, and $w_2^2/w_4$, a slightly different window factor than the one used to correct the spectrum amplitude (cf.~equation 17 in Ref.~\cite{hivon_master_ps_2002}). Simulations confirm that this factor results in the correct effective number of modes.

\begin{figure*}[t]
    \centering
    \includegraphics[width=1\linewidth]{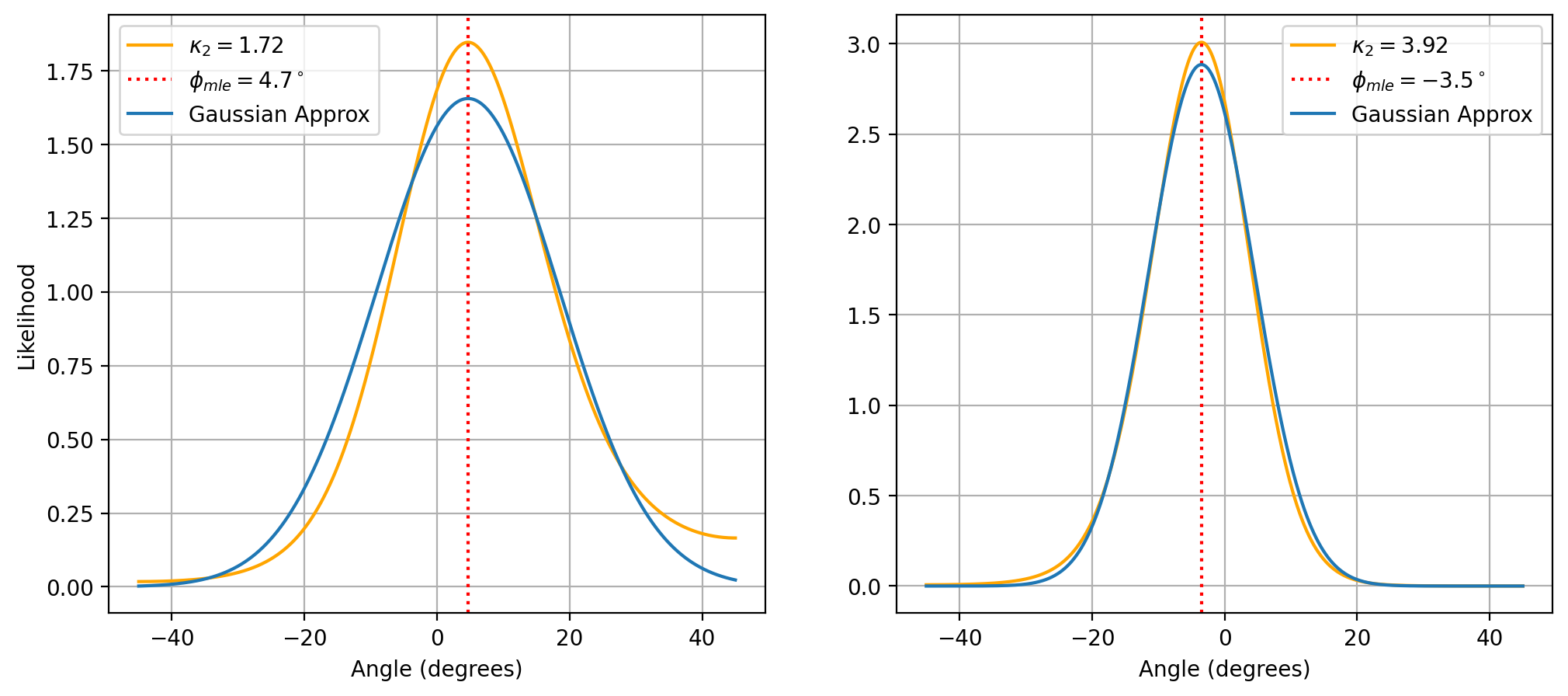}
    \caption{Example von Mises likelihoods are shown with their Gaussian approximations. The left panel shows a distribution that is near the Gaussianity cut while the right panel shows a likelihood that is nearly Gaussian. In most cases, the Gaussian approximation needs increased variance to account for the wide tails of the von Mises distribution.}
    \label{fig:likelihoods}
\end{figure*}

\subsection{Angle Extraction}
\label{sec:angle_extraction}

The EB power spectrum estimator and the error bars are used to construct a likelihood for the best fit angle,
\begin{equation}
    \label{eq:ang_lik}
    \mathcal{L}_{\text{angle}} \propto e^{-\frac{1}{2}\chi_{angle}^2}
\end{equation}
with
\begin{equation}
    \label{eq:angle_chi2}
    \chi_{\text{angle}}^2 = \sum_b\frac{(\Delta\widehat{C}_{b}^{EB}-C_b^{EE} \sin{(2\phi}))^2}{\widehat{\Theta}_b},
\end{equation}
where $\Delta\widehat{C}_{b}^{EB}$ denotes bin $b$ of the binned estimator,  $C_b^{EE}$ is a best-fit Planck 2018 EE spectrum from CAMB,\footnote{Calculated using the planck\_2018\_acc.ini file on the CAMB GitHub repository for the baseline Planck 2018 LCDM model using base\_plikHM\_TTTEEE\_lowl\_lowE\_lensing.} $\phi \equiv \theta_1-\theta_2$ is the angle we seek to measure, and $\widehat{\Theta}_b$ is the diagonal element of the covariance for bin $b$. %We will only ever be concerned with relative likelihoods, so we drop the normalization factor in front of the likelihood.

We note that the likelihood in Eq.~\ref{eq:ang_lik} forms a Gaussian in $\sin(2\phi)$, leading to asymmetric and non-Gaussian behavior in $\phi$. To characterize angle likelihoods that may deviate from a Gaussian and to define well-motivated data cuts, we first reparameterize the likelihoods in the form of
\begin{equation}
    \mathcal{L}_{\text{angle}}(\phi) \propto\exp\left(-\frac{(\sin(2\phi) - \mu)^2}{\sigma^2}\right),
\end{equation}
for some arbitrary $\mu$ and $\sigma^2$. We then expand the quadratic and obtain the analytical expression of the angle likelihood in the form
\begin{equation}
    \mathcal{L}_{\text{angle}}(\phi) \propto\exp\left(\kappa_1\sin(2\phi) + \kappa_2\cos(4\phi) \right),
\end{equation}
where we have $\kappa_1 = \mu / \sigma^2$ and $\kappa_2 = 1/(4\sigma^2)$, which relate to the asymmetrical and non-Gaussian behavior of the distribution respectively. 
%\draftnote{still don't follow this above. let's discuss it and/or write it much more simply. the likelihood itself written as an exponential of $-\chi^2$ in the current form is ``gaussian''. we should only say the necessary info in writing eq 8. Perhaps it's something like this: the likelihood to extract the angle is gaussian in sin(2phi) rather than phi, and to capture the asymmetric / non-gaussian / wrapping(?) behavior of the likelihood vs phi, we express it in the form given by eq 8 (then say mathematically, why we can write 8 from 6.} 
This is part of a wider family of second-order von-Mises distributions~\cite{von_mises_2007}. Maps with high signal-to-noise ratio have $\kappa_2 \gg 1$, which is the limit in which the von Mises distribution becomes a Gaussian.  We evaluate this likelihood at many points between $-45^\circ$ and $45^\circ$, the principal value domain of the $\sin(2\phi)$ in our estimator, and fit the likelihood to determine our $\kappa_1$ and $\kappa_2$ parameters.  %Many of our depth-1 maps yield such likelihoods, while others have lower signal-to-noise. %As noted in N25, the depth-1 maps are less processed than the large maps used in the ACT DR6 cosmology analyses, so some of the depth-1 maps may have beam, pointing, and calibration issues. Given the highly variable nature of the maps, some of them are also quite small or noisy relative to the others.

With the depth-1 maps spanning a wide range of sizes, shapes, and noise properties, some maps have low signal-to-noise and produce poorly constrained or unphysical likelihoods, which we remove using various criteria. Some likelihoods peak outside the principal range of $\sin(2\phi)$ with $|\mu|>1$, leading to invalid estimates of the angle; we cut these maps from our analysis. Other likelihoods peak at some value $\phi_{mle} \in [-45^\circ,45^\circ]$ but have a central peak with a tail on one side. For these, we use $\kappa_2$ as an indicator of Gaussianity and impose a threshold as described below. We do not impose cuts on $\kappa_1$ to limit asymmetrical behavior, since cuts on non-Gaussian behavior also constrain highly asymmetrical likelihoods. Example likelihoods are shown in Fig.~\ref{fig:likelihoods}. For the error bars, we require the Gaussian approximation to have the same variance as the von Mises distribution and find a relatively good fit for likelihoods with $\kappa_2 \gtrsim 1.75$. The full set of cuts used to eliminate maps from our final timestream is described below and summarized in Table~\ref{tab:map_cuts}.

\begin{enumerate}
    \item Some of the maps are entirely cut by our galaxy mask. No spectra can be calculated for these, so they are not included in the analysis.
    \item For a small number of maps, the fitting tools cannot find a satisfactory fit for a von Mises to the likelihood. Upon examination, these were usually likelihoods with unusual shapes that do not constrain a peak, so we exclude these.
    \item To remove most of the maps that did not form likelihoods with peaks within the principal range, we cut all maps with $|\mu|>1$.
    \item To enforce Gaussianity on our data points, we exclude maps below a certain $\kappa_2 < 1.75$ which deviates significantly from Gaussian. 
    \item We identified a small population of maps with unusually long durations. Some of these maps span a day or more and contain large gaps in the data. Since long durations can wash out the signal and the gaps raise concerns about data quality, we remove all maps longer than 14 hours.
\end{enumerate}

\begin{table}[b]
\centering
\renewcommand{\arraystretch}{1}
\begin{tabular*}{1\linewidth}{@{\extracolsep{\fill}} lccc}
\hline \hline
 & \textbf{f150} & \textbf{f090} & \textbf{Total} \\
\hline
Total & 13,918 & 8,663 & 22,581 \\
Galaxy Mask & $-$736 & $-$427 & $-$1,163 \\
No Fit & $-$2 & 0 & $-$2 \\
$|\mu| < 1$ & $-$2,360 & $-$999 & $-$3,359 \\
$\kappa_2 < 1.75$ & $-$5,231 & $-$2,646 & $-$7,877 \\
Duration $< 14$ hr & $-$60 & $-$16 & $-$76 \\
Total Cuts & $-$8,389 & $-$4,088 & $-$12,477 \\
Total Pass & 5,529 & 4,575 & 10,104 \\
\hline\hline
\end{tabular*}
\caption{Summary of map cuts for f150 and f090.}
\label{tab:map_cuts}
\end{table}

Low signal-to-noise maps produce likelihoods with low $\kappa_2$, which exhibit non-Gaussian behavior. To perform meaningful hypothesis tests in Sections~\ref{sec:nulls} and \ref{sec:results} using Gaussian statistics, we use the $\kappa_2$ cut to remove noisy maps that produce likelihoods that deviate significantly from Gaussian. The specific threshold was chosen by analyzing the total variation (TV) distance between the von Mises distribution and its Gaussian approximation as a function of $\kappa_2$. The TV distance reaches a local maximum at $\kappa_2 \approx 1.75$. For smaller values of $\kappa_2$, the residuals decrease toward a local minimum as the sidelobes of the von Mises distribution rise rapidly, signaling a transition from a Gaussian-like distribution toward a more uniform one. For larger values of $\kappa_2$, the residuals decrease monotonically, indicating that the central bell-shaped region of the distribution becomes increasingly well described by a Gaussian. Therefore, the local maximum in the TV distance near $\kappa_2 \approx 1.75$ marks the transition between the uniform-like and Gaussian-like regimes of the von Mises distribution, making it a natural cutoff for the validity of the Gaussian approximation.

\begin{figure*}[t]
\includegraphics[width=1\linewidth]{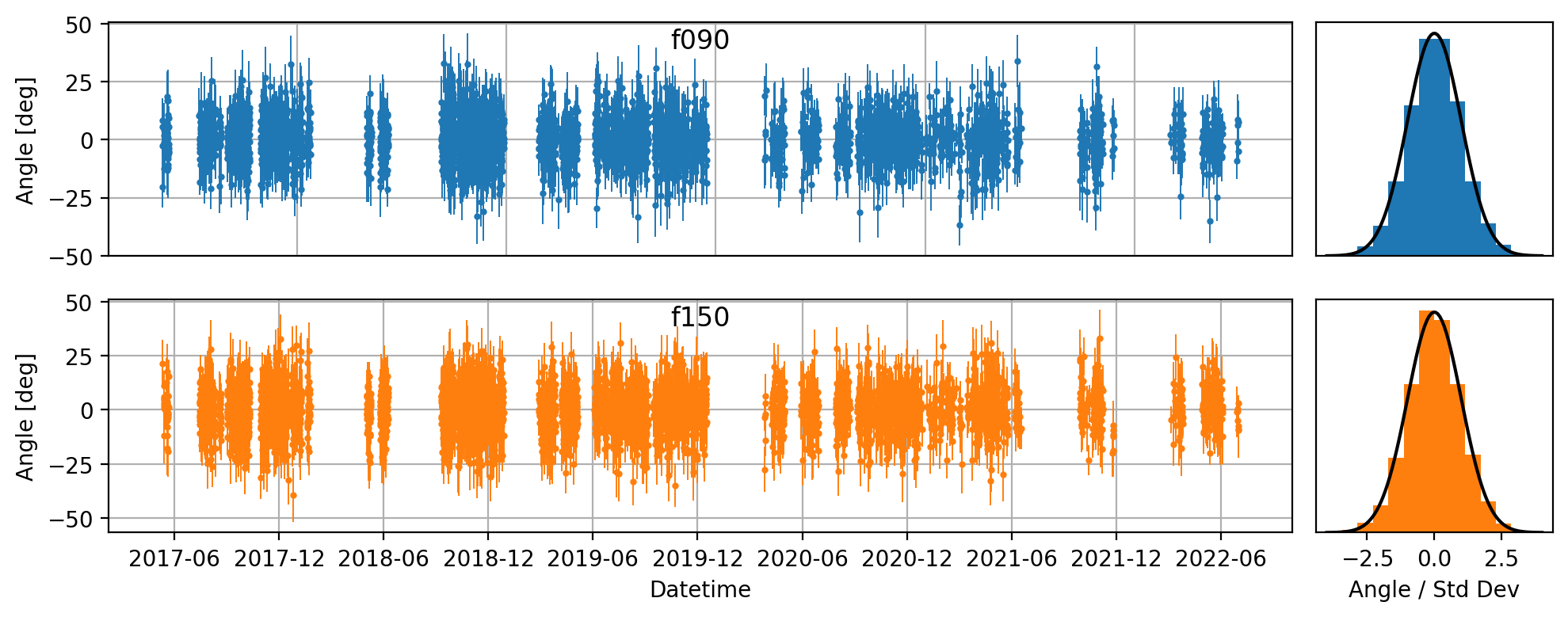}
\caption{The timestream of angles for f090 and f150, ACT DR6 after cutting low SNR maps with $\kappa_2 > 1.75$. This includes a total of 4,575 and 5,529 data points for 90 GHz and 150 GHz respectively. The length of the timestream is 1878 days, with a total observation time of 1996 days spread across 5 arrays. The angle measurements are described in Sections~\ref{sec:angle_extraction} and~\ref{sec:error_calibration}.}
\label{fig:timestream}
\end{figure*}

The final timestreams for f090 and f150 are shown in Fig.~\ref{fig:timestream}. Of the 13,918 f150 depth-1 maps, 5,529 (39.7\%) pass these cuts. Of the 8,663 f090 depth-1 maps, 4575 (52.8\%) pass these cuts. The large gaps in the timestream correspond to periods when the telescope was not operating due to adverse weather, repairs, or upgrades to the telescope and instrument between observing seasons. %The figure includes the effects of inflating covariances by 4\% which are discussed later in Section~\ref{sec:model_cons}.
%draftnote{Reminder:Add more detail on null tests revealing this uniofrm error inflation. Also, some words about analytic power spectrum error bars being underestimated citing with it later(Atkins et al. 2025 and Camphuis et al. 2022). Keeping it concise}

\subsection{Timestream Fitting}
\label{sec:timestream_fitting}

Once the timestream has been generated, we use a maximum likelihood estimator to calculate the best fit amplitude for sine waves of different frequencies. Following P23, we account for the finite duration of the measurement when calculating the average angle of each depth-1 map, $\langle\theta_i\rangle$, by integrating Eq.~\ref{eq:sine_parameterization} over the length of the observation:

\begin{align}
    \label{eq:sin_model}
    \langle \theta_i \rangle &= \frac{1}{t^{end}_i - t^{start}_i} \int^{t^{end}_i}_{t^{start}_i} dtA\sin(2\pi ft + \alpha) \nonumber \\
    &= A\sinc(\pi f \Delta t_i)\sin(2\pi f \langle t_i \rangle + \alpha).
\end{align}
Here, $\Delta t_i = t^{end}_i - t^{start}_i$ is the duration of the depth-1 observation inside of our non-apodized region used for the power spectra calculations, and $\langle t_i \rangle = \frac{t^{end}_i + t^{start}_i}{2}$ is the mean time of the same region.

By construction, the uncertainties on each angle are Gaussian or approximated as Gaussian, so we use the following chi-squared variable and likelihood
\begin{equation}
    \label{eq:ts_chi2}
    \chi_{ts}^2(A,f,\alpha) = \sum_{i}\frac{(\phi_i - \langle \theta_i \rangle)^2}{\sigma_i^2}
\end{equation}
and 
\begin{equation}
    \label{eq:ts_lik}
    \mathcal{L}_{ts}(A,f,\alpha) \propto e^{-\frac{1}{2}\chi_{ts}^2},
\end{equation}
where $\phi_i$ is the measured angle for a single depth-1 map, $\sigma_i$ is the uncertainty for the angle, and $\langle\theta_i\rangle$ is the expected observed angle from an oscillating sinusoid defined in Eq. \ref{eq:sin_model}. Here, ``ts'' denotes the timestream. The sum is over all f090 and f150 observations that pass cuts. Because we are concerned only with relative likelihoods, we omit the normalization factor.

In order to evaluate $\chi_{ts}^2$ quickly for a large grid of amplitudes, frequencies, and phases, we follow BXII (Section IV.B) in writing Eq.~\ref{eq:ts_chi2} as six sums by expanding the square in the numerator and using the sine sum formula to separate the sine in Eq.~\ref{eq:sin_model}, 
\begin{align}
    \chi_{ts}^2(A,f,\alpha) = &S_1(f) + S_2(f) A \cos{\alpha} \nonumber \\ 
    + &S_3(f) A \sin{\alpha} + S_4(f) A^2 \cos^2{\alpha} \nonumber \\
    + &S_5(f) A^2 \sin{\alpha}\cos{\alpha} + S_6(f) A^2 \sin^2{\alpha}.
\end{align}
Each resulting sum, $S_i$, depends only on the frequency and quantities calculated from the depth-1 maps in Sec.~\ref{sec:angle_extraction}: the angles, the uncertainities, durations, and mean times. For each frequency, these sums are calculated once to save time and memory when computing the full chi-squared variable for a large grid of amplitudes and phases to find the values that maximize the likelihood in this parameter space.

\subsubsection{Frequency Range}
\label{sec:freq_range}
A key consideration in this analysis is determining which oscillation frequencies to sample. With our current depth-1-map-based estimator, we are not sensitive to frequencies corresponding to periods shorter than the duration of our depth-1 maps, which are generally less than 10 hours of data. The maps could be split into shorter time intervals, extending the time resolution to approximately 4 minutes, but we do not explore this here. The irregular cadence of our observations effectively convolves any regularly sampled oscillation with a complicated window function with structure at periods less than 1 day. To avoid introducing possible oscillatory systematics related to this cadence, we restrict our minimum period to 24 hours; this sets the maximum frequency for our range at 1 inverse day or 0.042 inverse hours. The minimum frequency is bounded by the length of our timestream.

If our timestream were regularly sampled, as for a standard discrete Fourier transform, the frequency spacing required to extract all available frequency information would be the inverse of the total length of time of the timestream. For an irregularly sampled timestream, the window function introduces correlations between neighboring frequencies that could cause oscillatory signals to be missed if the frequency grid is the same as the discrete Fourier transform (see Section 7.1 of Ref.~\cite{vanderplas_lombscargle_2018} for a discussion of such considerations). To mitigate this issue, we oversample by a factor of 5: $\Delta f = f_{min}/5$. The expected virial broadening, $\Delta f_{\rm vir}/f\sim10^{-6}$, is negligible compared to our frequency resolution over this mass range. Our final frequency range spans from $5.3\times 10^{-4}~\mathrm{days}^{-1}$ to 1 $~\mathrm{days}^{-1}$ with a spacing of $1.06\times 10^{-4}~\mathrm{days}^{-1}$ for a total of 9387 frequencies. This corresponds to a mass range of $2.54\times 10^{-23} \text{ eV} \leq m_\phi \leq  4.78\times 10^{-20}\text{ eV}$ with a mass spacing of $5.1 \times 10^{-24}$ eV.

\subsection{Uncertainty Calibration}
\label{sec:error_calibration}

\begin{figure}[t]
\centering
\includegraphics[width=1.0\linewidth]{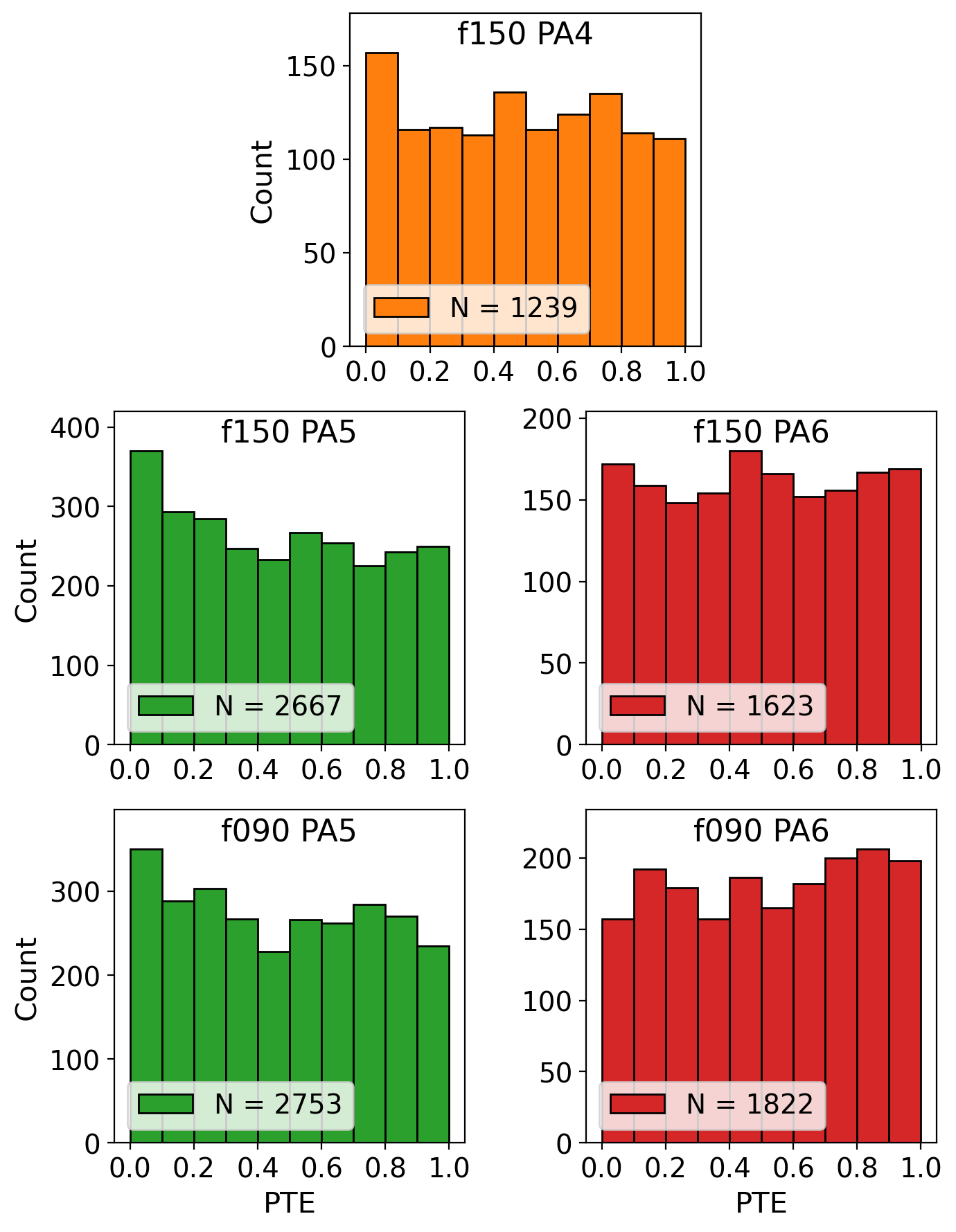}
\caption{PTE distribution ($\mathrm{dof}=4$) for the difference between the estimator and the theoretical estimator evaluated at the best-fit angle for f150 and f090 for all arrays. A flat PTE distribution indicates that the model is generally fitting the data well. }
\label{fig:combined_model_fit_ptes}
\end{figure}

Noting that analytic power spectrum uncertainties can be slightly underestimated \cite{Atkins_2025, Camphuis2022}, particularly given the large variation in the sizes, shapes, and noise properties of our depth-1 maps, we derive a small correction factor directly using timestream-level null tests constructed from randomized splits of the data. Using the timestream fitting procedure described above, we perform this test for multiple randomized data splits. We find that increasing the power spectrum covariance by 4\% allows such randomized splits to all pass the timestream-level null tests at PTE of at least 0.05. This correction is applied throughout the analysis, including to all error bars shown in Fig.~\ref{fig:timestream} and covariances in Figs.~\ref{fig:combined_model_fit_ptes} and \ref{fig:combined_array_diff_null_ptes}.

\subsection{Model Consistency}
\label{sec:model_cons}

For each map, we can assess the goodness of fit of the model to the estimator by constructing a chi-squared variable from the difference between the estimator and the model evaluated at the best-fit angle (Eq.~\ref{eq:estimator} and Eq.~\ref{eq:angle_chi2}). %The covariance is the sum of the estimator covariance and the calculated variance of the model (in terms of the theory EE spectrum, the best-fit angle, and the best-fit error bar on the angle) \draftnote{isn't it just the denominator in eq 7? I am not sure what is meant by adding the variance of the model...}. 
For each map that passes our data cuts, we calculate the probability to exceed (PTE) this chi-squared variable for $n_{bins}-n_{params} = 5-1 = 4$ degrees of freedom. The PTE distributions, including the covariance correction described above, are shown in Fig.~\ref{fig:combined_model_fit_ptes}. We note the skew seen in the PA5 distributions may indicate a slight underestimation of the uncertainties ($\sim$5\%), while the PA4 and PA6 distributions are consistent with statistical fluctuations. No obvious cause was found for this slight skew, so we move on to evaluate our overall data fidelity and search for possible systematic effects through internal consistency and null tests in the next section.

\section{Null Tests}
\label{sec:nulls}
We perform two complementary sets of tests for possible systematic effects that could bias our results. The first set compares simultaneous observations of the same sky region from different arrays, while the second set is applied at the timestream level, where fits are performed and the fitted parameters are nulled to assess consistency.
%\draftnote{we should relabel this section as ``null tests'' and move this discussion of whether the model fits the data in eq 7 to the previous section where the angle extraction is discussed.}
\subsection{Inter-Array Consistency Tests}
\label{sec:per_obs_tests}

\begin{figure}[t]
    \centering
    \includegraphics[width=1.0\linewidth]{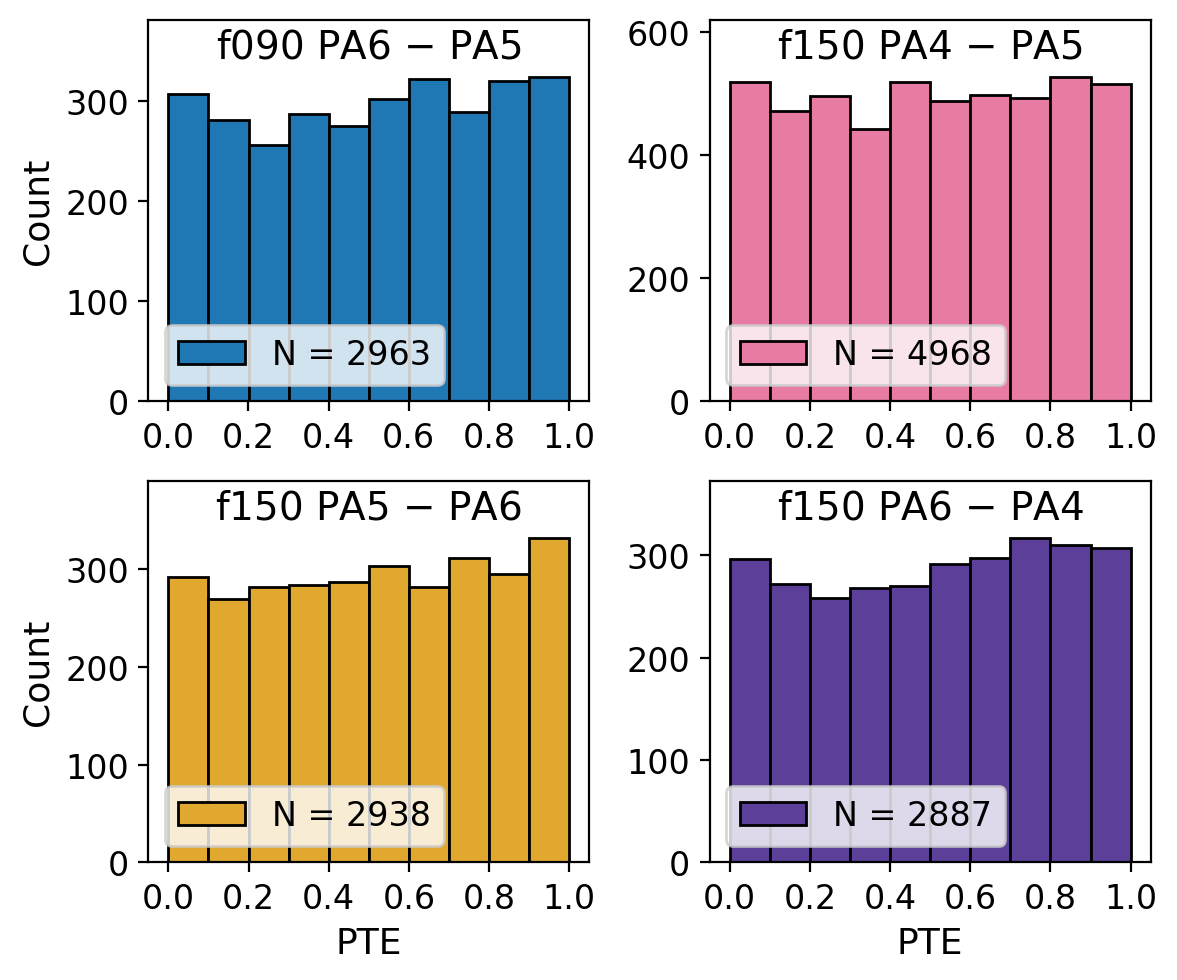}
    \caption{PTE distributions ($\mathrm{dof}=5$) for the difference of estimators for measurements of the same area of the sky made by a different arrays at the same time for f150 and f090 for all array pairs. }
    \label{fig:combined_array_diff_null_ptes}
\end{figure}

We check for consistency between measurements from different arrays that observe similar sky footprints at the same time. These measurements should be consistent, so the difference between their estimators should be consistent with zero. For each pair of maps $(i,j)$ that passes our data cuts, we construct a chi-squared variable.

\begin{equation}
    \chi_{null,ij}^2 = \sum_b\frac{(\Delta\widehat{C}_{b,i}^{EB}-\Delta\widehat{C}_{b,j}^{EB})^2}{(\widehat{\Theta}_{b,i}+\widehat{\Theta}_{b,j} -2\widehat{\Theta}_{b,i\times j})},
\end{equation}
where $\Delta\widehat{C}_{b,i}^{EB}$, $\Delta\widehat{C}_{b,j}^{EB}$ are the estimators for maps $(i,j)$ and $\widehat{\Theta}_{b,i}$, $\widehat{\Theta}_{b,j}$, and $\widehat{\Theta}_{b,i\times j}$ being the covariances and cross covariance for maps $(i,j)$  respectively. The cross covariance is derived in Appendix \ref{sec:cross_cov}. In practice, we find the cross covariance terms to be four orders of magnitude smaller than the other terms and hence omit them from our calculations. 
%\draftnote{how did we get the denominator here? I think there is suppose to be some cross terms from having common EE in the two terms in the numerator... do you know how to compute this and check? it's probably small enough but we should say so if so}
With this chi-squared variable, we then calculate the probability to exceed that chi-squared value for $n_{bins}=5$ degrees of freedom. The PTE distributions are shown in Fig.~\ref{fig:combined_array_diff_null_ptes} and indicate consistency between different instrument pairs. %\draftnote{maybe could add KS numbers here as well.}

\subsection{Timestream Null Tests}
\label{sec:timestream_nulls}

If there are periodic changes in the measurement of polarization due to astrophysical sources or aspects of telescope operations, these could mimic the ALP signal. To test for such systematics, we perform a set of null tests by comparing the best fit amplitudes and phases for splits of the full timestream, including both frequencies. The set of these splits includes: 

\begin{enumerate}
    \item A temporal split into roughly two halves. Following N25, we choose to split our timestream into before and after Feburary 2019. This relatively uneven split is to account for the telescope focus change in May 2018.
    \item Daytime vs.~nighttime splits to check for systematics related to the diurnal cycle.
    \item Moon up vs.~moon down splits to check for systematics related to the lunar cycle.
    \item An elevation split between low and high elevation scans such that each subset contains approximately half of the total statistical power. This checks for elevation-dependent systematics, including atmospheric effects or changes in telescope response. 
    \item Azimuthal split, comparing scans centered between $[0,\pi)$ and $[\pi, 2\pi)$ azimuths. This checks for systematics regarding ground pickups or scan strategy.
    \item Spatial isotropy split to check for systematics between observations above and below galactic plane.
    \item Frequency split (f150 vs.~f090) comparing best fit sinusoids from the two observing bands. This provides a consistency check on the per-observation null test and tests for frequency-dependent systematics, including possible effects from observations that do not have a paired counterpart in the other frequency band.
\end{enumerate}

% \subsubsection{Methodology}
% \label{sec:null_method}
For each test, we perform the method described in Section~\ref{sec:timestream_fitting} independently for each split to determine the best-fit amplitude and phase. To evaluate the null between the two sinusoids, accounting for differences in amplitude, phase, or both, we reparameterize the AC oscillation effect in Eq.~\ref{eq:sine_parameterization} as 
\begin{equation}
\label{eq:reparameterization}
    \theta = X\cos(m_\phi t)+Y\sin(m_\phi t)
\end{equation}
where $X$ and $Y$ are the Cartesian components of the sinusoid, while its amplitude and phase correspond to polar coordinates in this two-dimensional space. This representation naturally accounts for the periodicity of the phase and makes the model linear in its parameters, such that the likelihood in Eq.~\ref{eq:ts_lik} takes the form of a two-dimensional Gaussian. At each oscillation frequency, we fit this likelihood for each split and form a standard $\chi^2$-like test statistic with two degrees of freedom
\begin{equation}
    T_{null}(f) = (\vec\mu_1 - \vec\mu_2)^\top(\Sigma_1 + \Sigma_2)^{-1}(\vec\mu_1 - \vec\mu_2)
\end{equation}
where $\vec\mu_i$ and $\Sigma_i$ are the mean and covariance matrix of the likelihood of the $i$th split using the parameters in Eq.~\ref{eq:reparameterization}. This statistic measures the two-dimensional separation between the fitted sinusoids, weighted by their covariance, and is therefore sensitive to differences in amplitude, phase, or both. Similarly to P23, we use $T_{null}$ to form three different null test pass criteria, each optimized to search for different systematics. 
%\draftnote{let's call out PB or whoever we are following for this, by saying ``similarly to blah,''at the beginning}
The set of PTEs are

\begin{enumerate}
    \item $\max_f T_{null}$: The maximum $T_{null}$ across all frequencies, which tests for a spurious ALP signal.
    \item $\sum_f T_{null}$: The sum of $T_{null}$ across all frequencies, which tests for noise misestimation.
    \item $1-\text{KS}_f(T_{null})$: $1 -p$, where p is the p-value from the Kolmogorov-Smirnov (KS) test of uniformity for the PTE distribution $T_{null}$ across frequencies using $\chi^2_{\text{dof}=2}$, which tests for non-sinusoidal systematics that might be spread out across many frequencies that make $T_{null}$ deviate from chi-squared.
\end{enumerate}

For each pass criterion, we compute a PTE value by comparing the data null statistic to a distribution from simulations. Each simulated timestream is generated by drawing every data point from a Gaussian centered at zero with the same uncertainty as the corresponding angle measurement.
%\draftnote{what is this setting of the man to zero step? if you drew from a zero mean gaussian, it should have a zero mean, right? I don't think there should be any small offsets put in manually afterwards..}
The simulated timestreams are then split in the same way as the data null. These timestreams have similar statistical properties to the data, but any periodic signal would be removed. The fraction of simulated null test criterion values that exceed the data value at each frequency defines the PTE for that frequency.

It is important to note that while $T_{null}$ constrained at a single frequency is $\chi^2$-like distributed and independent across different simulated timestreams, $T_{null}$ across frequency for a single timestream is not because nearby frequencies are correlated. Thus, similarly to the KS tests done by P23, we generally expect the PTE distribution of $T_{null}$ to deviate from uniformity. Therefore, in criteria 3, we are comparing the p-value of the KS test statistic across simulated timestreams instead of just looking at the p-value for the data timestream, which by itself is statistically insignificant. We require the PTE for each criterion to exceed the testing threshold of $0.05/n_{\rm tests}$, where $n_{\rm tests} = 7$.

% \subsubsection{Null Test Results}
% \label{sec:null_results}

The null tests were run using 1023 simulations. As shown in Table~\ref{tab:null_test_ptes}, all null tests pass the predetermined $0.05/n_{\rm tests}$ threshold. We therefore proceed to the main results.%All of the null tests pass the $0.05/n_{tests}$ threshold. However, we observe a slight low PTE skew especially from the PTE values calculated from criteria 2 while the other criteria PTEs are more uniformly distributed. This suggests a possibly low degree of noise misestimation, which we had not fully corrected by inflating our covariances in section~\ref{sec:angle_extraction}. Despite this, the PTE values nonetheless pass  our predetermined criteria, thus we can continue to unblinding.

\begin{table}[h]
\centering
\begin{tabular}{lccc}
\hline
\hline
Split Type & Max $T_{\mathrm{null}}$ & Sum $T_{\mathrm{null}}$ & KS-PTE \\
\hline
Temporal & 0.075 & 0.162 & 0.516 \\
Diurnal & 0.631 & 0.367 & 0.834 \\
Lunar & 0.238 & 0.022 & 0.061 \\
Elevation & 0.462 & 0.296 & 0.706 \\
Azimuth & 0.683 & 0.091 & 0.216 \\
Isotropy & 0.325 & 0.099 & 0.347 \\
Frequency & 0.488 & 0.074 & 0.358 \\
% Temporal (PA7 Installation) & 0.211 & 0.023 & 0.054 \\
\hline
\hline
\end{tabular}
\caption{Results for timestream-level null tests.}
\label{tab:null_test_ptes}
\end{table}

\section{Results}
\label{sec:results}

After passing the null tests, we search for a significant detection of an oscillation that could be caused by ALP-photon interactions. If the data are not consistent with the background-only model, we can use the best fit amplitude to quantify the ALP-photon coupling constant at the mass corresponding to the frequency of the oscillation. If they are consistent with the background-only model, we instead integrate the marginalized likelihood to set Bayesian 95\% upper limits on the amplitude of the oscillation at each frequency.

\subsection{Search Results}
\label{sec:detection}

The method of Section~\ref{sec:timestream_fitting} was applied to the full timestream to identify the best fit amplitude and phase at each frequency and their respective $\chi^2_{ts}$ values from Eq.~\ref{eq:ts_chi2}. Following BXII, we form the statistic 
\begin{equation}
    \Delta\chi^2_{ts}(f) = \chi^2_{ts}(0,f,\alpha) - \chi^2_{ts}(A_{mle},f, \alpha_{mle}) 
\end{equation}
to assess the significance of a possible signal relative to the null hypothesis ($A=0$) at a single frequency, where the ``mle'' denotes the maximum likelihood estimate. Due to the high volume of frequencies we sample, there is bound to be some high $\chi^2$ value at some frequencies. Thus, we define a global detection statistic 
\begin{equation}
\label{eq:detect_chi2}
    \chi^2 = \max_f \Delta\chi^2_{ts}
\end{equation}
which uses the frequency that deviates the most from the background to evaluate whether there is a detection. We calibrate this statistic relative to 1023 simulations formed by drawing new angles from the likelihood centered at zero for each map in the timestream just as in Section~\ref{sec:timestream_nulls}. We set a standard threshold of $3\sigma$ for possible evidence of detection. Simulations verified that an injected signal with $A\approx0.55^\circ$ would be detected at $3\sigma$ significance, providing an estimate of our experimental sensitivity. 

\begin{figure}[htbp]
    \centering
    \includegraphics[width=1\linewidth]{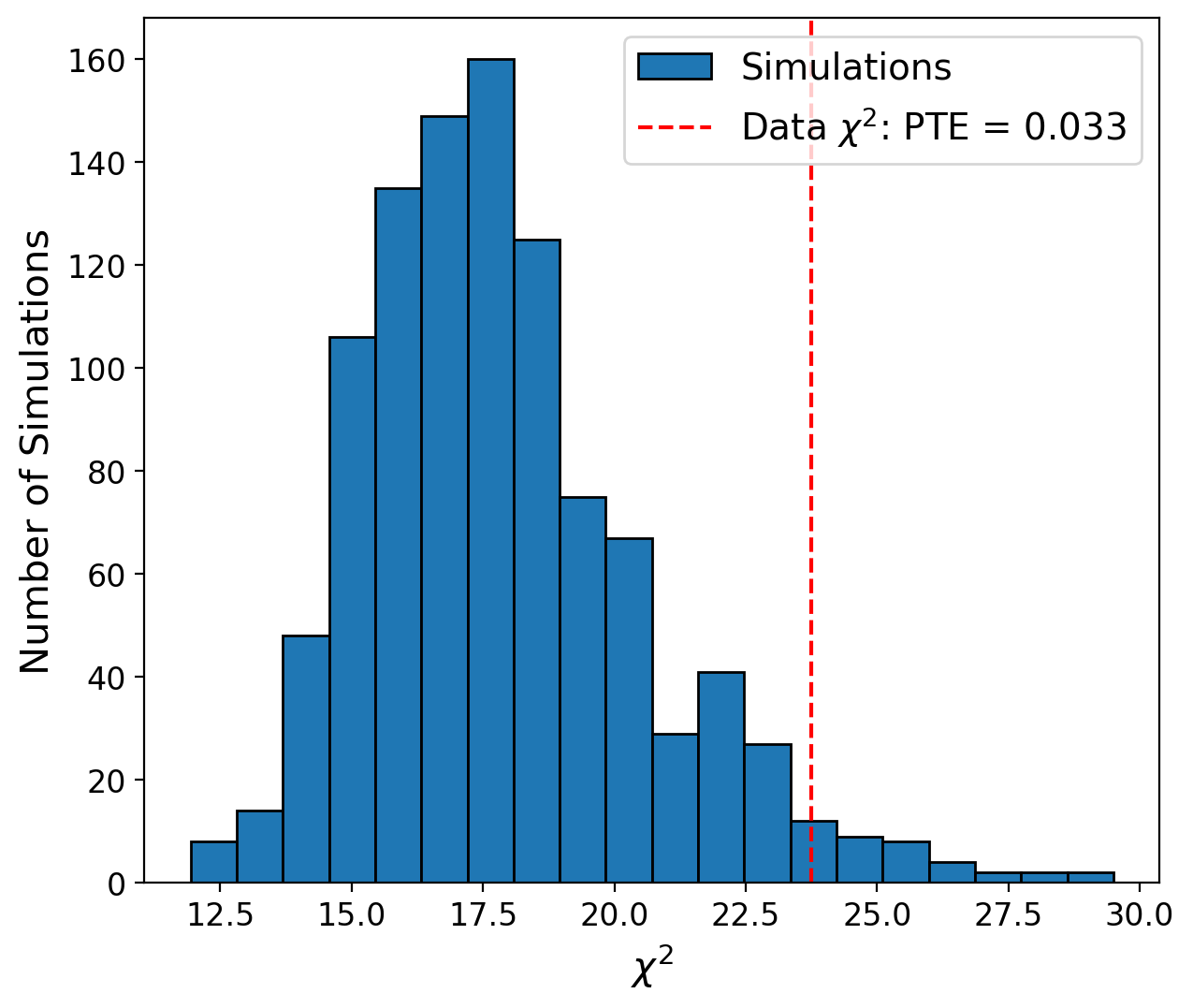}
    \caption{Distribution of the $\chi^2$-like variable defined in Eq.~\ref{eq:detect_chi2} for 1023 background only simulations shown along with the data $\chi^2$ corresponding to the most significant deviation, at an oscillation frequency of $0.87~\mathrm{days}^{-1}$. %\draftnote{update sig fig for PTE, and remove ``real'' from legend. also remove ``vaule'' in x label. }
    } 
    \label{fig:detection}
\end{figure}

Our results indicate that the most significant deviation from the background is found at a frequency of $0.87~\mathrm{days}^{-1}$, with a PTE of 0.033, corresponding to a significance of approximately $1.85\sigma$ (Fig.~\ref{fig:detection}). This is consistent with the null hypothesis and with the level of statistical fluctuations in the data.

\subsection{ALP Limits}
\label{sec:limits}

We present two sets of upper limits. The first assumes a deterministic ALP field, in which the local ALP field amplitude is fixed by the local dark matter density. The second treats the ALP field amplitude as a stochastic variable, as expected for virialized ultralight dark matter in the Galactic halo. For the deterministic case, we choose uniform priors over phase and amplitude while in the stochastic case we use the Berger-Bernardo reference priors \cite{centers_stochastic_alp_2021}. 
%\draftnote{add the relevant reference here}. 
Our primary result in this section will be the stochastic limits while we include the deterministic limits for a better comparison with results from other experiments. We discuss the choice of our priors in more detail in Section~\ref{sec:conclusion}.

\subsubsection{Deterministic}

In the deterministic case, the polarization rotation signal is described by an amplitude $A$, phase $\alpha$, and frequency $f$. For each frequency, we marginalize the timestream likelihood over the unknown phase, and form a cumulative distribution function on the amplitude. For both phase and amplitude, we follow S22, BXII, and BXIV in selecting a uniform prior: 

\begin{equation}
\label{eq:marg_lik}
    \mathcal L(A, f) = \int_0^{2\pi} P(\alpha)\mathcal{L}_{ts}(A',f,\alpha) \, d\alpha
\end{equation}

\begin{equation}
    \text{CDF}(A,f) = \frac{1}{N}\int_0^A P(A') \mathcal{L}_{ts}(A',f) \,dA',
\end{equation}

\begin{figure*}
    
    \includegraphics[width=1\linewidth]{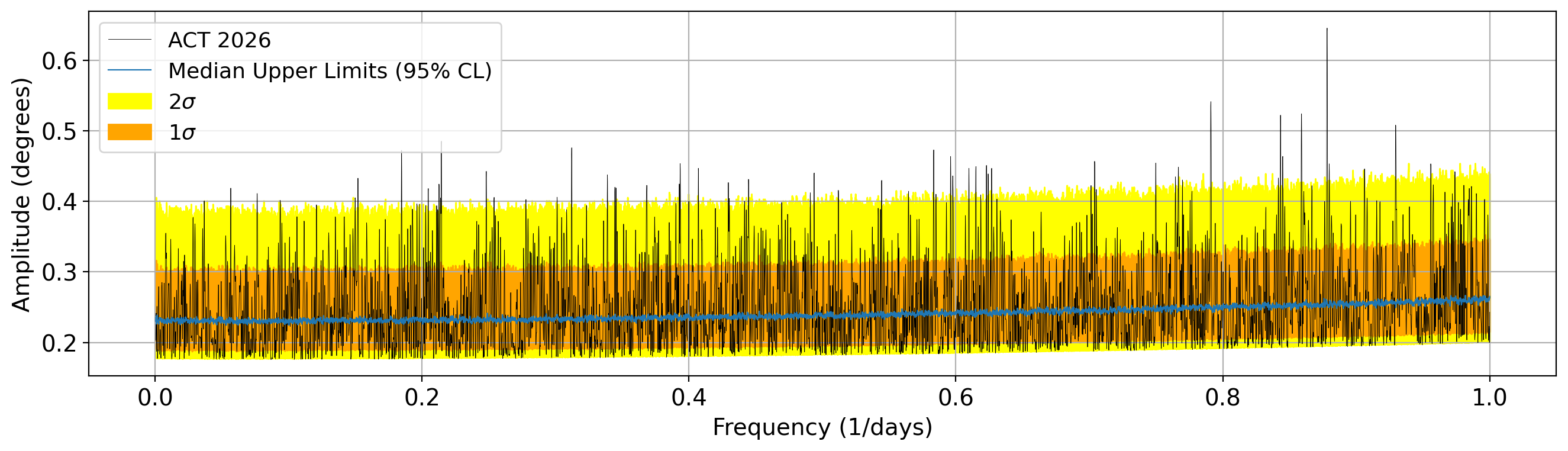}
    \caption{95\% confidence upper limit on the amplitude of the oscillation in the polarization rotation angle. The $1\sigma$, $2\sigma$ and median upper limits are calculated from 1023 background-only simulations at each frequency. We observe a slight degradation of the limits at higher frequencies, due to reduced sensitivity when the observation times become comparable to the ALP oscillation period. Some frequencies lie outside the nominal $1\sigma$ or $2\sigma$ expected bands, but these bands are defined locally at each frequency and do not account for the large number of frequencies scanned. Given the $\mathcal{O}(10^4)$ frequencies probed, such local excursions are expected to occur from statistical fluctuations alone. The statistic in Eq.~\ref{eq:detect_chi2} accounts for this effect when searching for detection; since the data only provides an $1.85\sigma$ excursion, there is not significant evidence for a signal.}
    \label{fig:angle_uplims}
\end{figure*}

\noindent where $P(\alpha) = 1/{2\pi}$ is the uniform prior and $N$ is a normalization constant. The 95\% upper limit is the value of $A$ for which $\text{CDF}(A,f) = 0.95$ at each frequency. We also apply this analysis to 1023 background-only simulations and obtain the expected limits, with the results shown in Fig.~\ref{fig:angle_uplims}. The data amplitudes are consistent with the background fluctuations from simulations. The slight degradation of limits at higher frequencies is caused by the finite duration of our observations. We find a median 95\% upper limit across the searched frequencies of $A < 0.24^\circ$. With the standard assumption that ALPs comprise all of the dark matter, adopting a local dark matter density of $\rho_0 = 0.3~\text{GeV}/\text{cm}^3$, and using 
\begin{equation}
    A = \frac{g_{\phi\gamma}\phi_0}{2}
\end{equation}
from Eq.~\ref{eq:rotation_angle} and Eq.~\ref{eq:sine_parameterization} and Eq. 47 in F19,
\begin{equation}
    \rho_0 = \frac{1}{2} \phi_0^2 m_{\phi}^2,
\end{equation}
this median limit corresponds to a 95\% confidence-level limit on the coupling of 
\begin{equation}
\label{eq:det_limit}
g_{\phi\gamma} < 3.9 \times 10^{-12}\left(\frac{m_\phi}{10^{-21}\,\text{eV}}\right)\,\text{GeV}^{-1}.
\end{equation}

% We set a median limit of $A < 0.24^\circ$ which translates into a deteministic limit of
% \begin{equation}
%     \label{eq:det_limit}
%     g_{\phi\gamma} < 3.92 \times 10^{-12}\,\,\text{GeV}^{-1} \times \frac{m_\phi}{1.0 \times 10^{21}\, \text{eV}}
% \end{equation}
% on the axion-photon coupling constant. For this conversion from amplitude to coupling, we have used the standard assumption that axions comprise the full fraction of dark matter and a local dark matter density of $\rho_0 = 0.3\, \text{GeV}/\text{cm}^3$.

\subsubsection{Stochastic}

The calculated limits in Eq.~\ref{eq:det_limit} rely on a critical assumption that the local dark matter field amplitude is constant. However, as indicated in Ref.~\cite{centers_stochastic_alp_2021}, ultralight dark matter, such as ALPs in the mass range being probed here, should be virialized in the Galactic potential. This local field amplitude should therefore be treated a stochastic variable and marginalized over. Given that the minimum coherence time of the field is much longer than our entire observation window ($\tau_c \sim 2700\,\text{yrs} \gg 5\,\text{yrs}$) the field amplitude probability can be taken as a Rayleigh distribution
\begin{equation}
    P(\phi_0) = \frac{2\phi_0}{\phi_{DM}^2}e^{-\frac{\phi_o^2}{\phi_{DM}^2}},
\end{equation}
where $\phi_{DM} = \sqrt {2\rho_0} / m_\phi$. Using this distribution, we can now form the likelihood on $g_{\phi\gamma}$ using the marginalized likelihood in Eq.~\ref{eq:marg_lik} 
\begin{equation}
    \label{eq:g_likelihood}
    \mathcal{L}(g_{\phi\gamma},f) =  \int_0^{\infty}d\phi_0 \,P(\phi_0) \mathcal{L}(\frac{g_{\phi\gamma}\phi_0}{2}, f).
\end{equation}

For our prior on $g_{\phi\gamma}$, we follow Ref.~\cite{centers_stochastic_alp_2021} and use the Berger-Bernardo reference prior. After marginalizing over the stochastic local ALP-field amplitude, a prior that is flat in $g_{\phi\gamma}$ would not in general be flat under an equivalent reparameterization, such as $g_{\phi\gamma}^2$, and could therefore lead to a different inferred limit. The Berger-Bernardo prior is constructed to be invariant under reparameterization, so that the resulting constraint does not depend on the particular choice of parameter used to describe the coupling. Furthermore, it maximizes the information gained by the data and in the one parameter case conveniently reduces to Jeffreys prior 
\begin{equation}
    \label{eq:reference_prior}
    P(g_{\phi\gamma}) \propto \sqrt{\mathcal{I}(g_{\phi\gamma})},
\end{equation}
where $\mathcal{I}$ is the Fisher information \cite{reference_priors}. Ref.~\cite{centers_stochastic_alp_2021} advocates for this prior because it is invariant under reparameterization and yields results that closely agree with frequentist methods.

\begin{figure*}
\includegraphics[width=1\linewidth]{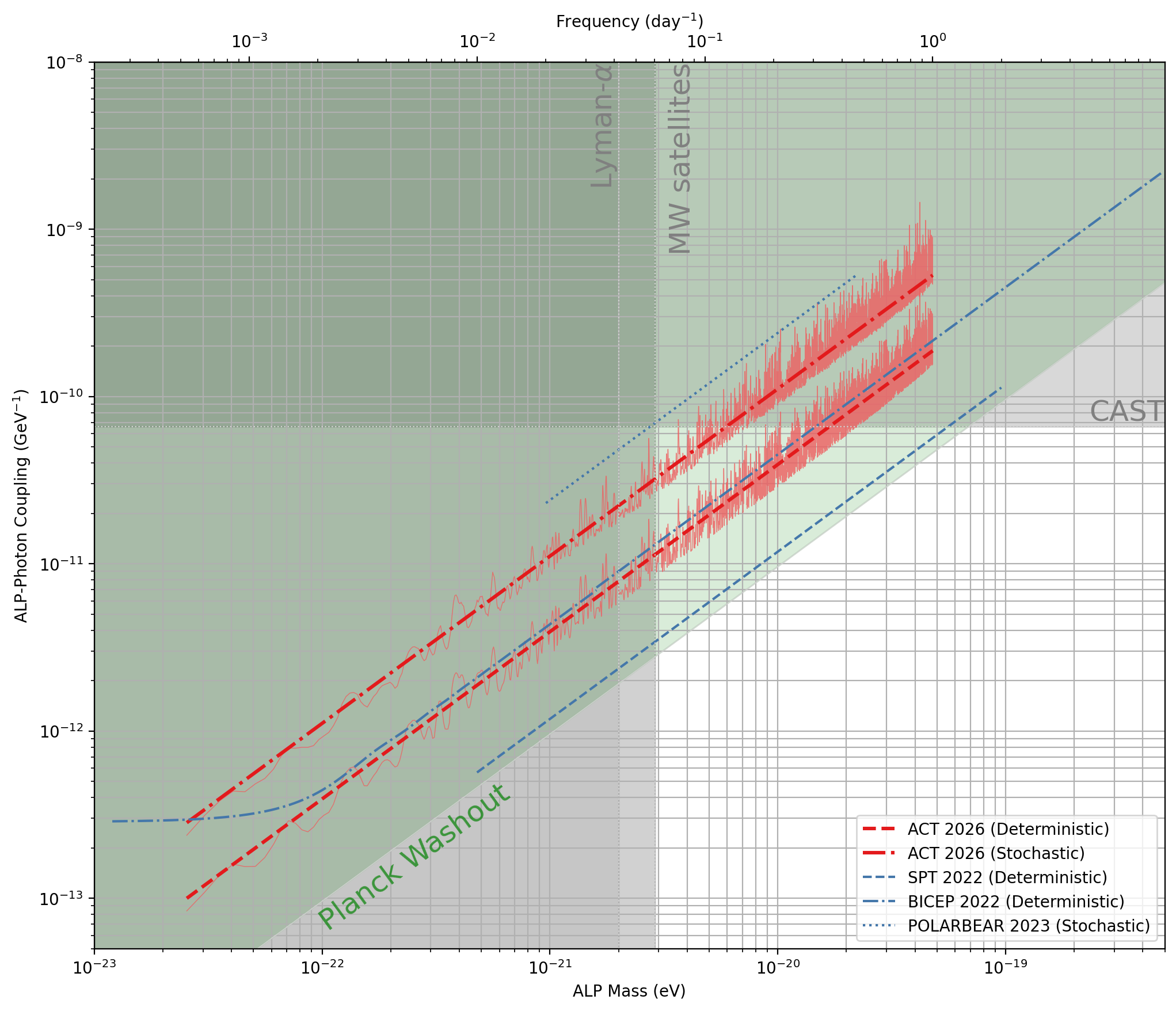}%

\caption{
Constraints on the ALP-photon coupling constant as a function of ALP mass. Our deterministic and stochastic constraints are shown in red, with their smoothed curves shown as dashed and dot-dashed lines, respectively. The blue curves show constraints from AC oscillation searches undertaken by other CMB experiments. The S22 (SPT 2022) and BXIV (BICEP 2022) constraints are deterministic Bayesian limits using uniform priors on $g_{\phi\gamma}$, while P23 (POLARBEAR 2023) reports stochastic limits. Mass ranges reflect each analysis's observation span and cadence. The green exclusion region comes from the constraints due to the washout effect explored in Ref.~\cite{fedderke_axion_2019} with Planck data. Gray lines and exclusion regions refer to other astrophysical limits (vertical lines) \cite{des_alp_milkywaysatellites_2021,irsic_alp_lymanalpha_2017} or direct detection limits (horizontal line) \cite{anastassopoulos_castlimit_2017}. Not shown are time-varying polarization rotation methods that use a different source than the CMB. 
%\draftnote{check the other constraints on here, especially BICEP's. Let's update the legend to call out the years and deterministic/stochastic consistently across all}%\todo{Add our constraints when we have them. Should check that these are the most up-to-date CAST, Lyman-alpha, and DES exclusion limits, but they probably are since I didn't find more up to date papers than PB cited.}
}
\label{fig:exclusion_plot}
\end{figure*}

Because it is challenging to obtain an analytical expression for the likelihood, we empirically estimate the Fisher information using Monte-Carlo methods with $\mathcal O(10^3)$ simulations per grid point in $g_{\phi\gamma}$ for each frequency. The background noise for the simulations is generated using the same method as the timestream-level null-test simulations described in Section~\ref{sec:timestream_nulls}. A detailed treatment of calculating the Fisher information is described in Appendix~\ref{sec:fisher_info}. While there is no analytical expression for the Fisher information, we find that our empirical estimate has a form similar to the analytic results derived in the supplementary material of Ref.~\cite{centers_stochastic_alp_2021}. Using the prior in Eq.~\ref{eq:reference_prior} and likelihood in Eq.~\ref{eq:g_likelihood}, we obtain a stochastic limit at the 95\% confidence level of %95\% upper bound of 
\begin{equation}
    \label{eq:sto_limit}
    g_{\phi\gamma} < 1.1 \times 10^{-11} \left(\frac{m_\phi}{10^{-21}\,\text{eV}}\right)\,\text{GeV}^{-1},
\end{equation}
which is degraded by a factor of $\sim2.8$ compared to the deterministic case in Eq.~\ref{eq:det_limit}. This degradation is expected since there is a non-zero probability that our observation window lies within a relative minimum of the field density due to its stochastic fluctuation. The long tail of the Rayleigh-distributed stochastic amplitude accounts for this effect, causing the posterior for the coupling constant to be stretched and resulting in a weaker limit. Both limits are compared in Fig.~\ref{fig:exclusion_plot}. 

Although much of the parameter space probed here is already constrained by other existing bounds, this work establishes a pipeline for searching large sets of time-sliced, small-angular-scale CMB polarization maps for coherent polarization-angle oscillations. The substantially greater data volume and sensitivity expected from ongoing and future surveys should significantly improve the reach of this approach.

\section{Discussion/Conclusion}
\label{sec:conclusion}
In this work, we searched for time-dependent CMB polarization rotation using approximately 10,000 ACT depth-1 maps spanning over five years of observations. We found no significant evidence for an oscillating signal and placed constraints on the ALP-photon coupling. Our primary result is the upper bound in Eq.~\ref{eq:sto_limit} on the axion-photon coupling constant, obtained after marginalizing over a Rayleigh-distributed local field amplitude. This represents the strongest reported CMB constraint on $g_{\phi\gamma}$ that accounts for the stochastic nature of the local ALP dark matter field. While S22, BXII, and BXIV report deterministic Bayesian limits, and P23 provides the first stochastic constraint using a frequentist approach, our analysis reports a Bayesian stochastic limit. A fully frequentist Neyman construction, as used in P23 \cite{workman_PDG_2022}, would require generating confidence intervals from simulations at many injected signal amplitudes. This is computationally demanding for our data set, which contains roughly 20 times more timestream samples and uses a frequency grid with approximately 6 times more frequencies than P23. We therefore adopt the Bayesian procedure described above, using the Berger-Bernardo reference prior motivated in Ref.~\cite{centers_stochastic_alp_2021}. This choice provides a more appropriate comparison with P23, since the reference prior gives limits that are less sensitive to arbitrary parameterization choices and are more closely aligned with the frequentist treatment \cite{centers_stochastic_alp_2021}. 

%\draftnote{Either argue more clearly why we should use BB prior instead of uniform or remove description of using uniform.} We also tested a uniform prior on $g_{\phi\gamma}$; however, this produced substantially weaker limits than P23, despite the larger data volume used in this work. We therefore interpret the uniform-prior result as being strongly prior-dependent rather than reflecting the constraining power of the data. For this reason, we report the stochastic limit obtained using the Berger-Bernardo reference prior.

Because Bayesian upper limits are strongly sensitive to choice of priors, we highlight the importance of choosing sensible priors when comparing with previous analysis. Previous works such as S22, BXII, and BXIV have reported Bayesian upper limits for only the deterministic case, all of which use uniform priors on $g_{\phi\gamma}$. While our primary result is the stochastic limit in Eq.~\ref{eq:sto_limit}, we include the deterministic limits as a comparison with the previous work. Given the dependence of the results on the choice of priors, we choose the same uniform priors as S22, BXII, and BXIV to provide a fair comparison. Our deterministic limits are slightly stronger than BXIV's and $\sim 3$ times weaker than SPT's. %While our timestream in Fig.~\ref{fig:timestream} has more datapoints than the timestream from S22, the noisy nature of depth-1 maps has resulted in large uncertainties causing our limits to be weaker. 

Unlike the Planck washout constraint from F19, the strength of the constraints on $g_{\phi\gamma}$ is not limited by cosmic variance and is expected to improve with more observations. Following the decommissioning of ACT, new ground based experiments including the Simons Observatory (SO) \cite{SO_forecasts, SO_LAT}, the CCAT Observatory \cite{CCAT}, and the South Pole Observatory \cite{spo2026} are expected to collect more data using improved polarization sensitive instruments. The pipeline for the analysis described in this work can easily be adapted for time-sliced (depth-1) maps from these observatories. With this in mind, we expect future bounds on $g_{\phi\gamma}$ to better the Planck washout bound.
%set by F19. 

% \todo{Should expand on the possibility of using this method or adapted versions of this for the SO LAT depth-1 maps. It could also be particularly interesting if the SO SATs could be used to do a measurement using this method or the method of PB/BICEP/SPT3G on a smaller patch in tandem with this method on a larger patch for the SO LAT.}

\begin{acknowledgments}
ZBH acknowledges support from a NASA Space Technology Graduate Research Opportunities Award. SKC acknowledges support from NASA award 80NSSC25K7821. SN acknowledges that this work was supported by a grant from the Simons
Foundation (CCA 918271, PBL).

Support for ACT was through the U.S.~National Science Foundation through awards AST-0408698, AST-0965625, and AST-1440226 for the ACT project, as well as awards PHY-0355328, PHY-0855887 and PHY-1214379. Funding was also provided by Princeton University, the University of Pennsylvania, and a Canada Foundation for Innovation (CFI) award to UBC. ACT operated in the Parque Astron\'omico Atacama in northern Chile under the auspices of the Agencia Nacional de Investigaci\'on y Desarrollo (ANID). The development of multichroic detectors and lenses was supported by NASA grants NNX13AE56G and NNX14AB58G. Detector research at NIST was supported by the NIST Innovations in Measurement Science program. Computing for ACT was performed using the Princeton Research Computing resources at Princeton University, the National Energy Research Scientific Computing Center (NERSC), and the Niagara supercomputer at the SciNet HPC Consortium. SciNet is funded by the CFI under the auspices of Compute Canada, the Government of Ontario, the Ontario Research Fund–Research Excellence, and the University of Toronto. We thank the Republic of Chile for hosting ACT in the northern Atacama, and the local indigenous Licanantay communities whom we follow in observing and learning from the night sky. 

We gratefully acknowledge the many publicly available software packages that were essential for parts of this analysis. They include numpy \cite{harris_numpy_2020}, scipy \cite{2020SciPy-NMeth}, matplotlib \cite{Hunter_matplotlib_2007}, camb \cite{Lewis_camb}, healpy \cite{zonca_healpy_2019}, HEALPix \cite{gorski/etal:2005}, and pixell.

\end{acknowledgments}

\appendix

\section{Estimator and Covariance Derivations}
\label{sec:derivations}
\subsection{Estimator Derivation}
\label{sec:estimator_deriv}
The derivation of our power spectrum angle estimator (Eq.~\ref{eq:estimator}) is as follows:
The rotation of the E and B coefficients by a nonzero EB mixing angle is
\begin{align}
    a_{\ell}^{\prime E} &= \quad \cos{(2\theta)} a_{\ell}^{E} + \sin{(2\theta)} a_{\ell}^{B} \\
    a_{\ell}^{\prime B} &= -\sin{(2\theta)} a_{\ell}^{E} + \cos{(2\theta)} a_{\ell}^{B}
\end{align}
and the conversion to the power spectrum is
\begin{equation} 
    C_{\ell}^{EE} = \langle a_{\ell}^{\prime E} a_{\ell}^{\prime E} \rangle
\end{equation}
where E can be replaced by T, B, Q, or U as needed.

Expanding out the two power spectra from our estimator in terms of the unrotated (unprimed) coefficients,
\begin{widetext}
\begin{align*}
    C_{\ell}^{' E_{1} \times B_{2}} = &\langle a_{\ell}^{\prime E_{1}} a_{\ell}^{\prime B_{2}} \rangle \\
    = &\langle [\cos{(2\theta_1)} a_{\ell}^{E_1} + \sin{(2\theta_1)} a_{\ell}^{B_1}] [-\sin{(2\theta_2)} a_{\ell}^{E_2} + \cos{(2\theta_2)} a_{\ell}^{B_2}]  \rangle \\
    = &\langle -\sin{(2\theta_2)} \cos{(2\theta_1)} a_{\ell}^{E_1} a_{\ell}^{E_2} - \sin{(2\theta_1)} \sin{(2\theta_2)} a_{\ell}^{E_2} a_{\ell}^{B_1} + \\ & \quad \cos{(2\theta_1)} \cos{(2\theta_2)} a_{\ell}^{E_1} a_{\ell}^{B_2} + \sin{(2\theta_1)} \cos{(2\theta_2)} a_{\ell}^{B_1} a_{\ell}^{B_2} \rangle
\end{align*}
and
\begin{align*}
    C_{\ell}^{' E_{2} \times B_{1}} = &\langle a_{\ell}^{\prime E_{2}} a_{\ell}^{\prime B_{1}} \rangle \\
    = &\langle [\cos{(2\theta_2)} a_{\ell}^{E_2} + \sin{(2\theta_2)} a_{\ell}^{B_2}] [-\sin{(2\theta_1)} a_{\ell}^{E_1} + \cos{(2\theta_1)} a_{\ell}^{B_1}]  \rangle \\
    = &\langle -\sin{(2\theta_1)}\cos{(2\theta_2)} a_{\ell}^{E_2} a_{\ell}^{E_1} - \sin{(2\theta_2)} \sin{(2\theta_1)} a_{\ell}^{E_1} a_{\ell}^{B_2} + \\ & \cos{(2\theta_2)} \cos{(2\theta_1)} a_{\ell}^{E_2} a_{\ell}^{B_1} + \sin{(2\theta_2)} \cos{(2\theta_1)} a_{\ell}^{B_2} a_{\ell}^{B_1} \rangle
\end{align*}
Taking the difference between these two expressions and allowing the angle brackets to convert products of coefficients back into power spectra gives
\begin{align*}
    C_{\ell}^{' E_{1} \times B_{2}} - C_{\ell}^{' E_{2} \times B_{1}} = &-\sin{(2\theta_2)} \cos{(2\theta_1)} C_{\ell}^{E_{1} \times E_{2}} - \sin{(2\theta_1)} \sin{(2\theta_2)} C_{\ell}^{E_{2} \times B_{1}} \\
    &+\cos{(2\theta_1)} \cos{(2\theta_2)} C_{\ell}^{E_{1} \times B_{2}} + \sin{(2\theta_1)} \cos{(2\theta_2)} C_{\ell}^{B_{1} \times B_{2}} \\
    &+\sin{(2\theta_1)} \cos{(2\theta_2)} C_{\ell}^{E_{1} \times E_{2}} + \sin{(2\theta_1)} \sin{(2\theta_2)} C_{\ell}^{E_{1} \times B_{2}} \\
    &-\cos{(2\theta_1)} \cos{(2\theta_2)} C_{\ell}^{E_{2} \times B_{1}} - \sin{(2\theta_2)} \cos{(2\theta_1)} C_{\ell}^{B_{1} \times B_{2}}.
\end{align*}
Combining like terms,
\begin{align*}
    C_{\ell}^{' E_{1} \times B_{2}} - C_{\ell}^{' E_{2} \times B_{1}} = \quad &[\sin{(2\theta_1)} \cos{(2\theta_2)} - \cos{(2\theta_1)} \sin{(2\theta_2)}] C_{\ell}^{E_{1} \times E_{2}} \\
    +&[\sin{(2\theta_1)} \cos{(2\theta_2)} - \cos{(2\theta_1)} \sin{(2\theta_2)}] C_{\ell}^{B_{1} \times B_{2}} \\
    +&[\sin{(2\theta_1)} \sin{(2\theta_2)} + \cos{(2\theta_1)} \cos{(2\theta_2)}] C_{\ell}^{E_{1} \times B_{2}} \\
    -&[\sin{(2\theta_1)} \sin{(2\theta_2)} + \cos{(2\theta_1)} \cos{(2\theta_2)}] C_{\ell}^{E_{2} \times B_{1}}.
\end{align*}
\end{widetext}
Using the difference formulas for sine and cosine, these can be rewritten as

\begin{align*}
    C_{\ell}^{' E_{1} \times B_{2}} - C_{\ell}^{' E_{2} \times B_{1}} = \quad &\sin{[2(\theta_1-\theta_2)]} C_{\ell}^{E_{1} \times E_{2}} \\
    +&\sin{[2(\theta_1-\theta_2)]} C_{\ell}^{B_{1} \times B_{2}} \\
    +&\cos{[2(\theta_1-\theta_2)]} C_{\ell}^{E_{1} \times B_{2}} \\
    -&\cos{[2(\theta_1-\theta_2)]} C_{\ell}^{E_{2} \times B_{1}}.
\end{align*}

These spectra are now unrotated ``primordial" quantities, so we expect $C_{\ell}^{E_{1} \times B_{2}} = C_{\ell}^{E_{2} \times B_{1}} = 0$. In $\Lambda CDM$, $C_{\ell}^{E_{1} \times E_{2}} >> C_{\ell}^{B_{1} \times B_{2}}$, so we neglect $C_{\ell}^{B_{1} \times B_{2}}$.

Under these standard cosmological assumptions,our estimate reduces to Eq.~\ref{eq:estimator} above. Both quantities on the left hand side of the equation can be estimated directly from the depth-1 maps and corresponding full-season coadds, while the right hand side is either the angle that we are trying to measure or the $\Lambda CDM$ prediction for $C_{\ell}^{EE}$.

\subsection{Covariance Derivation}
\label{sec:cov_deriv}

To estimate the uncertainty in this estimator, we construct the covariance matrix as follows.
\begin{align}
    \Theta = \langle &[C_{\ell}^{' E_{1} \times B_{2}} - C_{\ell}^{' E_{2} \times B_{1}} - \langle C_{\ell}^{' E_{1} \times B_{2}} - C_{\ell}^{' E_{2} \times B_{1}} \rangle] \times\nonumber \\
    &[C_{\ell}^{' E_{1} \times B_{2}} - C_{\ell}^{' E_{2} \times B_{1}} - \langle C_{\ell}^{' E_{1} \times B_{2}} - C_{\ell}^{' E_{2} \times B_{1}} \rangle] \rangle
\end{align}

Expanding out the product and rearranging terms, this longer covariance term can be written as the sum of three simpler ones.
\begin{widetext}
\begin{align*}
    \Theta = \langle &C_{\ell}^{' E_{1} \times B_{2}}C_{\ell}^{' E_{1} \times B_{2}} - 2 C_{\ell}^{' E_{1} \times B_{2}}C_{\ell}^{' E_{2} \times B_{1}} + C_{\ell}^{' E_{2} \times B_{1}}C_{\ell}^{' E_{2} \times B_{1}} + \langle C_{\ell}^{' E_{1} \times B_{2}} \rangle^2 - 2 \langle C_{\ell}^{' E_{1} \times B_{2}} \rangle \langle C_{\ell}^{' E_{2} \times B_{1}} \rangle + \langle C_{\ell}^{' E_{2} \times B_{1}} \rangle^2 \\
    & - 2 C_{\ell}^{' E_{1} \times B_{2}} \langle C_{\ell}^{' E_{1} \times B_{2}} \rangle + 2 C_{\ell}^{' E_{1} \times B_{2}} \langle C_{\ell}^{' E_{2} \times B_{1}} \rangle + 2 C_{\ell}^{' E_{2} \times B_{1}} \langle C_{\ell}^{' E_{1} \times B_{2}} \rangle - 2 C_{\ell}^{' E_{2} \times B_{1}} \langle C_{\ell}^{' E_{2} \times B_{1}} \rangle \rangle
\end{align*}

\begin{align*}
    \Theta = \quad & \langle C_{\ell}^{' E_{1} \times B_{2}}C_{\ell}^{' E_{1} \times B_{2}} - 2 C_{\ell}^{' E_{1} \times B_{2}} \langle C_{\ell}^{' E_{1} \times B_{2}} \rangle + \langle C_{\ell}^{' E_{1} \times B_{2}} \rangle^2 \rangle \\
    + &\langle C_{\ell}^{' E_{2} \times B_{1}}C_{\ell}^{' E_{2} \times B_{1}} - 2 C_{\ell}^{' E_{2} \times B_{1}} \langle C_{\ell}^{' E_{2} \times B_{1}} \rangle + \langle C_{\ell}^{' E_{2} \times B_{1}} \rangle^2 \rangle \\
    -2 &\langle C_{\ell}^{' E_{1} \times B_{2}}C_{\ell}^{' E_{2} \times B_{1}} - C_{\ell}^{' E_{1} \times B_{2}} \langle C_{\ell}^{' E_{2} \times B_{1}} \rangle - C_{\ell}^{' E_{2} \times B_{1}} \langle C_{\ell}^{' E_{1} \times B_{2}} \rangle + \langle C_{\ell}^{' E_{1} \times B_{2}} \rangle \langle C_{\ell}^{' E_{2} \times B_{1}} \rangle \rangle
\end{align*}
\end{widetext}
\begin{align*}
    \Theta = \quad & \langle [C_{\ell}^{' E_{1} \times B_{2}} - \langle C_{\ell}^{' E_{1} \times B_{2}} \rangle] \times [C_{\ell}^{' E_{1} \times B_{2}} - \langle C_{\ell}^{' E_{1} \times B_{2}} \rangle] \rangle\\
    + &\langle [C_{\ell}^{' E_{2} \times B_{1}} - \langle C_{\ell}^{' E_{2} \times B_{1}} \rangle] \times [C_{\ell}^{' E_{2} \times B_{1}} - \langle C_{\ell}^{' E_{2} \times B_{1}} \rangle] \rangle \\
    -2 &\langle [C_{\ell}^{' E_{1} \times B_{2}} - \langle C_{\ell}^{' E_{1} \times B_{2}} \rangle] \times [C_{\ell}^{' E_{2} \times B_{1}} - \langle C_{\ell}^{' E_{2} \times B_{1}} \rangle] \rangle
\end{align*}
Using the notation of Appendix A (the unlabeled equation between A4 and A5) in Ref.~\cite{das_ACT2008ps_2011},
\begin{equation}
    \Theta = \Theta^{E_{1} \times B_{2}; E_{1} \times B_{2}} + \Theta^{E_{2} \times B_{1}; E_{2} \times B_{1}} - 2 \Theta^{E_{1} \times B_{2}; E_{2} \times B_{1}}.
\end{equation}

Each of these expressions for the covariance matrix can be simplified by using equation A7 from Ref.~\onlinecite{das_ACT2008ps_2011}.
\begin{align}
    \Theta = \frac{1}{\nu_b} [&(\langle C_{\ell}^{' E_{1} \times E_{1}} \rangle \langle C_{\ell}^{' B_{2} \times B_{2}} \rangle + \langle C_{\ell}^{' E_{1} \times B_{2}} \rangle^2) \nonumber \\ 
    +&(\langle C_{\ell}^{' E_{2} \times E_{2}} \rangle \langle C_{\ell}^{' B_{1} \times B_{1}} \rangle + \langle C_{\ell}^{' E_{2} \times B_{1}} \rangle^2) \nonumber \\ 
    -2&(\langle C_{\ell}^{' E_{1} \times E_{2}} \rangle \langle C_{\ell}^{' B_{2} \times B_{1}} \rangle + \langle C_{\ell}^{' E_{1} \times B_{1}} \rangle \langle C_{\ell}^{' E_{2} \times B_{2}} \rangle)]
\end{align}
where $\nu_b$ is the effective number of pixels in the annulus of Fourier k-space for bin b. The $\nu_b$ term needs to be corrected for masking and filtering by multiplying by $w_2^2/w_4$ (cf.~equation 17 in Ref.~\cite{hivon_master_ps_2002}), which accounts for both $f_{sky}$ and the inverse variance weighting, and the filtering transfer function, $t_b$, as in the text after Eq. 1 in Ref.~\cite{Choi2020}. Symbolically, the final $\nu_b$ is given by $\nu_b = \nu_{b,i} t_b \frac{w_2^2}{w_4}$.

Any expectation value of a spectrum that contains the same map (any auto-spectrum) will generate a noise term in addition to the signal term, while any cross-spectra should have independent noise and yield only a signal term. All of the quantities here are spectra involving only the rotated maps, so they can be directly calculated from our maps using the flat-sky pseudo-$C_{\ell}$ method.

\subsection{Cross Covariance Derivation}
\label{sec:cross_cov}
Below we calculate the covariance for the inter-array consistency test. Suppose we are calculating the covariance for the difference of estimators between maps $i=1$ and $j=2$, let us denote the estimator for map 1 and map 2 to be $\Delta C_1$ and $\Delta C_2$ respectively. Their covariances will be denoted as $\Theta_1$ and $\Theta_2$ respectively. The covariance for the difference between $\Delta C_1$ and $\Delta C_2$ is
\begin{align*}
    \Theta =\, & \langle \left[(\Delta C_1 - \Delta C_2) - (\langle\Delta C_1\rangle - \langle \Delta C_2\rangle)\right]^2\rangle\\
    =\, & \langle (\Delta C_1 - \langle\Delta C_1\rangle)^2\rangle + \langle (\Delta C_2 - \langle\Delta C_2\rangle)^2\rangle\\
     & -2 \langle (\Delta C_1 - \langle\Delta C_1\rangle)  (\Delta C_1 - \langle\Delta C_1\rangle)\rangle\\
     = \, & \Theta_1 + \Theta_2 -2\Theta_{1\times2},
\end{align*}
where we have defined the cross covariance $\Theta_{1\times2}$ to be $\langle (\Delta C_1 - \langle\Delta C_1\rangle)  (\Delta C_1 - \langle\Delta C_1\rangle)\rangle$. We expand the estimators in the cross covariance using Eq.~\ref{eq:estimator} and apply Wick's Theorem to get
\begin{align*}
    \Theta_{1\times2} =\, & \frac{1}{\nu_b} \big[ C_\ell^{\prime E_1\times E_2}C_\ell^{\prime B_R\times B_R}
                            + C_\ell^{\prime E_1\times B_R}C_\ell^{\prime E_2\times B_R}\\
                            & - \;\, C_\ell^{\prime E_1\times E_R}C_\ell^{\prime B_2\times B_R}
                            - C_\ell^{\prime E_1\times B_2}C_\ell^{\prime E_R\times B_R} \\
                            &- \;\,C_\ell^{\prime E_2\times E_R}C_\ell^{\prime B_1\times B_R}
                            - C_\ell^{\prime E_2\times B_1}C_\ell^{\prime E_R\times B_R} \\
                            &+\;\, C_\ell^{\prime E_R\times E_R}C_\ell^{\prime B_1\times B_2}
                            + C_\ell^{\prime E_R\times B_1}C_\ell^{\prime E_R\times B_2}
                            \big],
\end{align*}
where we use R to denote the reference map that all angles are calculated with respect to. We observe that the largest term in the cross covariance is four orders of magnitude smaller than both $\Theta_1$ and $\Theta_2$. Thus, we omit the cross covariance term entirely in our calculations. 
\section{Estimating Fisher Information}\
\label{sec:fisher_info}
We present the method used to estimate the Fisher information for the stochastic likelihood using Monte-Carlo simulations. The quantity required for the Berger-Bernardo reference prior in Eq.~\ref{eq:reference_prior} is
\begin{equation}
    \mathcal I(g) = \left\langle \left[\frac{\partial}{\partial g}\log \mathcal L(g)\right]^2 \right\rangle,
\end{equation}
where $g\equiv g_{\phi\gamma}$.  The expectation value is taken over repeated realizations of the data generated at fixed $g$. In the following we derive an  expression of the score, $\partial \log\mathcal L/\partial g$, as a function of observed data, which can be evaluated quickly for taking the Monte-Carlo estimate of the expectation value. 

Let $\bm{\phi}$ denote the vector of measured polarization rotation angles in the timestream, with components $\phi_i$, and let the data covariance be
\begin{equation}
    C = \mathrm{diag}(\sigma_1^2,\sigma_2^2,\ldots,\sigma_N^2),
\end{equation}
for independent angle measurements. For a fixed frequency $f$, we use Eq.~\ref{eq:sin_model} as our model and to shorten notation we define
\begin{equation}
    w_i(f) = \sinc(\pi f\Delta t_i),
\end{equation}
where $\Delta t_i=t_i^{\rm end}-t_i^{\rm start}$. Same as the parameterization in Eq.~\ref{eq:reparameterization}, the sinusoidal model can then be parameterized as a linear model into in-phase and quadrature components. Recall that the amplitude of the sinusoidal model is the product of the coupling constant and field amplitude, both of which are random variables in the stochastic case. We can let the field component be factored into in-phase and quadrature components, while the coupling constant remains an overall amplitude:
\begin{equation}
    \langle\theta_i\rangle = \frac{g_{\phi\gamma}}{2}\left[x w_i \sin(2\pi f\langle t_i\rangle) + y w_i \cos(2\pi f\langle t_i\rangle)\right],
\end{equation}
where $x$ is the in phase component and $y$ is the quadrature component with the field amplitude absorbed into the new variables. Equivalently,
\begin{equation}
    \langle\bm{\theta}\rangle = \frac{g_{\phi\gamma}}{2}M\bm\beta,
    \qquad
    \bm\beta = 
    \begin{pmatrix}
        x \\
        y
    \end{pmatrix},
\end{equation}
with the matrix defined as
\begin{equation}
    M =
    \begin{pmatrix}
        w_1 \sin(2\pi f\langle t_1\rangle) & w_1 \cos(2\pi f\langle t_1\rangle) \\
        \vdots & \vdots \\
        w_N \sin(2\pi f\langle t_N\rangle) & w_N \cos(2\pi f\langle t_N\rangle)
    \end{pmatrix}.
\end{equation}
Using these new variables, Eq.~\ref{eq:ts_chi2} can then be written as
\begin{align}
    \chi^2(\bm\beta) &= (\bm\phi-\frac{g_{\phi\gamma}}{2}M\bm\beta)^T C^{-1}(\bm\phi-\frac{g_{\phi\gamma}}{2}M\bm\beta)\\
    &= (\frac{g_{\phi\gamma}}{2})^2\bm\beta^T F\bm\beta -2\frac{g_{\phi\gamma}}{2}\bm\beta^T\bm q + \bm\phi^T C^{-1}\bm\phi,
\end{align}
where
\begin{equation}
    F = M^T C^{-1}M,
    \qquad
    \bm q = M^T C^{-1}\bm\phi.
\end{equation}

We can see that as mentioned in Section~\ref{sec:timestream_nulls}, the likelihood in these parameters indeed forms a multivariate Gaussian as it is possible to complete the square. We then use uniform priors in phase and Rayleigh prior in field amplitude which the joint prior reparameterized in $x,y$ conveniently takes the form
\begin{equation}
    P_{x,y}(x,y)
    =
    \frac{1}{\pi \phi_{\rm DM}^2}
    e^{-(x^2+y^2)/\phi_{\rm DM}^2}.
\end{equation}

We can marginalize over $x$ and $y$ using this Gaussian prior, which is equivalent to marginalizing over $\phi_0$ and $\alpha$. The final likelihood in $g_{\phi\gamma}$ is an integral over the product of two multivariate Gaussians. This yields a log-likelihood of 

\begin{widetext}
\begin{equation}
    \log \mathcal L_{\rm marg}(g)
    =
    -\frac12
    \log\det
    \left(
    I_2+\frac{g^2\phi_{\rm DM}^2}{8}F
    \right)
    +
    \frac12 g^2
    \bm q^T
    \left(
    g^2F+\frac{8}{\phi_{\rm DM}^2}I_2
    \right)^{-1}
    \bm q
    + \mathrm{constants}.
\end{equation}
To compute the score $S$, we take the derivative of the log-likelihood with respect to the coupling constant and find
\begin{equation}
\label{eq:score}
    S(g_{\phi\gamma})
    =
    \frac{g_{\phi\gamma}\phi_{DM}^2}{8}
    \left[
    \bm q^T
    \left(I_2+\frac{g_{\phi\gamma}^2\phi_{DM}^2}{8} F\right)^{-2}
    \bm q
    -
    \mathrm{Tr}
    \left[
    \left(I_2+\frac{g_{\phi\gamma}^2\phi^2_{DM}}{8} F\right)^{-1}F
    \right]
    \right].
\end{equation}
\end{widetext}
The calculation of the score can be done relatively quickly without numerically evaluating the likelihood for each grid point in the amplitude-phase and marginalizing over phase and field amplitude. Furthermore, for each point in $g_{\phi\gamma}$ at a fixed frequency, all factors except for $\bm q$ are independent of the simulated data, thus only needs to be calculated once for all simulations.

We used 1000 background simulations to calculate the Fisher information. For each simulation, a random phase and field amplitude is generated from their respective priors. At each frequency, we iterate through a grid-space of $g_{\phi\gamma}$ and for each $g_{\phi\gamma}$ we inject an ALP signal into the background simulation with that specific coupling constant with the generated phase and field amplitude paired with the background simulation. We use the injected signal with noise as the simulated dataset to calculate the score at the injected $g_{\phi\gamma}$ using Eq.~\ref{eq:score}. We then calculate the variance of the score over simulations which provides the value of the Fisher information at that point. The process is repeated over different $g_{\phi\gamma}$ to get the Fisher information at that frequency and repeat over all frequencies. 

\bibliographystyle{act_titles} % Uses Sigurd's style file to print both journal entries and arXiv links
\bibliography{bibliography}% Produces the bibliography via BibTeX.

@ARTICLE{orlowski2024,
       author = {{Orlowski-Scherer}, John and {Venterea}, Ricco C. and {Battaglia}, Nicholas and {Naess}, Sigurd and {Bhandarkar}, Tanay and {Biermann}, Emily and {Calabrese}, Erminia and {Devlin}, Mark and {Dunkley}, Jo and {Herv{\'\i}as-Caimapo}, Carlos and {Gallardo}, Patricio A. and {Hilton}, Matt and {Hincks}, Adam D. and {Knowles}, Kenda and {Li}, Yaqiong and {McMahon}, Jeffrey J. and {Niemack}, Michael D. and {Page}, Lyman A. and {Partridge}, Bruce and {Salatino}, Maria and {Sievers}, Jonathan and {Sif{\'o}n}, Crist{\'o}bal and {Staggs}, Suzanne and {van Engelen}, Alexander and {Vargas}, Cristian and {Vavagiakis}, Eve M. and {Wollack}, Edward J.},
        title = "{The Atacama Cosmology Telescope: Millimeter Observations of a Population of Asteroids or: ACTeroids}",
      journal = {\apj},
         year = 2024,
        month = apr,
       volume = {964},
       number = {2},
          eid = {138},
        pages = {138},
          doi = {10.3847/1538-4357/ad21fe},
archivePrefix = {arXiv},
       eprint = {2306.05468},
 primaryClass = {astro-ph.EP},
       adsurl = {https://ui.adsabs.harvard.edu/abs/2024ApJ...964..138O}
}

@ARTICLE{biermann2025,
       author = {{Biermann}, Emily and {Li}, Yaqiong and {Naess}, Sigurd and {Choi}, Steve K. and {Clark}, Susan E. and {Devlin}, Mark and {Dunkley}, Jo and {Gallardo}, P.~A. and {Guan}, Yilun and {Foster}, Allen and {Hasselfield}, Matthew and {Herv{\'\i}as-Caimapo}, Carlos and {Hilton}, Matt and {Hincks}, Adam D. and {Ho}, Anna Y.~Q. and {Hood}, John C. and {Huffenberger}, Kevin M. and {Kosowsky}, Arthur and {Niemack}, Michael D. and {Orlowski-Scherer}, John and {Page}, Lyman and {Partridge}, Bruce and {Salatino}, Maria and {Sif{\'o}n}, Crist{\'o}bal and {Staggs}, Suzanne T. and {Vargas}, Cristian and {Wollack}, Edward J.},
        title = "{The Atacama Cosmology Telescope: Systematic Transient Search of Single Observation Maps}",
      journal = {\apj},
         year = 2025,
        month = jun,
       volume = {986},
       number = {1},
          eid = {7},
        pages = {7},
          doi = {10.3847/1538-4357/adce70},
archivePrefix = {arXiv},
       eprint = {2409.08429},
 primaryClass = {astro-ph.HE},
       adsurl = {https://ui.adsabs.harvard.edu/abs/2025ApJ...986....7B}
}

@ARTICLE{kramer2026,
       author = {{Kramer}, Darby M. and {van Engelen}, Alexander and {Qu}, Frank J. and {Cain}, Christopher and {Abril-Cabezas}, Irene and {Bond}, J. Richard and {Gluscevic}, Vera and {Herv{\'\i}as-Caimapo}, Carlos and {MacCrann}, Niall and {Madhavacheril}, Mathew S. and {McMahon}, Jeff and {Namikawa}, Toshiya and {Partridge}, Bruce and {Schaan}, Emmanuel and {Sehgal}, Neelima and {Sif{\'o}n}, Crist{\'o}bal and {Wollack}, Ed},
        title = "{The Atacama Cosmology Telescope: Constraints on the anisotropic screening and birefringence effects with DR6}",
      journal = {arXiv e-prints},
         year = 2026,
        month = aug,
          eid = {arXiv:2608.27458},
        pages = {arXiv:2608.27458},
          doi = {10.48550/arXiv.2608.27458},
archivePrefix = {arXiv},
       eprint = {2608.27458},
 primaryClass = {astro-ph.CO},
       adsurl = {https://ui.adsabs.harvard.edu/abs/2026arXiv260827458K}
}

@ARTICLE{spo2026,
       author = {{Natoli}, T. and {Ahmed}, Z. and {Athreya}, H. and {Austermann}, J.~E. and {Bae}, K. and {Bapat}, A. and {Barron}, D.~R. and {Barry}, P.~S. and {Bender}, A.~N. and {Benson}, B.~A. and {Bleem}, L.~E. and {Carlstrom}, J.~E. and {Cecil}, T.~W. and {Chang}, C.~L. and {Cisneros}, S. and {Coerver}, A. and {Cornelison}, J. and {Crawford}, T.~M. and {Datta}, R. and {Dibert}, K.~R. and {Dominguez}, W. and {Duff}, S.~M. and {Fichman}, K. and {Filippini}, J.~P. and {Gades}, L. and {Gallardo}, P.~A. and {Galli}, S. and {Halverson}, N.~W. and {Hao}, Q. and {Henderson}, S. and {Herbst}, R. and {Holzapfel}, W.~L. and {Hryciuk}, A. and {Hubmayr}, J. and {Jones}, D. and {Kabra}, V. and {Karkare}, K.~S. and {King}, C. and {Koc}, M.~A. and {Kofman}, A.~M. and {Lew}, R.~A. and {Link}, M.~J. and {Lucas}, T.~J. and {Mangu}, A. and {Martsen}, E.~S. and {McMahon}, J.~J. and {Montgomery}, J. and {Nagy}, J.~M. and {Nguyen}, H. and {Novosad}, V. and {Padin}, S. and {Pinsonneault-Marotte}, T. and {Raghunathan}, S. and {Rahlin}, A.~S. and {Reichardt}, C.~L. and {Ruckman}, L. and {Ruhl}, J.~E. and {Sarwar}, M.~S. and {Simon}, S. and {Singh}, R. and {Sobrin}, J.~A. and {Stark}, A.~A. and {Taylor}, Q. and {Tucker}, C. and {Ullom}, J. and {Van Lanen}, J. and {Vieira}, J.~D. and {Vieregg}, A.~G. and {Vissers}, M.~R. and {Wang}, G. and {Wu}, W.~L.~K. and {Yefremenko}, V. and {Yilmaz}, E. and {Yu}, C. and {Zivick}, J.},
        title = "{SPT-3G+: A Cosmic Microwave Background Experiment for the South Pole Telescope}",
      journal = {arXiv e-prints},
         year = 2026,
        month = aug,
          eid = {arXiv:2608.20236},
        pages = {arXiv:2608.20236},
          doi = {10.48550/arXiv.2608.20236},
archivePrefix = {arXiv},
       eprint = {2608.20236},
 primaryClass = {astro-ph.CO},
       adsurl = {https://ui.adsabs.harvard.edu/abs/2026arXiv260820236N}
}

@ARTICLE{DalalKravtsov2022,
       author = {{Dalal}, Neal and {Kravtsov}, Andrey},
        title = "{Excluding fuzzy dark matter with sizes and stellar kinematics of ultrafaint dwarf galaxies}",
      journal = {\prd},
         year = 2022,
        month = sep,
       volume = {106},
       number = {6},
          eid = {063517},
        pages = {063517},
          doi = {10.1103/PhysRevD.106.063517},
archivePrefix = {arXiv},
       eprint = {2203.05750},
 primaryClass = {astro-ph.CO},
       adsurl = {https://ui.adsabs.harvard.edu/abs/2022PhRvD.106f3517D}
}

@ARTICLE{Li2021,
       author = {{Li}, Yaqiong and {Austermann}, Jason E. and {Beall}, James A. and {Bruno}, Sarah Marie and {Choi}, Steve K. and {Cothard}, Nicholas F. and {Crowley}, Kevin T. and {Duff}, Shannon M. and {Ho}, Shuay-Pwu Patty and {Golec}, Joseph E. and {Hilton}, Gene C. and {Hasselfield}, Matthew and {Hubmayr}, Johannes and {Koopman}, Brian J. and {Lungu}, Marius and {McMahon}, Jeff and {Niemack}, Michael D. and {Page}, Lyman A. and {Salatino}, Maria and {Simon}, Sara M. and {Staggs}, Suzanne T. and {Stevens}, Jason R. and {Ullom}, Joel N. and {Vavagiakis}, Eve M. and {Wang}, Yuhan and {Wollack}, Edward J. and {Xu}, Zhilei},
        title = "{In Situ Performance of the Low Frequency Array for Advanced ACTPol}",
      journal = {IEEE Transactions on Applied Superconductivity},
         year = 2021,
        month = aug,
       volume = {31},
       number = {5},
          eid = {TASC.2021},
        pages = {TASC.2021},
          doi = {10.1109/TASC.2021.3063334},
archivePrefix = {arXiv},
       eprint = {2101.02658},
 primaryClass = {astro-ph.IM},
       adsurl = {https://ui.adsabs.harvard.edu/abs/2021ITAS...3163334L}
}

@inproceedings{Ho2017,
author = {Shuay-Pwu Patty Ho and Jason Austermann and James A. Beall and Steve K. Choi and Nicholas F. Cothard and Kevin T. Crowley and Rahul Datta and Mark J. Devlin and Shannon M. Duff and Patricio A. Gallardo and Matthew Hasselfield and Shawn W. Henderson and Gene Hilton and Johannes Hubmayr and Brian J. Koopman and Yaqiong Li and Jeffrey McMahon and Michael D. Niemack and Maria Salatino and Sara M. Simon and Suzanne T. Staggs and Jonathan T. Ward and Joel N. Ullom and Eve M. Vavagiakis and Edward J. Wollack},
title = {{Highly uniform 150 mm diameter multichroic polarimeter array deployed for CMB detection}},
volume = {9914},
booktitle = {Millimeter, Submillimeter, and Far-Infrared Detectors and Instrumentation for Astronomy VIII},
editor = {Wayne S. Holland and Jonas Zmuidzinas},
organization = {International Society for Optics and Photonics},
publisher = {SPIE},
pages = {991418},
year = {2017},
doi = {10.1117/12.2233113},
URL = {https://doi.org/10.1117/12.2233113}
}

@ARTICLE{Choi2018,
       author = {{Choi}, S.~K. and {Austermann}, J. and {Beall}, J.~A. and {Crowley}, K.~T. and {Datta}, R. and {Duff}, S.~M. and {Gallardo}, P.~A. and {Ho}, S.~P. and {Hubmayr}, J. and {Koopman}, B.~J. and {Li}, Y. and {Nati}, F. and {Niemack}, M.~D. and {Page}, L.~A. and {Salatino}, M. and {Simon}, S.~M. and {Staggs}, S.~T. and {Stevens}, J. and {Ullom}, J. and {Wollack}, E.~J.},
        title = "{Characterization of the Mid-Frequency Arrays for Advanced ACTPol}",
      journal = {Journal of Low Temperature Physics},
         year = 2018,
        month = nov,
       volume = {193},
       number = {3-4},
        pages = {267-275},
          doi = {10.1007/s10909-018-1982-4},
archivePrefix = {arXiv},
       eprint = {1711.04841},
 primaryClass = {astro-ph.IM},
       adsurl = {https://ui.adsabs.harvard.edu/abs/2018JLTP..193..267C}
}

@ARTICLE{Henderson2016,
       author = {{Henderson}, S.~W. and {Allison}, R. and {Austermann}, J. and {Baildon}, T. and {Battaglia}, N. and {Beall}, J.~A. and {Becker}, D. and {De Bernardis}, F. and {Bond}, J.~R. and {Calabrese}, E. and {Choi}, S.~K. and {Coughlin}, K.~P. and {Crowley}, K.~T. and {Datta}, R. and {Devlin}, M.~J. and {Duff}, S.~M. and {Dunkley}, J. and {D{\"u}nner}, R. and {van Engelen}, A. and {Gallardo}, P.~A. and {Grace}, E. and {Hasselfield}, M. and {Hills}, F. and {Hilton}, G.~C. and {Hincks}, A.~D. and {Hloẑek}, R. and {Ho}, S.~P. and {Hubmayr}, J. and {Huffenberger}, K. and {Hughes}, J.~P. and {Irwin}, K.~D. and {Koopman}, B.~J. and {Kosowsky}, A.~B. and {Li}, D. and {McMahon}, J. and {Munson}, C. and {Nati}, F. and {Newburgh}, L. and {Niemack}, M.~D. and {Niraula}, P. and {Page}, L.~A. and {Pappas}, C.~G. and {Salatino}, M. and {Schillaci}, A. and {Schmitt}, B.~L. and {Sehgal}, N. and {Sherwin}, B.~D. and {Sievers}, J.~L. and {Simon}, S.~M. and {Spergel}, D.~N. and {Staggs}, S.~T. and {Stevens}, J.~R. and {Thornton}, R. and {Van Lanen}, J. and {Vavagiakis}, E.~M. and {Ward}, J.~T. and {Wollack}, E.~J.},
        title = "{Advanced ACTPol Cryogenic Detector Arrays and Readout}",
      journal = {Journal of Low Temperature Physics},
         year = 2016,
        month = aug,
       volume = {184},
       number = {3-4},
        pages = {772-779},
          doi = {10.1007/s10909-016-1575-z},
archivePrefix = {arXiv},
       eprint = {1510.02809},
 primaryClass = {astro-ph.IM},
       adsurl = {https://ui.adsabs.harvard.edu/abs/2016JLTP..184..772H}
}

@ARTICLE{Chathirathas2025,
       author = {{Chathirathas}, Kierthika and {Schwetz}, Thomas},
        title = "{How light can ALP dark matter be?}",
      journal = {arXiv e-prints},
         year = 2025,
        month = nov,
          eid = {arXiv:2511.15790},
        pages = {arXiv:2511.15790},
          doi = {10.48550/arXiv.2511.15790},
archivePrefix = {arXiv},
       eprint = {2511.15790},
 primaryClass = {hep-ph},
       adsurl = {https://ui.adsabs.harvard.edu/abs/2025arXiv251115790C}
}

@ARTICLE{GarciaGallego2026,
       author = {{Garcia-Gallego}, Olga and {Ir{\v{s}}i{\v{c}}}, Vid and {Viel}, Matteo and {Haehnelt}, Martin G. and {Bolton}, James S.},
        title = "{Post-inflationary axion constraints from the Lyman-$\alpha$ forest}",
      journal = {arXiv e-prints},
         year = 2026,
        month = mar,
          eid = {arXiv:2603.04401},
        pages = {arXiv:2603.04401},
          doi = {10.48550/arXiv.2603.04401},
archivePrefix = {arXiv},
       eprint = {2603.04401},
 primaryClass = {astro-ph.CO},
       adsurl = {https://ui.adsabs.harvard.edu/abs/2026arXiv260304401G}
}

@article{Nadler2026,
doi = {10.3847/2041-8213/ae8b9f},
url = {https://doi.org/10.3847/2041-8213/ae8b9f},
year = {2026},
month = {jul},
publisher = {The American Astronomical Society},
volume = {1006},
number = {2},
pages = {L65},
author = {Nadler, Ethan O. and Amin, Mustafa A. and Wechsler, Risa H. and Delos, M. Sten and Benson, Andrew and Gluscevic, Vera},
title = {Warm, Not Fuzzy: Generalized Ultralight Dark Matter Limits from Milky Way Satellites},
journal = {The Astrophysical Journal Letters}
}

@ARTICLE{Bernardeau2011,
       author = {{Bernardeau}, Francis and {Pitrou}, Cyril and {Uzan}, Jean-Philippe},
        title = "{CMB spectra and bispectra calculations: making the flat-sky approximation rigorous}",
      journal = {\jcap},
         year = 2011,
        month = feb,
       volume = {2011},
       number = {2},
          eid = {015},
        pages = {015},
          doi = {10.1088/1475-7516/2011/02/015},
archivePrefix = {arXiv},
       eprint = {1012.2652},
 primaryClass = {astro-ph.CO},
       adsurl = {https://ui.adsabs.harvard.edu/abs/2011JCAP...02..015B}
}

@ARTICLE{calabrese_actdr6_models_2025,
       author = {{Calabrese}, Erminia and {Hill}, J. Colin and {Jense}, Hidde T. and {La Posta}, Adrien and {Abril-Cabezas}, Irene and {Addison}, Graeme E. and {Ade}, Peter A.~R. and {Aiola}, Simone and {Alford}, Tommy and {Alonso}, David and {Amiri}, Mandana and {An}, Rui and {Atkins}, Zachary and {Austermann}, Jason E. and {Barbavara}, Eleonora and {Barbieri}, Nicola and {Battaglia}, Nicholas and {Battistelli}, Elia Stefano and {Beall}, James A. and {Bean}, Rachel and {Beheshti}, Ali and {Beringue}, Benjamin and {Bhandarkar}, Tanay and {Biermann}, Emily and {Bolliet}, Boris and {Bond}, J. Richard and {Capalbo}, Valentina and {Carrero}, Felipe and {Chen}, Shi-Fan and {Chesmore}, Grace and {Cho}, Hsiao-mei and {Choi}, Steve K. and {Clark}, Susan E. and {Cothard}, Nicholas F. and {Coughlin}, Kevin and {Coulton}, William and {Crichton}, Devin and {Crowley}, Kevin T. and {Darwish}, Omar and {Devlin}, Mark J. and {Dicker}, Simon and {Duell}, Cody J. and {Duff}, Shannon M. and {Duivenvoorden}, Adriaan J. and {Dunkley}, Jo and {Dunner}, Rolando and {Embil Villagra}, Carmen and {Fankhanel}, Max and {Farren}, Gerrit S. and {Ferraro}, Simone and {Foster}, Allen and {Freundt}, Rodrigo and {Fuzia}, Brittany and {Gallardo}, Patricio A. and {Garrido}, Xavier and {Gerbino}, Martina and {Giardiello}, Serena and {Gill}, Ajay and {Givans}, Jahmour and {Gluscevic}, Vera and {Goldstein}, Samuel and {Golec}, Joseph E. and {Gong}, Yulin and {Guan}, Yilun and {Halpern}, Mark and {Harrison}, Ian and {Hasselfield}, Matthew and {He}, Adam and {Healy}, Erin and {Henderson}, Shawn and {Hensley}, Brandon and {Herv{\'\i}as-Caimapo}, Carlos and {Hilton}, Gene C. and {Hilton}, Matt and {Hincks}, Adam D. and {Hlo{\v{z}}ek}, Ren{\'e}e and {Ho}, Shuay-Pwu Patty and {Hood}, John and {Hornecker}, Erika and {Huber}, Zachary B. and {Hubmayr}, Johannes and {Huffenberger}, Kevin M. and {Hughes}, John P. and {Ikape}, Margaret and {Irwin}, Kent and {Isopi}, Giovanni and {Joshi}, Neha and {Keller}, Ben and {Kim}, Joshua and {Knowles}, Kenda and {Koopman}, Brian J. and {Kosowsky}, Arthur and {Kramer}, Darby and {Kusiak}, Aleksandra and {Lagu{\"e}}, Alex and {Lakey}, Victoria and {Lattanzi}, Massimiliano and {Lee}, Eunseong and {Li}, Yaqiong and {Li}, Zack and {Limon}, Michele and {Lokken}, Martine and {Louis}, Thibaut and {Lungu}, Marius and {MacCrann}, Niall and {MacInnis}, Amanda and {Madhavacheril}, Mathew S. and {Maldonado}, Diego and {Maldonado}, Felipe and {Mallaby-Kay}, Maya and {Marques}, Gabriela A. and {van Marrewijk}, Joshiwa and {McCarthy}, Fiona and {McMahon}, Jeff and {Mehta}, Yogesh and {Menanteau}, Felipe and {Moodley}, Kavilan and {Morris}, Thomas W. and {Mroczkowski}, Tony and {Naess}, Sigurd and {Namikawa}, Toshiya and {Nati}, Federico and {Nerval}, Simran K. and {Newburgh}, Laura and {Nicola}, Andrina and {Niemack}, Michael D. and {Nolta}, Michael R. and {Orlowski-Scherer}, John and {Pagano}, Luca and {Page}, Lyman A. and {Pandey}, Shivam and {Partridge}, Bruce and {Perez Sarmiento}, Karen and {Prince}, Heather and {Puddu}, Roberto and {Qu}, Frank J. and {Ragavan}, Damien C. and {Ried Guachalla}, Bernardita and {Rogers}, Keir K. and {Rojas}, Felipe and {Sakuma}, Tai and {Schaan}, Emmanuel and {Schmitt}, Benjamin L. and {Sehgal}, Neelima and {Shaikh}, Shabbir and {Sherwin}, Blake D. and {Sierra}, Carlos and {Sievers}, Jon and {Sif{\'o}n}, Crist{\'o}bal and {Simon}, Sara and {Sonka}, Rita and {Spergel}, David N. and {Staggs}, Suzanne T. and {Storer}, Emilie and {Surrao}, Kristen and {Switzer}, Eric R. and {Tampier}, Niklas and {Thiele}, Leander and {Thornton}, Robert and {Trac}, Hy and {Tucker}, Carole and {Ullom}, Joel and {Vale}, Leila R. and {Van Engelen}, Alexander and {Van Lanen}, Jeff and {Vargas}, Cristian and {Vavagiakis}, Eve M. and {Wagoner}, Kasey and {Wang}, Yuhan and {Wenzl}, Lukas and {Wollack}, Edward J. and {Zheng}, Kaiwen and {The Atacama Cosmology Telescope collaboration}},
        title = "{The Atacama Cosmology Telescope: DR6 constraints on extended cosmological models}",
      journal = {\jcap},
         year = 2025,
        month = nov,
       volume = {2025},
       number = {11},
          eid = {063},
        pages = {063},
          doi = {10.1088/1475-7516/2025/11/063},
archivePrefix = {arXiv},
       eprint = {2503.14454},
 primaryClass = {astro-ph.CO},
       adsurl = {https://ui.adsabs.harvard.edu/abs/2025JCAP...11..063C}
}

@ARTICLE{louis_DR6spectra_params_2025,
       author = {{Louis}, Thibaut and {La Posta}, Adrien and {Atkins}, Zachary and {Jense}, Hidde T. and {Abril-Cabezas}, Irene and {Addison}, Graeme E. and {Ade}, Peter A.~R. and {Aiola}, Simone and {Alford}, Tommy and {Alonso}, David and {Amiri}, Mandana and {An}, Rui and {Austermann}, Jason E. and {Barbavara}, Eleonora and {Battaglia}, Nicholas and {Battistelli}, Elia Stefano and {Beall}, James A. and {Bean}, Rachel and {Beheshti}, Ali and {Beringue}, Benjamin and {Bhandarkar}, Tanay and {Biermann}, Emily and {Bolliet}, Boris and {Bond}, J. Richard and {Calabrese}, Erminia and {Capalbo}, Valentina and {Carrero}, Felipe and {Chen}, Shi-Fan and {Chesmore}, Grace and {Cho}, Hsiao-mei and {Choi}, Steve K. and {Clark}, Susan E. and {Cothard}, Nicholas F. and {Coughlin}, Kevin and {Coulton}, William and {Crichton}, Devin and {Crowley}, Kevin T. and {Darwish}, Omar and {Devlin}, Mark J. and {Dicker}, Simon and {Duell}, Cody J. and {Duff}, Shannon M. and {Duivenvoorden}, Adriaan J. and {Dunkley}, Jo and {Dunner}, Rolando and {Embil Villagra}, Carmen and {Fankhanel}, Max and {Farren}, Gerrit S. and {Ferraro}, Simone and {Foster}, Allen and {Freundt}, Rodrigo and {Fuzia}, Brittany and {Gallardo}, Patricio A. and {Garrido}, Xavier and {Gerbino}, Martina and {Giardiello}, Serena and {Gill}, Ajay and {Givans}, Jahmour and {Gluscevic}, Vera and {Goldstein}, Samuel and {Golec}, Joseph E. and {Gong}, Yulin and {Guan}, Yilun and {Halpern}, Mark and {Harrison}, Ian and {Hasselfield}, Matthew and {Healy}, Erin and {Henderson}, Shawn and {Hensley}, Brandon and {Herv{\'\i}as-Caimapo}, Carlos and {Hill}, J. Colin and {Hilton}, Gene C. and {Hilton}, Matt and {Hincks}, Adam D. and {Hlo{\v{z}}ek}, Ren{\'e}e and {Ho}, Shuay-Pwu Patty and {Hood}, John and {Hornecker}, Erika and {Huber}, Zachary B. and {Hubmayr}, Johannes and {Huffenberger}, Kevin M. and {Hughes}, John P. and {Ikape}, Margaret and {Irwin}, Kent and {Isopi}, Giovanni and {Joshi}, Neha and {Keller}, Ben and {Kim}, Joshua and {Knowles}, Kenda and {Koopman}, Brian J. and {Kosowsky}, Arthur and {Kramer}, Darby and {Kusiak}, Aleksandra and {Lagu{\"e}}, Alex and {Lakey}, Victoria and {Lee}, Eunseong and {Li}, Yaqiong and {Li}, Zack and {Limon}, Michele and {Lokken}, Martine and {Lungu}, Marius and {MacCrann}, Niall and {MacInnis}, Amanda and {Madhavacheril}, Mathew S. and {Maldonado}, Diego and {Maldonado}, Felipe and {Mallaby-Kay}, Maya and {Marques}, Gabriela A. and {van Marrewijk}, Joshiwa and {McCarthy}, Fiona and {McMahon}, Jeff and {Mehta}, Yogesh and {Menanteau}, Felipe and {Moodley}, Kavilan and {Morris}, Thomas W. and {Mroczkowski}, Tony and {Naess}, Sigurd and {Namikawa}, Toshiya and {Nati}, Federico and {Nerval}, Simran K. and {Newburgh}, Laura and {Nicola}, Andrina and {Niemack}, Michael D. and {Nolta}, Michael R. and {Orlowski-Scherer}, John and {Pagano}, Luca and {Page}, Lyman A. and {Pandey}, Shivam and {Partridge}, Bruce and {Perez Sarmiento}, Karen and {Prince}, Heather and {Puddu}, Roberto and {Qu}, Frank J. and {Ragavan}, Damien C. and {Ried Guachalla}, Bernardita and {Rogers}, Keir K. and {Rojas}, Felipe and {Sakuma}, Tai and {Schaan}, Emmanuel and {Schmitt}, Benjamin L. and {Sehgal}, Neelima and {Shaikh}, Shabbir and {Sherwin}, Blake D. and {Sierra}, Carlos and {Sievers}, Jon and {Sif{\'o}n}, Crist{\'o}bal and {Simon}, Sara and {Sonka}, Rita and {Spergel}, David N. and {Staggs}, Suzanne T. and {Storer}, Emilie and {Surrao}, Kristen and {Switzer}, Eric R. and {Tampier}, Niklas and {Thornton}, Robert and {Trac}, Hy and {Tucker}, Carole and {Ullom}, Joel and {Vale}, Leila R. and {Van Engelen}, Alexander and {Van Lanen}, Jeff and {Vargas}, Cristian and {Vavagiakis}, Eve M. and {Wagoner}, Kasey and {Wang}, Yuhan and {Wenzl}, Lukas and {Wollack}, Edward J. and {Zheng}, Kaiwen and {The Atacama Cosmology Telescope collaboration}},
        title = "{The Atacama Cosmology Telescope: DR6 power spectra, likelihoods and {\ensuremath{\Lambda}}CDM parameters}",
      journal = {\jcap},
         year = 2025,
        month = nov,
       volume = {2025},
       number = {11},
          eid = {062},
        pages = {062},
          doi = {10.1088/1475-7516/2025/11/062},
archivePrefix = {arXiv},
       eprint = {2503.14452},
 primaryClass = {astro-ph.CO},
       adsurl = {https://ui.adsabs.harvard.edu/abs/2025JCAP...11..062L}
}

@ARTICLE{naess_actdr6_maps_2025,
       author = {{Naess}, Sigurd and {Guan}, Yilun and {Duivenvoorden}, Adriaan J. and {Hasselfield}, Matthew and {Wang}, Yuhan and {Abril-Cabezas}, Irene and {Addison}, Graeme E. and {Ade}, Peter A.~R. and {Aiola}, Simone and {Alford}, Tommy and {Alonso}, David and {Amiri}, Mandana and {An}, Rui and {Atkins}, Zachary and {Austermann}, Jason E. and {Barbavara}, Eleonora and {Battaglia}, Nicholas and {Battistelli}, Elia Stefano and {Beall}, James A. and {Bean}, Rachel and {Beheshti}, Ali and {Beringue}, Benjamin and {Bhandarkar}, Tanay and {Biermann}, Emily and {Bolliet}, Boris and {Bond}, J. Richard and {Calabrese}, Erminia and {Capalbo}, Valentina and {Carrero}, Felipe and {Chen}, Stephen and {Chesmore}, Grace and {Cho}, Hsiao-mei and {Choi}, Steve K. and {Clark}, Susan E. and {Rosado}, Rodrigo Cordova and {Cothard}, Nicholas F. and {Coughlin}, Kevin and {Coulton}, William and {Crichton}, Devin and {Crowley}, Kevin T. and {Devlin}, Mark J. and {Dicker}, Simon and {Duell}, Cody J. and {Duff}, Shannon M. and {Dunkley}, Jo and {Dunner}, Rolando and {Embil Villagra}, Carmen and {Fankhanel}, Max and {Farren}, Gerrit S. and {Ferraro}, Simone and {Foster}, Allen and {Freundt}, Rodrigo and {Fuzia}, Brittany and {Gallardo}, Patricio A. and {Garrido}, Xavier and {Giardiello}, Serena and {Gill}, Ajay and {Givans}, Jahmour and {Gluscevic}, Vera and {Golec}, Joseph E. and {Gong}, Yulin and {Halpern}, Mark and {Harrison}, Ian and {Healy}, Erin and {Henderson}, Shawn and {Hensley}, Brandon and {Herv{\'\i}as-Caimapo}, Carlos and {Hill}, J. Colin and {Hilton}, Gene C. and {Hilton}, Matt and {Hincks}, Adam D. and {Hlo{\v{z}}ek}, Ren{\'e}e and {Ho}, Shuay-Pwu Patty and {Hood}, John and {Hornecker}, Erika and {Huber}, Zachary B. and {Hubmayr}, Johannes and {Huffenberger}, Kevin M. and {Hughes}, John P. and {Ikape}, Margaret and {Irwin}, Kent and {Isopi}, Giovanni and {Jense}, Hidde T. and {Joshi}, Neha and {Keller}, Ben and {Kim}, Joshua and {Knowles}, Kenda and {Koopman}, Brian J. and {Kosowsky}, Arthur and {Kramer}, Darby and {Kusiak}, Aleksandra and {La Posta}, Adrien and {Lagu{\"e}}, Alex and {Lakey}, Victoria and {Lee}, Eunseong and {Li}, Yaqiong and {Li}, Zack and {Limon}, Michele and {Lokken}, Martine and {Louis}, Thibaut and {Lungu}, Marius and {MacCrann}, Niall and {MacInnis}, Amanda and {Madhavacheril}, Mathew S. and {Maldonado}, Diego and {Maldonado}, Felipe and {Mallaby-Kay}, Maya and {Marques}, Gabriela A. and {van Marrewijk}, Joshiwa and {McCarthy}, Fiona and {McMahon}, Jeff and {Mehta}, Yogesh and {Menanteau}, Felipe and {Moodley}, Kavilan and {Morris}, Thomas W. and {Mroczkowski}, Tony and {Namikawa}, Toshiya and {Nati}, Federico and {Nerval}, Simran K. and {Newburgh}, Laura and {Nicola}, Andrina and {Niemack}, Michael D. and {Nolta}, Michael R. and {Orlowski-Scherer}, John and {Page}, Lyman A. and {Pandey}, Shivam and {Partridge}, Bruce and {Perez Sarmiento}, Karen and {Prince}, Heather and {Puddu}, Roberto and {Qu}, Frank J. and {Ragavan}, Damien C. and {Ried Guachalla}, Bernardita and {Rogers}, Keir K. and {Rojas}, Felipe and {Sakuma}, Tai and {Schaan}, Emmanuel and {Schmitt}, Benjamin L. and {Sehgal}, Neelima and {Shaikh}, Shabbir and {Sherwin}, Blake D. and {Sierra}, Carlos and {Sievers}, Jon and {Sif{\'o}n}, Crist{\'o}bal and {Simon}, Sara and {Sonka}, Rita and {London}, Alexander Spencer and {Spergel}, David N. and {Staggs}, Suzanne T. and {Storer}, Emilie and {Surrao}, Kristen and {Switzer}, Eric R. and {Tampier}, Niklas and {Thornton}, Robert and {Trac}, Hy and {Tucker}, Carole and {Ullom}, Joel and {Vale}, Leila R. and {Van Engelen}, Alexander and {Van Lanen}, Jeff and {Vargas}, Cristian and {Vavagiakis}, Eve M. and {Wagoner}, Kasey and {Wenzl}, Lukas and {Wollack}, Edward J. and {Zheng}, Kaiwen and {The Atacama Cosmology Telescope collaboration}},
        title = "{The Atacama Cosmology Telescope: DR6 maps}",
      journal = {\jcap},
         year = 2025,
        month = nov,
       volume = {2025},
       number = {11},
          eid = {061},
        pages = {061},
          doi = {10.1088/1475-7516/2025/11/061},
archivePrefix = {arXiv},
       eprint = {2503.14451},
 primaryClass = {astro-ph.CO},
       adsurl = {https://ui.adsabs.harvard.edu/abs/2025JCAP...11..061N}
}

@ARTICLE{das_ACT2008ps_2011,
       author = {{Das}, Sudeep and {Marriage}, Tobias A. and {Ade}, Peter A.~R. and {Aguirre}, Paula and {Amiri}, Mandana and {Appel}, John W. and {Barrientos}, L. Felipe and {Battistelli}, Elia S. and {Bond}, John R. and {Brown}, Ben and {Burger}, Bryce and {Chervenak}, Jay and {Devlin}, Mark J. and {Dicker}, Simon R. and {Bertrand Doriese}, W. and {Dunkley}, Joanna and {D{\"u}nner}, Rolando and {Essinger-Hileman}, Thomas and {Fisher}, Ryan P. and {Fowler}, Joseph W. and {Hajian}, Amir and {Halpern}, Mark and {Hasselfield}, Matthew and {Hern{\'a}ndez-Monteagudo}, Carlos and {Hilton}, Gene C. and {Hilton}, Matt and {Hincks}, Adam D. and {Hlozek}, Ren{\'e}e and {Huffenberger}, Kevin M. and {Hughes}, David H. and {Hughes}, John P. and {Infante}, Leopoldo and {Irwin}, Kent D. and {Baptiste Juin}, Jean and {Kaul}, Madhuri and {Klein}, Jeff and {Kosowsky}, Arthur and {Lau}, Judy M. and {Limon}, Michele and {Lin}, Yen-Ting and {Lupton}, Robert H. and {Marsden}, Danica and {Martocci}, Krista and {Mauskopf}, Phil and {Menanteau}, Felipe and {Moodley}, Kavilan and {Moseley}, Harvey and {Netterfield}, Calvin B. and {Niemack}, Michael D. and {Nolta}, Michael R. and {Page}, Lyman A. and {Parker}, Lucas and {Partridge}, Bruce and {Reid}, Beth and {Sehgal}, Neelima and {Sherwin}, Blake D. and {Sievers}, Jon and {Spergel}, David N. and {Staggs}, Suzanne T. and {Swetz}, Daniel S. and {Switzer}, Eric R. and {Thornton}, Robert and {Trac}, Hy and {Tucker}, Carole and {Warne}, Ryan and {Wollack}, Ed and {Zhao}, Yue},
        title = "{The Atacama Cosmology Telescope: A Measurement of the Cosmic Microwave Background Power Spectrum at 148 and 218 GHz from the 2008 Southern Survey}",
      journal = {\apj},
         year = 2011,
        month = mar,
       volume = {729},
       number = {1},
          eid = {62},
        pages = {62},
          doi = {10.1088/0004-637X/729/1/62},
archivePrefix = {arXiv},
       eprint = {1009.0847},
 primaryClass = {astro-ph.CO},
       adsurl = {https://ui.adsabs.harvard.edu/abs/2011ApJ...729...62D}
}

@ARTICLE{choi2020,
       author = {{Choi}, Steve K. and {Hasselfield}, Matthew and {Ho}, Shuay-Pwu Patty and {Koopman}, Brian and {Lungu}, Marius and {Abitbol}, Maximilian H. and {Addison}, Graeme E. and {Ade}, Peter A.~R. and {Aiola}, Simone and {Alonso}, David and {Amiri}, Mandana and {Amodeo}, Stefania and {Angile}, Elio and {Austermann}, Jason E. and {Baildon}, Taylor and {Battaglia}, Nick and {Beall}, James A. and {Bean}, Rachel and {Becker}, Daniel T. and {Bond}, J. Richard and {Bruno}, Sarah Marie and {Calabrese}, Erminia and {Calafut}, Victoria and {Campusano}, Luis E. and {Carrero}, Felipe and {Chesmore}, Grace E. and {Cho}, Hsiao-mei and {Clark}, Susan E. and {Cothard}, Nicholas F. and {Crichton}, Devin and {Crowley}, Kevin T. and {Darwish}, Omar and {Datta}, Rahul and {Denison}, Edward V. and {Devlin}, Mark J. and {Duell}, Cody J. and {Duff}, Shannon M. and {Duivenvoorden}, Adriaan J. and {Dunkley}, Jo and {D{\"u}nner}, Rolando and {Essinger-Hileman}, Thomas and {Fankhanel}, Max and {Ferraro}, Simone and {Fox}, Anna E. and {Fuzia}, Brittany and {Gallardo}, Patricio A. and {Gluscevic}, Vera and {Golec}, Joseph E. and {Grace}, Emily and {Gralla}, Megan and {Guan}, Yilun and {Hall}, Kirsten and {Halpern}, Mark and {Han}, Dongwon and {Hargrave}, Peter and {Henderson}, Shawn and {Hensley}, Brandon and {Hill}, J. Colin and {Hilton}, Gene C. and {Hilton}, Matt and {Hincks}, Adam D. and {Hlo{\v{z}}ek}, Ren{\'e}e and {Hubmayr}, Johannes and {Huffenberger}, Kevin M. and {Hughes}, John P. and {Infante}, Leopoldo and {Irwin}, Kent and {Jackson}, Rebecca and {Klein}, Jeff and {Knowles}, Kenda and {Kosowsky}, Arthur and {Lakey}, Vincent and {Li}, Dale and {Li}, Yaqiong and {Li}, Zack and {Lokken}, Martine and {Louis}, Thibaut and {MacInnis}, Amanda and {Madhavacheril}, Mathew and {Maldonado}, Felipe and {Mallaby-Kay}, Maya and {Marsden}, Danica and {Maurin}, Lo{\"\i}c and {McMahon}, Jeff and {Menanteau}, Felipe and {Moodley}, Kavilan and {Morton}, Tim and {Naess}, Sigurd and {Namikawa}, Toshiya and {Nati}, Federico and {Newburgh}, Laura and {Nibarger}, John P. and {Nicola}, Andrina and {Niemack}, Michael D. and {Nolta}, Michael R. and {Orlowski-Sherer}, John and {Page}, Lyman A. and {Pappas}, Christine G. and {Partridge}, Bruce and {Phakathi}, Phumlani and {Prince}, Heather and {Puddu}, Roberto and {Qu}, Frank J. and {Rivera}, Jesus and {Robertson}, Naomi and {Rojas}, Felipe and {Salatino}, Maria and {Schaan}, Emmanuel and {Schillaci}, Alessandro and {Schmitt}, Benjamin L. and {Sehgal}, Neelima and {Sherwin}, Blake D. and {Sierra}, Carlos and {Sievers}, Jon and {Sifon}, Cristobal and {Sikhosana}, Precious and {Simon}, Sara and {Spergel}, David N. and {Staggs}, Suzanne T. and {Stevens}, Jason and {Storer}, Emilie and {Sunder}, Dhaneshwar D. and {Switzer}, Eric R. and {Thorne}, Ben and {Thornton}, Robert and {Trac}, Hy and {Treu}, Jesse and {Tucker}, Carole and {Vale}, Leila R. and {Van Engelen}, Alexander and {Van Lanen}, Jeff and {Vavagiakis}, Eve M. and {Wagoner}, Kasey and {Wang}, Yuhan and {Ward}, Jonathan T. and {Wollack}, Edward J. and {Xu}, Zhilei and {Zago}, Fernando and {Zhu}, Ningfeng},
        title = "{The Atacama Cosmology Telescope: a measurement of the Cosmic Microwave Background power spectra at 98 and 150 GHz}",
      journal = {\jcap},
         year = 2020,
        month = dec,
       volume = {2020},
       number = {12},
          eid = {045},
        pages = {045},
          doi = {10.1088/1475-7516/2020/12/045},
archivePrefix = {arXiv},
       eprint = {2007.07289},
 primaryClass = {astro-ph.CO},
       adsurl = {https://ui.adsabs.harvard.edu/abs/2020JCAP...12..045C}
}

@article{fedderke_axion_2019,
	title = {Axion dark matter detection with {CMB} polarization},
	volume = {100},
	url = {https://link.aps.org/doi/10.1103/PhysRevD.100.015040},
	doi = {10.1103/PhysRevD.100.015040},
	number = {1},
	urldate = {2022-10-24},
	journal = {Phys. Rev. D},
	author = {Fedderke, Michael A. and Graham, Peter W. and Rajendran, Surjeet},
	month = jul,
	year = {2019},
	note = {Publisher: American Physical Society},
	pages = {015040},
}

@article{spt_alp_2022,
	title = {Searching for axionlike time-dependent cosmic birefringence with data from {SPT}-{3G}},
	volume = {106},
	url = {https://link.aps.org/doi/10.1103/PhysRevD.106.042011},
	doi = {10.1103/PhysRevD.106.042011},
	number = {4},
	urldate = {2022-10-24},
	journal = {Phys. Rev. D},
	author = {{SPT-3G Collaboration} and Ferguson, K. R. and Anderson, A. J. and Whitehorn, N. and Ade, P. A. R. and Archipley, M. and Avva, J. S. and Balkenhol, L. and Benabed, K. and Bender, A. N. and Benson, B. A. and Bianchini, F. and Bleem, L. E. and Bouchet, F. R. and Bryant, L. and Camphuis, E. and Carlstrom, J. E. and Cecil, T. W. and Chang, C. L. and Chaubal, P. and Chichura, P. M. and Chou, T.-L. and Crawford, T. M. and Cukierman, A. and Daley, C. and de Haan, T. and Dibert, K. R. and Dobbs, M. A. and Doussot, A. and Dutcher, D. and Everett, W. and Feng, C. and Foster, A. and Galli, S. and Gambrel, A. E. and Gardner, R. W. and Goeckner-Wald, N. and Gualtieri, R. and Guidi, F. and Guns, S. and Halverson, N. W. and Hivon, E. and Holder, G. P. and Holzapfel, W. L. and Hood, J. C. and Huang, N. and Knox, L. and Korman, M. and Kuo, C.-L. and Lee, A. T. and Lowitz, A. E. and Lu, C. and Millea, M. and Montgomery, J. and Natoli, T. and Noble, G. I. and Novosad, V. and Omori, Y. and Padin, S. and Pan, Z. and Paschos, P. and Prabhu, K. and Quan, W. and Rahlin, A. and Reichardt, C. L. and Rouble, M. and Ruhl, J. E. and Schiappucci, E. and Smecher, G. and Sobrin, J. A. and Stephen, J. and Suzuki, A. and Tandoi, C. and Thompson, K. L. and Thorne, B. and Tucker, C. and Umilta, C. and Vieira, J. D. and Wang, G. and Wu, W. L. K. and Yefremenko, V. and Young, M. R.},
	month = aug,
	year = {2022},
	note = {Publisher: American Physical Society},
	pages = {042011},
}

@article{bicepkeck_xiv_2022,
	title = {{BICEP}/\${Keck}\$ {XIV}: {Improved} constraints on axionlike polarization oscillations in the cosmic microwave background},
	volume = {105},
	shorttitle = {{BICEP}/\${Keck}\$ {XIV}},
	url = {https://link.aps.org/doi/10.1103/PhysRevD.105.022006},
	doi = {10.1103/PhysRevD.105.022006},
	number = {2},
	urldate = {2022-10-24},
	journal = {Phys. Rev. D},
	author = {{BICEP/Keck Collaboration} and Ade, P. A. R. and Ahmed, Z. and Amiri, M. and Barkats, D. and Basu Thakur, R. and Bischoff, C. A. and Beck, D. and Bock, J. J. and Boenish, H. and Bullock, E. and Buza, V. and Cheshire, J. R. and Connors, J. and Cornelison, J. and Crumrine, M. and Cukierman, A. and Denison, E. V. and Dierickx, M. and Duband, L. and Eiben, M. and Fatigoni, S. and Filippini, J. P. and Fliescher, S. and Goeckner-Wald, N. and Goldfinger, D. C. and Grayson, J. and Grimes, P. and Hall, G. and Halal, G. and Halpern, M. and Hand, E. and Harrison, S. and Henderson, S. and Hildebrandt, S. R. and Hilton, G. C. and Hubmayr, J. and Hui, H. and Irwin, K. D. and Kang, J. and Karkare, K. S. and Karpel, E. and Kefeli, S. and Kernasovskiy, S. A. and Kovac, J. M. and Kuo, C. L. and Lau, K. and Leitch, E. M. and Lennox, A. and Megerian, K. G. and Minutolo, L. and Moncelsi, L. and Nakato, Y. and Namikawa, T. and Nguyen, H. T. and O’Brient, R. and Ogburn, R. W. and Palladino, S. and Prouve, T. and Pryke, C. and Racine, B. and Reintsema, C. D. and Richter, S. and Schillaci, A. and Schwarz, R. and Schmitt, B. L. and Sheehy, C. D. and Soliman, A. and Germaine, T. St. and Steinbach, B. and Sudiwala, R. V. and Teply, G. P. and Thompson, K. L. and Tolan, J. E. and Tucker, C. and Turner, A. D. and Umiltà, C. and Vergès, C. and Vieregg, A. G. and Wandui, A. and Weber, A. C. and Wiebe, D. V. and Willmert, J. and Wong, C. L. and Wu, W. L. K. and Yang, H. and Yoon, K. W. and Young, E. and Yu, C. and Zeng, L. and Zhang, C. and Zhang, S.},
	month = {Jan},
	year = {2022},
	note = {Publisher: American Physical Society},
	pages = {022006},
}

@article{bicepkeck_xii_2022,
	title = {{BICEP} / {Keck} {XII}: {Constraints} on axion-like polarization oscillations in the cosmic microwave background},
	volume = {103},
	issn = {2470-0010, 2470-0029},
	shorttitle = {{BICEP} / {Keck} {XII}},
	url = {http://arxiv.org/abs/2011.03483},
	doi = {10.1103/PhysRevD.103.042002},
	number = {4},
	urldate = {2022-10-28},
	journal = {Phys. Rev. D},
	author = {Collaboration, BICEP/Keck and Ade, P. A. R. and Ahmed, Z. and Amiri, M. and Barkats, D. and Thakur, R. Basu and Bischoff, C. A. and Bock, J. J. and Boenish, H. and Bullock, E. and Buza, V. and Cheshire IV, J. R. and Connors, J. and Cornelison, J. and Crumrine, M. and Cukierman, A. and Dierickx, M. and Duband, L. and Fatigoni, S. and Filippini, J. P. and Fliescher, S. and Goeckner-Wald, N. and Grayson, J. and Hall, G. and Halpern, M. and Harrison, S. and Henderson, S. and Hildebrandt, S. R. and Hilton, G. C. and Hubmayr, J. and Hui, H. and Irwin, K. D. and Kang, J. and Karkare, K. S. and Karpel, E. and Keating, B. G. and Kefeli, S. and Kernasovskiy, S. A. and Kovac, J. M. and Kuo, C. L. and Lau, K. and Leitch, E. M. and Megerian, K. G. and Moncelsi, L. and Namikawa, T. and Netterfield, C. B. and Nguyen, H. T. and O'Brient, R. and Ogburn IV, R. W. and Palladino, S. and Prouve, T. and Pryke, C. and Racine, B. and Reintsema, C. D. and Richter, S. and Schillaci, A. and Schmitt, B. L. and Schwarz, R. and Sheehy, C. D. and Soliman, A. and Germaine, T. St and Steinbach, B. and Sudiwala, R. V. and Teply, G. and Thompson, K. L. and Tolan, J. E. and Tucker, C. and Turner, A. D. and Umilta, C. and Vieregg, A. G. and Wandui, A. and Weber, A. C. and Wiebe, D. V. and Willmert, J. and Wong, C. L. and Wu, W. L. K. and Yang, H. and Yoon, K. W. and Young, E. and Yu, C. and Zeng, L. and Zhang, C.},
	month = {Feb},
	year = {2021},
	note = {arXiv:2011.03483 [astro-ph, physics:hep-ex, physics:hep-ph]},
	pages = {042002},
}

@article{Hui_fdm_2017,
	title = {Ultralight scalars as cosmological dark matter},
	volume = {95},
	url = {https://link.aps.org/doi/10.1103/PhysRevD.95.043541},
	doi = {10.1103/PhysRevD.95.043541},
	number = {4},
	urldate = {2024-10-03},
	journal = {Phys. Rev. D},
	author = {Hui, Lam and Ostriker, Jeremiah P. and Tremaine, Scott and Witten, Edward},
	month = feb,
	year = {2017},
	note = {Publisher: American Physical Society},
	pages = {043541},
}

@article{Carroll1990,
	title = {Limits on a {Lorentz}- and parity-violating modification of electrodynamics},
	volume = {41},
	url = {https://link.aps.org/doi/10.1103/PhysRevD.41.1231},
	doi = {10.1103/PhysRevD.41.1231},
	number = {4},
	urldate = {2024-10-15},
	journal = {Phys. Rev. D},
	author = {Carroll, Sean M. and Field, George B. and Jackiw, Roman},
	month = feb,
	year = {1990},
	note = {Publisher: American Physical Society},
	pages = {1231--1240},
}

@article{polarbear_axionCMB_2023,
	title = {Constraints on axionlike polarization oscillations in the cosmic microwave background with {POLARBEAR}},
	volume = {108},
	url = {https://link.aps.org/doi/10.1103/PhysRevD.108.043017},
	doi = {10.1103/PhysRevD.108.043017},
	number = {4},
	urldate = {2024-11-08},
	journal = {Phys. Rev. D},
	author = {{POLARBEAR Collaboration} and Adachi, Shunsuke and Adkins, Tylor and Arnold, Kam and Baccigalupi, Carlo and Barron, Darcy and Cheung, Kolen and Chinone, Yuji and Crowley, Kevin T. and Errard, Josquin and Fabbian, Giulio and Feng, Chang and Flauger, Raphael and Fujino, Takuro and Green, Daniel and Hasegawa, Masaya and Hazumi, Masashi and Kaneko, Daisuke and Katayama, Nobuhiko and Keating, Brian and Kusaka, Akito and Lee, Adrian T. and Minami, Yuto and Nishino, Haruki and Reichardt, Christian L. and Siritanasak, Praween and Spisak, Jacob and Tajima, Osamu and Takakura, Satoru and Takatori, Sayuri and Teply, Grant Paul and Yamada, Kyohei},
	month = aug,
	year = {2023},
	note = {Publisher: American Physical Society},
	pages = {043017},
}

@article{marsh_axion_2016,
	series = {Axion cosmology},
	title = {Axion cosmology},
	volume = {643},
	issn = {0370-1573},
	url = {https://www.sciencedirect.com/science/article/pii/S0370157316301557},
	doi = {10.1016/j.physrep.2016.06.005},
	urldate = {2024-11-14},
	journal = {Physics Reports},
	author = {Marsh, David J. E.},
	month = jul,
	year = {2016},
	pages = {1--79},
}

@article{hlozek_cmb_2015,
	title = {A search for ultralight axions using precision cosmological data},
	volume = {91},
	url = {https://link.aps.org/doi/10.1103/PhysRevD.91.103512},
	doi = {10.1103/PhysRevD.91.103512},
	number = {10},
	urldate = {2024-11-14},
	journal = {Phys. Rev. D},
	author = {Hlozek, Renée and Grin, Daniel and Marsh, David J. E. and Ferreira, Pedro G.},
	month = may,
	year = {2015},
	note = {Publisher: American Physical Society},
	pages = {103512},
}

@article{hlozek_cmb_2018,
	title = {Using the full power of the cosmic microwave background to probe axion dark matter},
	volume = {476},
	issn = {0035-8711},
	url = {https://doi.org/10.1093/mnras/sty271},
	doi = {10.1093/mnras/sty271},
	number = {3},
	urldate = {2024-11-14},
	journal = {Monthly Notices of the Royal Astronomical Society},
	author = {Hložek, Renée and Marsh, David J E and Grin, Daniel},
	month = may,
	year = {2018},
	pages = {3063--3085},
}

@ARTICLE{polarbear_taua_2024,
       author = {{Adachi}, Shunsuke and {Adkins}, Tylor and {Baccigalupi}, Carlo and {Chinone}, Yuji and {Crowley}, Kevin T. and {Errard}, Josquin and {Fabbian}, Giulio and {Feng}, Chang and {Fujino}, Takuro and {Hasegawa}, Masaya and {Hazumi}, Masashi and {Jeong}, Oliver and {Kaneko}, Daisuke and {Keating}, Brian and {Kusaka}, Akito and {Lee}, Adrian T. and {Lonappan}, Anto I. and {Minami}, Yuto and {Murata}, Masaaki and {Piccirillo}, Lucio and {Reichardt}, Christian L. and {Siritanasak}, Praween and {Spisak}, Jacob and {Takakura}, Satoru and {Teply}, Grant P. and {Yamada}, Kyohei and {Polarbear Collaboration}},
        title = "{Exploration of the polarization angle variability of the Crab Nebula with POLARBEAR and its application to the search for axionlike particles}",
      journal = {\prd},
         year = 2024,
        month = sep,
       volume = {110},
       number = {6},
          eid = {063013},
        pages = {063013},
          doi = {10.1103/PhysRevD.110.063013},
archivePrefix = {arXiv},
       eprint = {2403.02096},
 primaryClass = {astro-ph.CO},
       adsurl = {https://ui.adsabs.harvard.edu/abs/2024PhRvD.110f3013A}
}

@ARTICLE{payez_sn1987a_gamma_2015,
       author = {{Payez}, Alexandre and {Evoli}, Carmelo and {Fischer}, Tobias and {Giannotti}, Maurizio and {Mirizzi}, Alessandro and {Ringwald}, Andreas},
        title = "{Revisiting the SN1987A gamma-ray limit on ultralight axion-like particles}",
      journal = {\jcap},
         year = 2015,
        month = feb,
       volume = {2015},
       number = {2},
        pages = {006-006},
          doi = {10.1088/1475-7516/2015/02/006},
archivePrefix = {arXiv},
       eprint = {1410.3747},
 primaryClass = {astro-ph.HE},
       adsurl = {https://ui.adsabs.harvard.edu/abs/2015JCAP...02..006P}
}

@article{anastassopoulos_castlimit_2017,
	title = {New {CAST} limit on the axion-photon interaction},
	volume = {13},
	copyright = {2017 The Author(s)},
	issn = {1745-2481},
	url = {https://www.nature.com/articles/nphys4109},
	doi = {10.1038/nphys4109},
	number = {6},
	urldate = {2025-02-05},
	journal = {Nature Phys},
	author = {Anastassopoulos, V. and Aune, S. and Barth, K. and Belov, A. and Bräuninger, H. and Cantatore, G. and Carmona, J. M. and Castel, J. F. and Cetin, S. A. and Christensen, F. and Collar, J. I. and Dafni, T. and Davenport, M. and Decker, T. A. and Dermenev, A. and Desch, K. and Eleftheriadis, C. and Fanourakis, G. and Ferrer-Ribas, E. and Fischer, H. and García, J. A. and Gardikiotis, A. and Garza, J. G. and Gazis, E. N. and Geralis, T. and Giomataris, I. and Gninenko, S. and Hailey, C. J. and Hasinoff, M. D. and Hoffmann, D. H. H. and Iguaz, F. J. and Irastorza, I. G. and Jakobsen, A. and Jacoby, J. and Jakovčić, K. and Kaminski, J. and Karuza, M. and Kralj, N. and Krčmar, M. and Kostoglou, S. and Krieger, Ch. and Lakić, B. and Laurent, J. M. and Liolios, A. and Ljubičić, A. and Luzón, G. and Maroudas, M. and Miceli, L. and Neff, S. and Ortega, I. and Papaevangelou, T. and Paraschou, K. and Pivovaroff, M. J. and Raffelt, G. and Rosu, M. and Ruz, J. and Chóliz, E. Ruiz and Savvidis, I. and Schmidt, S. and Semertzidis, Y. K. and Solanki, S. K. and Stewart, L. and Vafeiadis, T. and Vogel, J. K. and Yildiz, S. C. and Zioutas, K. and {CAST Collaboration}},
	month = jun,
	year = {2017},
	note = {Publisher: Nature Publishing Group},
	pages = {584--590},
}

@article{irsic_alp_lymanalpha_2017,
	title = {First {Constraints} on {Fuzzy} {Dark} {Matter} from {Lyman}-$\alpha$ {Forest} {Data} and {Hydrodynamical} {Simulations}},
	volume = {119},
	url = {https://link.aps.org/doi/10.1103/PhysRevLett.119.031302},
	doi = {10.1103/PhysRevLett.119.031302},
	number = {3},
	urldate = {2025-04-29},
	journal = {Phys. Rev. Lett.},
	author = {Iršič, Vid and Viel, Matteo and Haehnelt, Martin G. and Bolton, James S. and Becker, George D.},
	month = jul,
	year = {2017},
	note = {Publisher: American Physical Society},
	pages = {031302},
}

@article{des_alp_milkywaysatellites_2021,
	title = {Constraints on {Dark} {Matter} {Properties} from {Observations} of {Milky} {Way} {Satellite} {Galaxies}},
	volume = {126},
	url = {https://link.aps.org/doi/10.1103/PhysRevLett.126.091101},
	doi = {10.1103/PhysRevLett.126.091101},
	number = {9},
	urldate = {2025-04-29},
	journal = {Phys. Rev. Lett.},
	author = {{DES Collaboration} and Nadler, E. O. and Drlica-Wagner, A. and Bechtol, K. and Mau, S. and Wechsler, R. H. and Gluscevic, V. and Boddy, K. and Pace, A. B. and Li, T. S. and McNanna, M. and Riley, A. H. and García-Bellido, J. and Mao, Y.-Y. and Green, G. and Burke, D. L. and Peter, A. and Jain, B. and Abbott, T. M. C. and Aguena, M. and Allam, S. and Annis, J. and Avila, S. and Brooks, D. and Carrasco Kind, M. and Carretero, J. and Costanzi, M. and da Costa, L. N. and De Vicente, J. and Desai, S. and Diehl, H. T. and Doel, P. and Everett, S. and Evrard, A. E. and Flaugher, B. and Frieman, J. and Gerdes, D. W. and Gruen, D. and Gruendl, R. A. and Gschwend, J. and Gutierrez, G. and Hinton, S. R. and Honscheid, K. and Huterer, D. and James, D. J. and Krause, E. and Kuehn, K. and Kuropatkin, N. and Lahav, O. and Maia, M. A. G. and Marshall, J. L. and Menanteau, F. and Miquel, R. and Palmese, A. and Paz-Chinchón, F. and Plazas, A. A. and Romer, A. K. and Sanchez, E. and Scarpine, V. and Serrano, S. and Sevilla-Noarbe, I. and Smith, M. and Soares-Santos, M. and Suchyta, E. and Swanson, M. E. C. and Tarle, G. and Tucker, D. L. and Walker, A. R. and Wester, W.},
	month = mar,
	year = {2021},
	note = {Publisher: American Physical Society},
	pages = {091101},
}

@article{sisk-reynes_xray_quasar_2022,
	title = {New constraints on light axion-like particles using {Chandra} transmission grating spectroscopy of the powerful cluster-hosted quasar {H1821}+643},
	volume = {510},
	issn = {0035-8711},
	url = {https://doi.org/10.1093/mnras/stab3464},
	doi = {10.1093/mnras/stab3464},
	number = {1},
	urldate = {2025-04-29},
	journal = {Monthly Notices of the Royal Astronomical Society},
	author = {Sisk-Reynés, Júlia and Matthews, James H and Reynolds, Christopher S and Russell, Helen R and Smith, Robyn N and Marsh, M C David},
	month = feb,
	year = {2022},
	pages = {1264--1277},
}

@ARTICLE{marsh_xray_m87_2017,
       author = {{Marsh}, M.~C. David and {Russell}, Helen R. and {Fabian}, Andrew C. and {McNamara}, Brian R. and {Nulsen}, Paul and {Reynolds}, Christopher S.},
        title = "{A new bound on axion-like particles}",
      journal = {\jcap},
         year = 2017,
        month = dec,
       volume = {2017},
       number = {12},
          eid = {036},
        pages = {036},
          doi = {10.1088/1475-7516/2017/12/036},
archivePrefix = {arXiv},
       eprint = {1703.07354},
 primaryClass = {hep-ph},
       adsurl = {https://ui.adsabs.harvard.edu/abs/2017JCAP...12..036M}
}

@article{xue_ppta_alp_2024,
  title = {{Pulsar Polarization Array Limits on Ultralight Axionlike Dark Matter}},
  author = {Xue, Xiao and Dai, Shi and Luu, Hoang Nhan and Liu, Tao and Ren, Jing and Shu, Jing and Zhao, Yue and Zic, Andrew and Bhat, N. D. Ramesh and Chen, Zu-Cheng and Feng, Yi and Hobbs, George and Kapur, Agastya and Manchester, Richard N. and Mandow, Rami and Mishra, Saurav and Reardon, Daniel J. and Russell, Christopher J. and Shannon, Ryan M. and Wang, Shuangqiang and Zhang, Lei and Zhang, Songbo and Zhu, Xingjiang},
  collaboration = {PPTA Collaboration},
  journal = {Phys. Rev. Lett.},
  volume = {136},
  issue = {1},
  pages = {011001},
  numpages = {8},
  year = {2026},
  month = {Jan},
  publisher = {American Physical Society},
  doi = {10.1103/mptv-3x6g},
  url = {https://link.aps.org/doi/10.1103/mptv-3x6g}
}

@article{cast_axionlimit_2024,
	title = {New {Upper} {Limit} on the {Axion}-{Photon} {Coupling} with an {Extended} {CAST} {Run} with a {Xe}-{Based} {Micromegas} {Detector}},
	volume = {133},
	url = {https://link.aps.org/doi/10.1103/PhysRevLett.133.221005},
	doi = {10.1103/PhysRevLett.133.221005},
	number = {22},
	urldate = {2025-04-29},
	journal = {Phys. Rev. Lett.},
	author = {{CAST Collaboration} and Altenmüller, K. and Anastassopoulos, V. and Arguedas-Cuendis, S. and Aune, S. and Baier, J. and Barth, K. and Bräuninger, H. and Cantatore, G. and Caspers, F. and Castel, J. F. and Çetin, S. A. and Christensen, F. and Cogollos, C. and Dafni, T. and Davenport, M. and Decker, T. A. and Desch, K. and Díez-Ibáñez, D. and Döbrich, B. and Ferrer-Ribas, E. and Fischer, H. and Funk, W. and Galán, J. and García, J. A. and Gardikiotis, A. and Giomataris, I. and Golm, J. and Hailey, C. H. and Hasinoff, M. D. and Hoffmann, D. H. H. and Irastorza, I. G. and Jacoby, J. and Jakobsen, A. C. and Jakovčić, K. and Kaminski, J. and Karuza, M. and Kostoglou, S. and Krieger, C. and Lakić, B. and Laurent, J. M. and Luzón, G. and Malbrunot, C. and Margalejo, C. and Maroudas, M. and Miceli, L. and Mirallas, H. and Navarro, P. and Obis, L. and Özbey, A. and Özbozduman, K. and Papaevangelou, T. and Pérez, O. and Pivovaroff, M. J. and Rosu, M. and Ruiz-Chóliz, E. and Ruz, J. and Schmidt, S. and Schumann, M. and Semertzidis, Y. K. and Solanki, S. K. and Stewart, L. and Vafeiadis, T. and Vogel, J. K. and Zioutas, K.},
	month = nov,
	year = {2024},
	note = {Publisher: American Physical Society},
	pages = {221005},
}

@article{caputo_alp_pulsars_2019,
	title = {Constraints on millicharged dark matter and axionlike particles from timing of radio waves},
	volume = {100},
	url = {https://link.aps.org/doi/10.1103/PhysRevD.100.063515},
	doi = {10.1103/PhysRevD.100.063515},
	number = {6},
	urldate = {2025-04-29},
	journal = {Phys. Rev. D},
	author = {Caputo, Andrea and Sberna, Laura and Frías, Miguel and Blas, Diego and Pani, Paolo and Shao, Lijing and Yan, Wenming},
	month = sep,
	year = {2019},
	note = {Publisher: American Physical Society},
	pages = {063515},
}

@ARTICLE{ivanov_alp_agn_pol_2019,
       author = {{Ivanov}, M.~M. and {Kovalev}, Y.~Y. and {Lister}, M.~L. and {Panin}, A.~G. and {Pushkarev}, A.~B. and {Savolainen}, T. and {Troitsky}, S.~V.},
        title = "{Constraining the photon coupling of ultra-light dark-matter axion-like particles by polarization variations of parsec-scale jets in active galaxies}",
      journal = {\jcap},
         year = 2019,
        month = feb,
       volume = {2019},
       number = {2},
          eid = {059},
        pages = {059},
          doi = {10.1088/1475-7516/2019/02/059},
archivePrefix = {arXiv},
       eprint = {1811.10997},
 primaryClass = {astro-ph.CO},
       adsurl = {https://ui.adsabs.harvard.edu/abs/2019JCAP...02..059I}
}

@ARTICLE{castillo_alp_pulsars_2022,
       author = {{Castillo}, Andr{\'e}s and {Martin-Camalich}, Jorge and {Terol-Calvo}, Jorge and {Blas}, Diego and {Caputo}, Andrea and {G{\'e}nova Santos}, Ricardo Tanaus{\'u} and {Sberna}, Laura and {Peel}, Michael and {Rubi{\~n}o-Mart{\'\i}n}, Jose Alberto},
        title = "{Searching for dark-matter waves with PPTA and QUIJOTE pulsar polarimetry}",
      journal = {\jcap},
         year = 2022,
        month = jun,
       volume = {2022},
       number = {6},
          eid = {014},
        pages = {014},
          doi = {10.1088/1475-7516/2022/06/014},
archivePrefix = {arXiv},
       eprint = {2201.03422},
 primaryClass = {astro-ph.CO},
       adsurl = {https://ui.adsabs.harvard.edu/abs/2022JCAP...06..014C}
}

@article{bar_gal_rotation_curves_2022,
  title = {{Galactic rotation curves versus ultralight dark matter: A systematic comparison with SPARC data}},
  author = {Bar, Nitsan and Blum, Kfir and Sun, Chen},
  journal = {Phys. Rev. D},
  volume = {105},
  issue = {8},
  pages = {083015},
  numpages = {14},
  year = {2022},
  month = {Apr},
  publisher = {American Physical Society},
  doi = {10.1103/PhysRevD.105.083015},
  url = {https://link.aps.org/doi/10.1103/PhysRevD.105.083015}
}

@Article{Hunter_matplotlib_2007,
  Author    = {Hunter, J. D.},
  Title     = {Matplotlib: A 2D graphics environment},
  Journal   = {Computing in Science \& Engineering},
  Volume    = {9},
  Number    = {3},
  Pages     = {90--95},
  publisher = {IEEE COMPUTER SOC},
  doi       = {10.1109/MCSE.2007.55},
  year      = 2007
}

@Article{harris_numpy_2020,
 title         = {Array programming with {NumPy}},
 author        = {Charles R. Harris and K. Jarrod Millman and St{\'{e}}fan J.
                 van der Walt and Ralf Gommers and Pauli Virtanen and David
                 Cournapeau and Eric Wieser and Julian Taylor and Sebastian
                 Berg and Nathaniel J. Smith and Robert Kern and Matti Picus
                 and Stephan Hoyer and Marten H. van Kerkwijk and Matthew
                 Brett and Allan Haldane and Jaime Fern{\'{a}}ndez del
                 R{\'{i}}o and Mark Wiebe and Pearu Peterson and Pierre
                 G{\'{e}}rard-Marchant and Kevin Sheppard and Tyler Reddy and
                 Warren Weckesser and Hameer Abbasi and Christoph Gohlke and
                 Travis E. Oliphant},
 year          = {2020},
 month         = sep,
 journal       = {Nature},
 volume        = {585},
 number        = {7825},
 pages         = {357--362},
 doi           = {10.1038/s41586-020-2649-2},
 publisher     = {Springer Science and Business Media {LLC}},
 url           = {https://doi.org/10.1038/s41586-020-2649-2}
}

@ARTICLE{2020SciPy-NMeth,
  author  = {Virtanen, Pauli and Gommers, Ralf and Oliphant, Travis E. and
            Haberland, Matt and Reddy, Tyler and Cournapeau, David and
            Burovski, Evgeni and Peterson, Pearu and Weckesser, Warren and
            Bright, Jonathan and {van der Walt}, St{\'e}fan J. and
            Brett, Matthew and Wilson, Joshua and Millman, K. Jarrod and
            Mayorov, Nikolay and Nelson, Andrew R. J. and Jones, Eric and
            Kern, Robert and Larson, Eric and Carey, C J and
            Polat, {\.I}lhan and Feng, Yu and Moore, Eric W. and
            {VanderPlas}, Jake and Laxalde, Denis and Perktold, Josef and
            Cimrman, Robert and Henriksen, Ian and Quintero, E. A. and
            Harris, Charles R. and Archibald, Anne M. and
            Ribeiro, Ant{\^o}nio H. and Pedregosa, Fabian and
            {van Mulbregt}, Paul and {SciPy 1.0 Contributors}},
  title   = {{{SciPy} 1.0: Fundamental Algorithms for Scientific
            Computing in Python}},
  journal = {Nature Methods},
  year    = {2020},
  volume  = {17},
  pages   = {261--272},
  adsurl  = {https://rdcu.be/b08Wh},
  doi     = {10.1038/s41592-019-0686-2},
}

@ARTICLE{Lewis_camb,
       author = {{Lewis}, Antony and {Challinor}, Anthony and {Lasenby}, Anthony},
        title = "{Efficient Computation of Cosmic Microwave Background Anisotropies in Closed Friedmann-Robertson-Walker Models}",
      journal = {\apj},
         year = 2000,
        month = aug,
       volume = {538},
       number = {2},
        pages = {473-476},
          doi = {10.1086/309179},
archivePrefix = {arXiv},
       eprint = {astro-ph/9911177},
 primaryClass = {astro-ph}
}

@article{zonca_healpy_2019,
  doi = {10.21105/joss.01298},
  url = {https://doi.org/10.21105/joss.01298},
  year = {2019},
  month = mar,
  publisher = {The Open Journal},
  volume = {4},
  number = {35},
  pages = {1298},
  author = {Andrea Zonca and Leo Singer and Daniel Lenz and Martin Reinecke and Cyrille Rosset and Eric Hivon and Krzysztof Gorski},
  title = {healpy: equal area pixelization and spherical harmonics transforms for data on the sphere in Python},
  journal = {JOSS}}

@ARTICLE{gorski/etal:2005,
       author = {{G{\'o}rski}, K.~M. and {Hivon}, E. and {Banday}, A.~J. and {Wand
        elt}, B.~D. and {Hansen}, F.~K. and {Reinecke}, M. and {Bartelmann}, M.},
        title = "{HEALPix: A Framework for High-Resolution Discretization and Fast Analysis of Data Distributed on the Sphere}",
      journal = {\apj},
         year = "2005",
        month = "Apr",
       volume = {622},
       number = {2},
        pages = {759-771},
          doi = {10.1086/427976},
archivePrefix = {arXiv},
       eprint = {astro-ph/0409513},
 primaryClass = {astro-ph},
       adsurl = {https://ui.adsabs.harvard.edu/abs/2005ApJ...622..759G}
}

@article{namikawa_act_aniso_biref_2020,
	title = {Atacama {Cosmology} {Telescope}: {Constraints} on cosmic birefringence},
	volume = {101},
	shorttitle = {Atacama {Cosmology} {Telescope}},
	url = {https://link.aps.org/doi/10.1103/PhysRevD.101.083527},
	doi = {10.1103/PhysRevD.101.083527},
	number = {8},
	urldate = {2022-10-24},
	journal = {Phys. Rev. D},
	author = {Namikawa, Toshiya and Guan, Yilun and Darwish, Omar and Sherwin, Blake D. and Aiola, Simone and Battaglia, Nicholas and Beall, James A. and Becker, Daniel T. and Bond, J. Richard and Calabrese, Erminia and Chesmore, Grace E. and Choi, Steve K. and Devlin, Mark J. and Dunkley, Joanna and Dünner, Rolando and Fox, Anna E. and Gallardo, Patricio A. and Gluscevic, Vera and Han, Dongwon and Hasselfield, Matthew and Hilton, Gene C. and Hincks, Adam D. and Hložek, Renée and Hubmayr, Johannes and Huffenberger, Kevin and Hughes, John P. and Koopman, Brian J. and Kosowsky, Arthur and Louis, Thibaut and Lungu, Marius and MacInnis, Amanda and Madhavacheril, Mathew S. and Mallaby-Kay, Maya and Maurin, Loïc and McMahon, Jeffrey and Moodley, Kavilan and Naess, Sigurd and Nati, Federico and Newburgh, Laura B. and Nibarger, John P. and Niemack, Michael D. and Page, Lyman A. and Qu, Frank J. and Robertson, Naomi and Schillaci, Alessandro and Sehgal, Neelima and Sifón, Cristóbal and Simon, Sara M. and Spergel, David N. and Staggs, Suzanne T. and Storer, Emilie R. and van Engelen, Alexander and van Lanen, Jeff and Wollack, Edward J.},
	month = apr,
	year = {2020},
	note = {Publisher: American Physical Society},
	pages = {083527},
}

@article{Planck2018_params_2020,
   title={{Planck 2018 Results - VI. Cosmological Parameters}},
   volume={641},
   ISSN={1432-0746},
   url={http://dx.doi.org/10.1051/0004-6361/201833910},
   DOI={10.1051/0004-6361/201833910},
   journal={Astronomy \& Astrophysics},
   publisher={EDP Sciences},
   author={Aghanim, N. and Akrami, Y. and Ashdown, M. and Aumont, J. and Baccigalupi, C. and Ballardini, M. and Banday, A. J. and Barreiro, R. B. and Bartolo, N. and et al.},
   year={2020},
   month={Sep},
   pages={A6}
}

@article{Alam_BOSSDR12clustering_2017,
   title={{The clustering of galaxies in the completed SDSS-III Baryon Oscillation Spectroscopic Survey: cosmological analysis of the DR12 galaxy sample}},
   volume={470},
   ISSN={1365-2966},
   url={http://dx.doi.org/10.1093/mnras/stx721},
   DOI={10.1093/mnras/stx721},
   number={3},
   journal={Monthly Notices of the Royal Astronomical Society},
   publisher={Oxford University Press (OUP)},
   author={Alam, Shadab and Ata, Metin and Bailey, Stephen and Beutler, Florian and Bizyaev, Dmitry and Blazek, Jonathan A. and Bolton, Adam S. and Brownstein, Joel R. and Burden, Angela and Chuang, Chia-Hsun and Comparat, Johan and Cuesta, Antonio J. and Dawson, Kyle S. and et al.},
   year={2017},
   month={Mar},
   pages={2617–2652} }

@article{Hu_FDM_2000,
  title = {{Fuzzy Cold Dark Matter: The Wave Properties of Ultralight Particles}},
  author = {Hu, Wayne and Barkana, Rennan and Gruzinov, Andrei},
  journal = {Phys. Rev. Lett.},
  volume = {85},
  issue = {6},
  pages = {1158--1161},
  numpages = {0},
  year = {2000},
  month = {Aug},
  publisher = {American Physical Society},
  doi = {10.1103/PhysRevLett.85.1158},
  url = {https://link.aps.org/doi/10.1103/PhysRevLett.85.1158}
}

@article{Wilczek_QCDaxion_1978,
  title = {{Problem of Strong $P$ and $T$ Invariance in the Presence of Instantons}},
  author = {Wilczek, F.},
  journal = {Phys. Rev. Lett.},
  volume = {40},
  issue = {5},
  pages = {279--282},
  numpages = {0},
  year = {1978},
  month = {Jan},
  publisher = {American Physical Society},
  doi = {10.1103/PhysRevLett.40.279},
  url = {https://link.aps.org/doi/10.1103/PhysRevLett.40.279}
}

@article{Weinberg_QCDaxion_1978,
  title = {{A New Light Boson?}},
  author = {Weinberg, Steven},
  journal = {Phys. Rev. Lett.},
  volume = {40},
  issue = {4},
  pages = {223--226},
  numpages = {0},
  year = {1978},
  month = {Jan},
  publisher = {American Physical Society},
  doi = {10.1103/PhysRevLett.40.223},
  url = {https://link.aps.org/doi/10.1103/PhysRevLett.40.223}
}

@article{string_axiverse_2010,
  title = {String axiverse},
  author = {Arvanitaki, Asimina and Dimopoulos, Savas and Dubovsky, Sergei and Kaloper, Nemanja and March-Russell, John},
  journal = {Phys. Rev. D},
  volume = {81},
  issue = {12},
  pages = {123530},
  numpages = {23},
  year = {2010},
  month = {Jun},
  publisher = {American Physical Society},
  doi = {10.1103/PhysRevD.81.123530},
  url = {https://link.aps.org/doi/10.1103/PhysRevD.81.123530}
}

@article{Svrcek_Witten_axion_stringtheory_2006,
doi = {10.1088/1126-6708/2006/06/051},
url = {https://dx.doi.org/10.1088/1126-6708/2006/06/051},
year = {2006},
month = {jun},
volume = {2006},
number = {06},
pages = {051},
author = {Peter Svrcek and Edward Witten},
title = {Axions in string theory},
journal = {Journal of High Energy Physics},
}

@article{alps_lightthroughwall_2010,
title = {{New ALPS results on hidden-sector lightweights}},
journal = {Physics Letters B},
volume = {689},
number = {4},
pages = {149-155},
year = {2010},
issn = {0370-2693},
doi = {https://doi.org/10.1016/j.physletb.2010.04.066},
url = {https://www.sciencedirect.com/science/article/pii/S0370269310005526},
author = {Klaus Ehret and Maik Frede and Samvel Ghazaryan and Matthias Hildebrandt and Ernst-Axel Knabbe and Dietmar Kracht and Axel Lindner and Jenny List and Tobias Meier and Niels Meyer and Dieter Notz and Javier Redondo and Andreas Ringwald and Günter Wiedemann and Benno Willke},
}

@article{osqar_lightthoughwall_2015,
  title = {New exclusion limits on scalar and pseudoscalar axionlike particles from light shining through a wall},
  author = {Ballou, R. and Deferne, G. and Finger, M. and Finger, M. and Flekova, L. and Hosek, J. and Kunc, S. and Macuchova, K. and Meissner, K. A. and Pugnat, P. and Schott, M. and Siemko, A. and Slunecka, M. and Sulc, M. and Weinsheimer, C. and Zicha, J.},
  collaboration = {OSQAR Collaboration},
  journal = {Phys. Rev. D},
  volume = {92},
  issue = {9},
  pages = {092002},
  numpages = {6},
  year = {2015},
  month = {Nov},
  publisher = {American Physical Society},
  doi = {10.1103/PhysRevD.92.092002},
  url = {https://link.aps.org/doi/10.1103/PhysRevD.92.092002}
}

@article{admx_alp_cavity_2024,
  title = {{Axion Dark Matter eXperiment: Run 1A analysis details}},
  author = {Boutan, C. and LaRoque, B. H. and Lentz, E. and Oblath, N. S. and Taubman, M. S. and Tedeschi, J. and Yang, J. and Jones, A. M. and Braine, T. and Crisosto, N. and Rosenberg, L. J. and Rybka, G. and Will, D. and Zhang, D. and Kimes, S. and Ottens, R. and Bartram, C. and Bowring, D. and Cervantes, R. and Chou, A. S. and Knirck, S. and Mitchell, D. V. and Sonnenschein, A. and Wester, W. and Khatiwada, R. and Carosi, G. and Du, N. and Durham, S. and O'Kelley, S. R. and Woollett, N. and Duffy, L. D. and Bradley, R. and Clarke, J. and Siddiqi, I. and Agrawal, A. and Dixit, A. V. and Gleason, J. R. and Hipp, A. T. and Jois, S. and Sikivie, P. and Sullivan, N. S. and Tanner, D. B. and Buckley, J. H. and Gaikwad, C. and Henriksen, E. A. and Hoffman, J. and Murch, K. W. and Harrington, P. M. and Daw, E. J. and Perry, M. G. and Hilton, G. C.},
  collaboration = {ADMX Collaboration},
  journal = {Phys. Rev. D},
  volume = {109},
  issue = {1},
  pages = {012009},
  numpages = {26},
  year = {2024},
  month = {Jan},
  publisher = {American Physical Society},
  doi = {10.1103/PhysRevD.109.012009},
  url = {https://link.aps.org/doi/10.1103/PhysRevD.109.012009}
}

@article{ejlli_pvlas_2020,
	series = {The {PVLAS} experiment: {A} 25 year effort to measure vacuum magnetic birefringence},
	title = {The {PVLAS} experiment: {A} 25 year effort to measure vacuum magnetic birefringence},
	volume = {871},
	issn = {0370-1573},
	shorttitle = {The {PVLAS} experiment},
	url = {https://www.sciencedirect.com/science/article/pii/S0370157320302428},
	doi = {10.1016/j.physrep.2020.06.001},
	journal = {Physics Reports},
	author = {Ejlli, A. and Della Valle, F. and Gastaldi, U. and Messineo, G. and Pengo, R. and Ruoso, G. and Zavattini, G.},
	month = aug,
	year = {2020},
	pages = {1--74},
}

@article{vanderplas_lombscargle_2018,
	title = {Understanding the {Lomb}-{Scargle} {Periodogram}},
	volume = {236},
	issn = {0067-0049},
	url = {https://dx.doi.org/10.3847/1538-4365/aab766},
	doi = {10.3847/1538-4365/aab766},
	number = {1},
	urldate = {2025-05-02},
	journal = {ApJS},
	author = {VanderPlas, Jacob T.},
	month = may,
	year = {2018},
	note = {Publisher: The American Astronomical Society},
	pages = {16},
}

@article{hivon_master_ps_2002,
	title = {{MASTER} of the {Cosmic} {Microwave} {Background} {Anisotropy} {Power} {Spectrum}: {A} {Fast} {Method} for {Statistical} {Analysis} of {Large} and {Complex} {Cosmic} {Microwave} {Background} {Data} {Sets}},
	volume = {567},
	issn = {0004-637X},
	shorttitle = {{MASTER} of the {Cosmic} {Microwave} {Background} {Anisotropy} {Power} {Spectrum}},
	url = {https://iopscience.iop.org/article/10.1086/338126/meta},
	doi = {10.1086/338126},
	number = {1},
	urldate = {2025-02-06},
	journal = {ApJ},
	author = {Hivon, Eric and Górski, Krzysztof M. and Netterfield, C. Barth and Crill, Brendan P. and Prunet, Simon and Hansen, Frode},
	month = mar,
	year = {2002},
	note = {Publisher: IOP Publishing},
	pages = {2},
}

@article{centers_stochastic_alp_2021,
	title = {Stochastic fluctuations of bosonic dark matter},
	volume = {12},
	issn = {2041-1723},
	url = {https://www.nature.com/articles/s41467-021-27632-7},
	doi = {10.1038/s41467-021-27632-7},
	number = {1},
	urldate = {2025-05-02},
	journal = {Nature Commununications},
	author = {Centers, Gary P. and Blanchard, John W. and Conrad, Jan and Figueroa, Nataniel L. and Garcon, Antoine and Gramolin, Alexander V. and Kimball, Derek F. Jackson and Lawson, Matthew and Pelssers, Bart and Smiga, Joseph A. and Sushkov, Alexander O. and Wickenbrock, Arne and Budker, Dmitry and Derevianko, Andrei},
	month = dec,
	year = {2021},
	note = {Publisher: Nature Publishing Group},
	pages = {7321},
}

@article{von_mises_2007,
title = {The generalized von Mises distribution},
journal = {Statistical Methodology},
volume = {4},
number = {3},
pages = {341-353},
year = {2007},
issn = {1572-3127},
doi = {https://doi.org/10.1016/j.stamet.2006.11.003},
url = {https://www.sciencedirect.com/science/article/pii/S1572312706000530},
author = {Riccardo Gatto and Sreenivasa Rao Jammalamadaka}
}

@article{workman_PDG_2022,
    author = {Particle Data Group and Workman, R L and Burkert, V D and Crede, V and Klempt, E and Thoma, U and Tiator, L and Agashe, K and Aielli, G and Allanach, B C and Amsler, C and Antonelli, M and Aschenauer, E C and Asner, D M and Baer, H and Banerjee, Sw and Barnett, R M and Baudis, L and Bauer, C W and Beatty, J J and Belousov, V I and Beringer, J and Bettini, A and Biebel, O and Black, K M and Blucher, E and Bonventre, R and Bryzgalov, V V and Buchmuller, O and Bychkov, M A and Cahn, R N and Carena, M and Ceccucci, A and Cerri, A and Chivukula, R Sekhar and Cowan, G and Cranmer, K and Cremonesi, O and D'Ambrosio, G and Damour, T and de Florian, D and de Gouvêa, A and DeGrand, T and de Jong, P and Demers, S and Dobrescu, B A and D'Onofrio, M and Doser, M and Dreiner, H K and Eerola, P and Egede, U and Eidelman, S and El-Khadra, A X and Ellis, J and Eno, S C and Erler, J and Ezhela, V V and Fetscher, W and Fields, B D and Freitas, A and Gallagher, H and Gershtein, Y and Gherghetta, T and Gonzalez-Garcia, M C and Goodman, M and Grab, C and Gritsan, A V and Grojean, C and Groom, D E and Grünewald, M and Gurtu, A and Gutsche, T and Haber, H E and Hamel, Matthieu and Hanhart, C and Hashimoto, S and Hayato, Y and Hebecker, A and Heinemeyer, S and Hernández-Rey, J J and Hikasa, K and Hisano, J and Höcker, A and Holder, J and Hsu, L and Huston, J and Hyodo, T and Ianni, Al and Kado, M and Karliner, M and Katz, U F and Kenzie, M and Khoze, V A and Klein, S R and Krauss, F and Kreps, M and Križan, P and Krusche, B and Kwon, Y and Lahav, O and Laiho, J and Lellouch, L P and Lesgourgues, J and Liddle, A R and Ligeti, Z and Lin, C-J and Lippmann, C and Liss, T M and Littenberg, L and Lourenço, C and Lugovsky, K S and Lugovsky, S B and Lusiani, A and Makida, Y and Maltoni, F and Mannel, T and Manohar, A V and Marciano, W J and Masoni, A and Matthews, J and Meißner, U-G and Melzer-Pellmann, I-A and Mikhasenko, M and Miller, D J and Milstead, D and Mitchell, R E and Mönig, K and Molaro, P and Moortgat, F and Moskovic, M and Nakamura, K and Narain, M and Nason, P and Navas, S and Nelles, A and Neubert, M and Nevski, P and Nir, Y and Olive, K A and Patrignani, C and Peacock, J A and Petrov, V A and Pianori, E and Pich, A and Piepke, A and Pietropaolo, F and Pomarol, A and Pordes, S and Profumo, S and Quadt, A and Rabbertz, K and Rademacker, J and Raffelt, G and Ramsey-Musolf, M and Ratcliff, B N and Richardson, P and Ringwald, A and Robinson, D J and Roesler, S and Rolli, S and Romaniouk, A and Rosenberg, L J and Rosner, J L and Rybka, G and Ryskin, M G and Ryutin, R A and Sakai, Y and Sarkar, S and Sauli, F and Schneider, O and Schönert, S and Scholberg, K and Schwartz, A J and Schwiening, J and Scott, D and Sefkow, F and Seljak, U and Sharma, V and Sharpe, S R and Shiltsev, V and Signorelli, G and Silari, M and Simon, F and Sjöstrand, T and Skands, P and Skwarnicki, T and Smoot, G F and Soffer, A and Sozzi, M S and Spanier, S and Spiering, C and Stahl, A and Stone, S L and Sumino, Y and Syphers, M J and Takahashi, F and Tanabashi, M and Tanaka, J and Taševský, M and Terao, K and Terashi, K and Terning, J and Thorne, R S and Titov, M and Tkachenko, N P and Tovey, D R and Trabelsi, K and Urquijo, P and Valencia, G and Van de Water, R and Varelas, N and Venanzoni, G and Verde, L and Vivarelli, I and Vogel, P and Vogelsang, W and Vorobyev, V and Wakely, S P and Walkowiak, W and Walter, C W and Wands, D and Weinberg, D H and Weinberg, E J and Wermes, N and White, M and Wiencke, L R and Willocq, S and Wohl, C G and Woody, C L and Yao, W-M and Yokoyama, M and Yoshida, R and Zanderighi, G and Zeller, G P and Zenin, O V and Zhu, R-Y and Zhu, Shi-Lin and Zimmermann, F and Zyla, P A},
    title = {Review of Particle Physics},
    journal = {Progress of Theoretical and Experimental Physics},
    volume = {2022},
    number = {8},
    pages = {083C01},
    year = {2022},
    month = {08},
    issn = {2050-3911},
    doi = {10.1093/ptep/ptac097},
    url = {https://doi.org/10.1093/ptep/ptac097},
    eprint = {https://academic.oup.com/ptep/article-pdf/2022/8/083C01/49175539/ptac097.pdf},
}

@article{reference_priors,
author = {James O. Berger and Jos{\'e} M. Bernardo and Dongchu Sun},
title = {{The formal definition of reference priors}},
volume = {37},
journal = {The Annals of Statistics},
number = {2},
publisher = {Institute of Mathematical Statistics},
pages = {905 -- 938},
year = {2009},
doi = {10.1214/07-AOS587},
URL = {https://doi.org/10.1214/07-AOS587}
}

@article{CCAT,
doi = {10.3847/1538-4365/ac9838},
url = {https://doi.org/10.3847/1538-4365/ac9838},
year = {2022},
month = {dec},
publisher = {The American Astronomical Society},
volume = {264},
number = {1},
pages = {7},
author = {CCAT-Prime Collaboration and Aravena, Manuel and Austermann, Jason E. and Basu, Kaustuv and Battaglia, Nicholas and Beringue, Benjamin and Bertoldi, Frank and Bigiel, Frank and Bond, J. Richard and Breysse, Patrick C. and Broughton, Colton and Bustos, Ricardo and Chapman, Scott C. and Charmetant, Maude and Choi, Steve K. and Chung, Dongwoo T. and Clark, Susan E. and Cothard, Nicholas F. and Crites, Abigail T. and Dev, Ankur and Douglas, Kaela and Duell, Cody J. and Dünner, Rolando and Ebina, Haruki and Erler, Jens and Fich, Michel and Fissel, Laura M. and Foreman, Simon and Freundt, R. G. and Gallardo, Patricio A. and Gao, Jiansong and García, Pablo and Giovanelli, Riccardo and Golec, Joseph E. and Groppi, Christopher E. and Haynes, Martha P. and Henke, Douglas and Hensley, Brandon and Herter, Terry and Higgins, Ronan and Hložek, Renée and Huber, Anthony and Huber, Zachary and Hubmayr, Johannes and Jackson, Rebecca and Johnstone, Douglas and Karoumpis, Christos and Keating, Laura C. and Komatsu, Eiichiro and Li, Yaqiong and Magnelli, Benjamin and Matthews, Brenda C. and Mauskopf, Philip D. and McMahon, Jeffrey J. and Meerburg, P. Daniel and Meyers, Joel and Muralidhara, Vyoma and Murray, Norman W. and Niemack, Michael D. and Nikola, Thomas and Okada, Yoko and Puddu, Roberto and Riechers, Dominik A. and Rosolowsky, Erik and Rossi, Kayla and Rotermund, Kaja and Roy, Anirban and Sadavoy, Sarah I. and Schaaf, Reinhold and Schilke, Peter and Scott, Douglas and Simon, Robert and Sinclair, Adrian K. and Sivakoff, Gregory R. and Stacey, Gordon J. and Stutz, Amelia M. and Stutzki, Juergen and Tahani, Mehrnoosh and Thanjavur, Karun and Timmermann, Ralf A. and Ullom, Joel N. and Engelen, Alexander van and Vavagiakis, Eve M. and Vissers, Michael R. and Wheeler, Jordan D. and White, Simon D. M. and Zhu, Yijie and Zou, Bugao},
title = {CCAT-prime Collaboration: Science Goals and Forecasts with Prime-Cam on the Fred Young Submillimeter Telescope},
journal = {The Astrophysical Journal Supplement Series}
}

@ARTICLE{SO_LAT,
       author = {{Abitbol}, M. and {Abril-Cabezas}, I. and {Adachi}, S. and {Ade}, P. and {Adler}, A.~E. and {Agrawal}, P. and {Aguirre}, J. and {Ahmed}, Z. and {Aiola}, S. and {Alford}, T. and {Ali}, A. and {Alonso}, D. and {Alvarez}, M.~A. and {An}, R. and {Arnold}, K. and {Ashton}, P. and {Atkins}, Z. and {Austermann}, J. and {Azzoni}, S. and {Baccigalupi}, C. and {Baleato Lizancos}, A. and {Barron}, D. and {Barry}, P. and {Bartlett}, J. and {Battaglia}, N. and {Battye}, R. and {Baxter}, E. and {Bazarko}, A. and {Beall}, J.~A. and {Bean}, R. and {Beck}, D. and {Beckman}, S. and {Begin}, J. and {Beheshti}, A. and {Beringue}, B. and {Bhandarkar}, T. and {Bhimani}, S. and {Bianchini}, F. and {Biermann}, E. and {Biquard}, S. and {Bixler}, B. and {Boada}, S. and {Boettger}, D. and {Bolliet}, B. and {Bond}, J.~R. and {Borrill}, J. and {Borrow}, J. and {Braithwaite}, C. and {Brien}, T.~L.~R. and {Brown}, M.~L. and {Bruno}, S.~M. and {Bryan}, S. and {Bustos}, R. and {Cai}, H. and {Calabrese}, E. and {Calafut}, V. and {Carl}, F.~M. and {Carones}, A. and {Carron}, J. and {Challinor}, A. and {Chanial}, P. and {Chen}, N. and {Cheung}, K. and {Chiang}, B. and {Chinone}, Y. and {Chluba}, J. and {Cho}, H.~S. and {Choi}, S.~K. and {Chu}, M. and {Clancy}, J. and {Clark}, S.~E. and {Clarke}, P. and {Cleary}, J. and {Clements}, D.~L. and {Connors}, J. and {Contaldi}, C. and {Coppi}, G. and {Corbett}, L. and {Cothard}, N.~F. and {Coulton}, W. and {Crowley}, K.~D. and {Crowley}, K.~T. and {Cukierman}, A. and {D'Ewart}, J.~M. and {Dachlythra}, K. and {Datta}, R. and {Day-Weiss}, S. and {de Haan}, T. and {Devlin}, M. and {Di Mascolo}, L. and {Dicker}, S. and {Dober}, B. and {Doux}, C. and {Dow}, P. and {Doyle}, S. and {Duell}, C.~J. and {Duff}, S.~M. and {Duivenvoorden}, A.~J. and {Dunkley}, J. and {Dutcher}, D. and {D{\"u}nner}, R. and {Edenton}, M. and {El Bouhargani}, H. and {Errard}, J. and {Fabbian}, G. and {Fanfani}, V. and {Farren}, G.~S. and {Fergusson}, J. and {Ferraro}, S. and {Flauger}, R. and {Foster}, A. and {Freese}, K. and {Frisch}, J.~C. and {Frolov}, A. and {Fuller}, G. and {Galitzki}, N. and {Gallardo}, P.~A. and {Galvez Ghersi}, J.~T. and {Ganga}, K. and {Gao}, J. and {Garrido}, X. and {Gawiser}, E. and {Gerbino}, M. and {Gerras}, R. and {Giardiello}, S. and {Gill}, A. and {Gilles}, V. and {Giri}, U. and {Gleave}, E. and {Gluscevic}, V. and {Goeckner-Wald}, N. and {Golec}, J.~E. and {Gordon}, S. and {Gralla}, M. and {Gratton}, S. and {Green}, D. and {Groh}, J.~C. and {Groppi}, C. and {Guan}, Y. and {Gupta}, N. and {Gudmundsson}, J.~E. and {Hagstotz}, S. and {Hargrave}, P. and {Haridas}, S. and {Harrington}, K. and {Harrison}, I. and {Hasegawa}, M. and {Hasselfield}, M. and {Haynes}, V. and {Hazumi}, M. and {He}, A. and {Healy}, E. and {Henderson}, S.~W. and {Hensley}, B.~S. and {Hertig}, E. and {Herv{\'\i}as-Caimapo}, C. and {Higuchi}, M. and {Hill}, C.~A. and {Hill}, J.~C. and {Hilton}, G. and {Hilton}, M. and {Hincks}, A.~D. and {Hinshaw}, G. and {Hlo{\v{z}}ek}, R. and {Ho}, A.~Y.~Q. and {Ho}, S. and {Ho}, S.~P. and {Hoang}, T.~D. and {Hoh}, J. and {Hornecker}, E. and {Hornsby}, A.~L. and {Hotinli}, S.~C. and {Huang}, Z. and {Huber}, Z.~B. and {Hubmayr}, J. and {Huffenberger}, K. and {Hughes}, J.~P. and {Idicherian Lonappan}, A. and {Ikape}, M. and {Irwin}, K. and {Iuliano}, J. and {Jaffe}, A.~H. and {Jain}, B. and {Jense}, H.~T. and {Jeong}, O. and {Johnson}, A. and {Johnson}, B.~R. and {Johnson}, M. and {Jones}, M. and {Jost}, B. and {Kaneko}, D. and {Karpel}, E.~D. and {Kasai}, Y. and {Katayama}, N. and {Keating}, B. and {Keller}, B. and {Keskitalo}, R. and {Kim}, J. and {Kisner}, T. and {Kiuchi}, K.},
        title = "{The Simons Observatory: science goals and forecasts for the enhanced Large Aperture Telescope}",
      journal = {\jcap},
         year = 2025,
        month = aug,
       volume = {2025},
       number = {8},
          eid = {034},
        pages = {034},
          doi = {10.1088/1475-7516/2025/08/034},
archivePrefix = {arXiv},
       eprint = {2503.00636},
 primaryClass = {astro-ph.IM},
       adsurl = {https://ui.adsabs.harvard.edu/abs/2025JCAP...08..034A}
}

@article{SO_forecasts,
doi = {10.3847/1538-4357/ac5e36},
url = {https://doi.org/10.3847/1538-4357/ac5e36},
year = {2022},
month = {apr},
publisher = {The American Astronomical Society},
volume = {929},
number = {2},
pages = {166},
author = {Hensley, Brandon S. and Clark, Susan E. and Fanfani, Valentina and Krachmalnicoff, Nicoletta and Fabbian, Giulio and Poletti, Davide and Puglisi, Giuseppe and Coppi, Gabriele and Nibauer, Jacob and Gerasimov, Roman and Galitzki, Nicholas and Choi, Steve K. and Ashton, Peter C. and Baccigalupi, Carlo and Baxter, Eric and Burkhart, Blakesley and Calabrese, Erminia and Chluba, Jens and Errard, Josquin and Frolov, Andrei V. and Hervías-Caimapo, Carlos and Huffenberger, Kevin M. and Johnson, Bradley R. and Jost, Baptiste and Keating, Brian and McCarrick, Heather and Nati, Federico and Sathyanarayana Rao, Mayuri and van Engelen, Alexander and Walker, Samantha and Wolz, Kevin and Xu, Zhilei and Zhu, Ningfeng and Zonca, Andrea},
title = {The Simons Observatory: Galactic Science Goals and Forecasts},
journal = {The Astrophysical Journal}
}

@ARTICLE{Atkins_2025,
       author = {{Atkins}, Zachary and {Li}, Zack and {Alonso}, David and {Richard Bond}, J. and {Calabrese}, Erminia and {Duivenvoorden}, Adriaan J. and {Dunkley}, Jo and {Giardiello}, Serena and {Herv{\'\i}as-Caimapo}, Carlos and {Colin Hill}, J. and {Jense}, Hidde T. and {Kim}, Joshua and {Louis}, Thibaut and {Moodley}, Kavilan and {Morris}, Thomas W. and {Naess}, Sigurd and {Niemack}, Michael D. and {Page}, Lyman and {Posta}, Adrien La and {Sif{\'o}n}, Crist{\'o}bal and {Wollack}, Edward J.},
        title = "{The Atacama Cosmology Telescope: semi-analytic covariance matrices for the DR6 CMB power spectra}",
      journal = {\jcap},
         year = 2025,
        month = may,
       volume = {2025},
       number = {5},
          eid = {015},
        pages = {015},
          doi = {10.1088/1475-7516/2025/05/015},
archivePrefix = {arXiv},
       eprint = {2412.07068},
 primaryClass = {astro-ph.CO},
       adsurl = {https://ui.adsabs.harvard.edu/abs/2025JCAP...05..015A}
}

@article{Camphuis2022,
	author = {{Camphuis, E.} and {Benabed, K.} and {Galli, S.} and {Hivon, E.} and {Lilley, M.}},
	title = {Accurate cosmic microwave background covariance matrices: Exact calculation and approximations},
	DOI= "10.1051/0004-6361/202243948",
	url= "https://doi.org/10.1051/0004-6361/202243948",
	journal = {A\&A},
	year = 2022,
	volume = 668,
	pages = "A62",
}

\end{document}